\documentclass[12pt,preprint]{aastex}

\newcommand{\bdv}[1]{\mbox{\boldmath$#1$}}

\def\au{{\rm au}} 
 
\def\kms{{\rm km}\,{\rm s}^{-1}}
\def\masyr{{\rm mas}\,{\rm yr}^{-1}}
\def\kpc{{\rm kpc}}
\def\mas{{\rm mas}}

\def\muas{\mu{\rm as}}

\def\max{{\rm max}}
\def\min{{\rm min}}
\def\host{{\rm host}}
\def\rel{{\rm rel}}

\def\hel{{\rm hel}}

\def\e{{\rm E}}
\def\bpi{{\bdv\pi}}
\def\bmu{{\bdv\mu}}

\def\btheta{{\bdv\theta}}

\def\bv{{\bf v}}

\begin{document}
\title{MASADA: From Microlensing Planet Mass-Ratio Function to Planet Mass Function}

\author{\textsc{
Andrew Gould$^{1,2}$}}

\affil{$^{1}$Max-Planck-Institute for Astronomy, K\"{o}nigstuhl 17,
69117 Heidelberg, Germany}

\affil{$^{2}$Department of Astronomy, Ohio State University, 140 W.
18th Ave., Columbus, OH 43210, USA}

\begin{abstract}

Using current technology, gravitational microlensing is the only
method that can measure planet masses over the full parameter
space of planet and stellar-host masses and at a broad range of
planet-host separations.  I present a comprehensive program
to transform the $\sim 150$ planet/host mass ratio measurements
from the first 6 full seasons of the KMTNet survey into planet
mass measurements via late-time adaptive optics (AO) imaging
on 30m-class telescopes.  This program will enable measurements of the 
overall planet mass function, the planet frequency as a function of Galactic 
environment and the planet mass functions within different environments.
I analyze a broad range of discrete
and continuous degeneracies as well as various false positives
and false negatives, and I present a variety of methods to resolve these.
I analyze the propagation from measurement uncertainties to
mass and distance errors and show that these present the greatest
difficulties for host masses $0.13\la(M/M_\odot)\la 0.4$, i.e.,
fully convective stars supported by the ideal gas law, and for very
nearby hosts.  While work can begin later this decade using AO on current
telescopes, of order 90\% of the target sample must await 30m-class AO.
I present extensive tables with information that is useful to
plan observations of more than 100 of these planets and provide
additional notes for a majority of these.  Applying the same
approach to two earlier surveys with 6 and 8 planets, respectively,
I find that 11 of these 14 planets already have mass measurements
by a variety of techniques.  These provide suggestive evidence that
planet frequency may be higher for nearby stars, $D_L\la 4\,\kpc$
compared to those in or near the Galactic bulge.  Finally, I analyze
the prospects for making the planet mass-function measurement for the
case that current astronomical capabilities are seriously degraded.

\end{abstract}


\section{{Introduction}
\label{sec:intro}}

Microlensing is the only technique that can, using existing and
under-construction instruments, routinely deliver accurate planet mass
measurements over a broad range of planetary and host masses and a
broad range of orbital separations.  This statement may seem
surprising because microlensing light-curve analysis routinely returns
only the planet-host mass ratio, $q$, while light-curve based measurements of
the host mass, $M_{\rm host}$, (and so the planet mass $m = q M_{\rm
  host}$), are the rare exceptions.

On the other hand, several other techniques do yield many mass measurements.
Most notably, the transit technique routinely identifies
planet candidates (based on planet/host radius ratio measurements,
together with photometric estimates of the source radius), which then
permit, in favorable cases, spectroscopic followup that measures
the radial velocity (RV) curve and so yields the planet mass (after
taking account of the fact that the transit itself demonstrates that the
orbital inclination is close to $i\sim 90^\circ$).  However, due to
low RV signal, together with the rarity of transits at wide orbital
separations, this method rapidly loses sensitivity for low-mass hosts,
low-mass planets and orbits of more than a few tens of solar radii.

The RV technique can detect planets with periods ranging from of
order a day to several decades, i.e., a factor $\sim 500$ in semi-major
axis, making it the most versatile planet-detection technique in this
sense.  However, the quantity that it directly measures (as with microlensing
light curves) is not $m$ itself, but rather the product of this
quantity with something else, i.e., $m\sin i$, where $i$ is the inclination.
The inclination can be recovered for the small fraction of RV planets
that are transiting (so, $\sin i\simeq 1$, as noted above).  And for any
individual system, $i$ can in principle be measured from astrometric
follow-up observations.  Indeed, it is expected that {\it Gaia} astrometry
will yield $\sin i$ (and so $m$) measurements for some known RV planets,
and that more $m$ measurements will come from RV follow-up of {\it Gaia}
astrometric planet discoveries.   However, given {\it Gaia}'s precision,
these planet-mass measurement will be restricted primarily to Jovian
and super-Jovian planets orbiting F and G stars in one-to-few year orbits.
Unfortunately, while the technology exists to make
much more precise astrometric measurements, which would greatly extend this
technique to lower-mass hosts and planets \citep{unwin08}, there are no
active programs to implement this design.

By contrast, the overwhelming majority of microlensing planet
detections can be transformed into planet-mass measurements by the
``simple'' expedient of imaging the host after it has separated from
the microlensed source on the sky.  The reason that this ``simple''
idea has not been widely implemented is that, using current instruments,
the typical wait time to separately resolve the host is of
order 15 years.  Because only about a half dozen microlensing planets
were discovered as of 15 years ago, the total number of microlensing
planets with mass measurements using this technique is still quite
small, being mainly restricted to favorable cases rather than just the
planets that were discovered first.  Moreover, it would be difficult to
rigorously characterize the sample of early microlensing planets from
a statistical point of view, so that it is not completely clear what
one would learn from simply measuring the masses of all early
microlensing planets.

However, the situation is rapidly changing in three respects.  First,
the rate of planet detections (with good light-curve characterization)
has increased to about 25 per year, beginning with the inauguration
of regular operations of the Korea Microlensing Telescope Network
(KMTNet, \citealt{kmtnet}) experiment.  While regular operations began
in 2016, leading to annual discoveries of about 3000 events from the
EventFinder \citep{eventfinder}, and while microlensing alerts were being
generated in real time by 2018 \citep{alertfinder}, it was only in the last
year that the new AnomalyFinder system \citep{ob191053,af2} began to detect
planets in a uniform way.  In particular, the 2018 season has now been
fully analyzed \citep{2018prime,2018subpr}, and it has yielded 33 planets 
that are suitable for a mass-ratio function analysis.  Preliminary analysis of
other seasons indicates that this is likely to be somewhat above average.
In this paper, I assume a rate of 25 planets per year that will have 
well-characterized light curves.

Second, as just indicated, the planetary sample is being homogeneously 
selected.  In fact, the KMT sample that is the present focus is just the 
latest and largest of three homogeneously selected samples, which are already
being subjected to analyses to determine the planet-host mass-ratio
function \citep{shvartzvald16,suzuki16,zang23}.  
See Section~\ref{sec:other_stat-sample} for more details about earlier surveys.
These mass-ratio function studies have independent importance, not
only because they can be completed much sooner, but also because we do
not yet know whether mass or mass ratio is the more fundamental
quantity in planet formation.  The fact of homogeneous selection
means that once the planet masses are measured, the sample
can be subjected to rigorous statistical analysis for the planet mass
function (and also for planet frequency as a function of host mass
and Galactic environment).

Third, the next generation of telescopes, i.e., 
extremely large telescopes (ELTs), which are already under construction,
will have apertures of 25--39 meters, i.e., about 4 times larger than 
those of current ``large'' telescopes, which are 6.5--10 meters.
This ratio implies that the resolution is 4 times better, so the wait
times will be 4 times shorter.  That is, at adaptive optics (AO) first light
on these telescopes (perhaps 2030), 
the ``typical'' wait time will be about 4 years and
a ``conservative'' wait time will be about twice that, i.e., 8 years.
Hence, the great majority of planets discovered up through this year (2022),
can have mass measurements at ELT AO first light.  According to the above rate
estimate, this implies a sample of about 150 AnomalyFinder planets 2016--2022
(taking account of the fact that 2020 was mostly a ``lost year'').
Assuming that KMTNet continues to operate in a similar mode until 2028,
the sample would roughly double to 300, which could then mostly be accessible
to mass measurements by 2035.

In a substantial majority of cases, a single late-time image in a
single band will be sufficient to make a correct and reasonably precise 
mass measurement.
The technique itself was first applied by \citet{alcock01}, and
a closely related idea was already advanced by \citet{refsdal64}.
It has been successfully applied to 6 planetary events:
OGLE-2005-BLG-071 \citep{ob05071c},
OGLE-2005-BLG-169 \citep{ob05169bat,ob05169ben},
MOA-2007-BLG-400 \citep{mb07400b},
MOA-2009-BLG-319 \citep{mb09319b},
OGLE-2012-BLG-0950 \citep{ob120950b}, and
MOA-2013-BLG-220 \citep{mb13220b}.


However, as I will show, mass application of this technique (in its 
simplest form) will also lead to a significant number of incorrect
measurements, together with detection failures that arise from various
causes.  As I also show, it is straightforward to identify and correct
these incorrect measurements, but this requires additional observing
resources and additional time.  Given that the overall program already
requires substantial observing time on large and (still to be completed)
very large telescopes, it is essential to understand these potential
complications before the observing program is undertaken in earnest.
Similarly, it is essential to conduct the observations in such a way
as to minimize the non-detections of hosts that are (ultimately) detectable.

Here, I systematically address these issues in the context of a program
of mass measurements of planets that have been (or will be)
detected in the KMTNet survey.  Some aspects have been investigated
previously, in which case I present brief summaries and appropriate
references.  Other aspects have not been addressed or have been
investigated only partially.  In any case, the orientation here
is on giving a systematic treatment.

Although the paper is framed in terms of the KMTNet sample,
there are at least four other samples (three historic and one
prospective) to which the same methods and principles could be
applied.  I briefly discuss these as well.

\section{{The Basic Method}
\label{sec:basic}}

The basic method is simply to image the lens (host) and source after
they have moved sufficiently far from each other to separately resolve them.  As mentioned in Section~\ref{sec:intro}, this has
already been done for a total of 6 planetary events, 
using AO on the Keck telescope for all 6 and
also using the {\it Hubble Space Telescope (HST)} for several of these.
In this section, I will treat the ``basic method'' as implying
single-band imaging in some infrared (IR) band, for example, I will assume
$K$ band.

The results of these late-time measurements are combined with the 
microlensing timescale, $t_\e$, 
\begin{equation}
t_\e \equiv {\theta_\e\over \mu_\rel};
\qquad
\theta_\e \equiv \sqrt{\kappa M\pi_\rel};
\qquad
\kappa\equiv {4 G\over c^2\au}\simeq 8.14\,{\mas\over M_\odot},
\label{eqn:tedef}
\end{equation}
which is measured during the event.  Here, $(\pi_\rel,\bmu_\rel)$
are the lens-source relative (parallax, proper motion),
$\mu_\rel\equiv |\bmu_\rel|$, and $M$ is the lens mass.  The essence
of the method is that, at late times, one measures the vector
lens-source separation, $\Delta\btheta$, and then derives
\begin{equation}
\bmu_{\rel,\hel} = {\Delta\btheta\over \Delta t};
\qquad
\Delta t \equiv t_{\rm obs} - t_0,
\label{eqn:bmu_meas}
\end{equation}
where $t_{\rm obs}$ is the time of the observation and $t_0$ is the
time of closest approach of the source to the host during the event.
Note that in this equation, $\bmu_\rel$ is further subscripted by ``hel'',
indicating that what is measured here is the {\it heliocentric} 
proper motion.  By contrast, the timescale measured during the event
(i.e., $t_\e$ without subscript) is the {\it geocentric} timescale.
The heliocentric (with subscript) and geocentric (without subscript)
parameters are related by,
\begin{equation}
\theta_\e\equiv \mu_\rel t_\e = \mu_{\rel,\hel}t_{\e,\hel}
\qquad
\bmu_{\rel,\hel} = \bmu_\rel + {\pi_\rel\over \au}\bv_{\oplus,\perp},
\label{eqn:bmu_hel}
\end{equation}
where $\bv_{\oplus,\perp}$, which is known exactly, is Earth's
velocity relative to the Sun at $t_0$, projected on the sky.


For completeness, I note that Equation~(\ref{eqn:bmu_meas}) is strictly
correct only if the measurement is made at the same time of year as the
peak of the event, $t_0$.  Otherwise, there is a correction of
order $\sim \pi_\rel/\Delta\theta$, which is typically ${\cal O}(10^{-3})$,
i.e., far below the level of other errors in the measurement.  Hence,
although it is no trouble to include this correction in practical measurements,
I ignore it here in the interest of clear exposition.

There are then two measured quantities 
(proper motion $\bmu_{\rel,\hel}$ and the host $K$-band flux)
and two unknowns (the host mass $M_\host$ 
and the lens-source relative parallax $\pi_\rel$).  Ignoring for the
moment the possibility of false positives (see Section~\ref{sec:false_pos}),
the two measurements lead to two equations that relate $M_\host$ to $\pi_\rel$,
and so (in the great majority of cases, see below) to a unique measurement
of these two quantities.  Specifically,
\begin{equation}
(1+q)\kappa M_\host \pi_\rel = 
\biggl(\bmu_{\rel,\hel} - {\pi_\rel\over\au}\bv_{\oplus,\perp}\biggr)^2 t_\e^2,
\label{eqn:mpirel1}
\end{equation}
and
\begin{equation}
K = M_K(M_\host) - 5\log\biggl({\pi_L\over\mas}\biggr) + 10 + 
A_K(\pi_L); 
\qquad
\pi_L = \pi_\rel + \pi_S ,
\label{eqn:mpirel2}
\end{equation}
where $M_K(M_\host)$ is the $K$-band mass-luminosity relation of the host star,
$\pi_S$ is the source parallax, $\pi_L$ is the lens parallax, and 
$A_K(\pi_L)$ is the $K$-band extinction at the lens distance, $D_L = \au/\pi_L$. 
Note that while Equations~(\ref{eqn:mpirel1}) and (\ref{eqn:mpirel2}) contain
3 quantities that are either known exactly ($\bv_{\oplus,\perp}$) or whose
errors are often so small that they can be ignored ($K$ and $\bmu_{\rel,\hel}$),
they also contain four other terms whose uncertainties generally
need to be taken into account.  These are the Einstein timescale $t_\e$,
the source parallax $\pi_S$,  the mass-luminosity relation $M_K(M_\host)$, and
the extinction as a function of distance, $A_K(\pi_L)$.  Note also that I have
included the term ``$(1+q)$'' in Equation~(\ref{eqn:mpirel1}) for 
completeness, although
(because $q\ll 1$ for planets), it can be ignored at the conceptual level.




For any individual event, the usual procedure is to consider all possible
pairs of $(M_\host,\pi_\rel)$ and ask what is the total $\chi^2$ from
Equations~(\ref{eqn:mpirel1}) and (\ref{eqn:mpirel2}), taking
account of the measurement errors in $t_\e$, $\bmu_{\rel,\hel}$ and $K$,
as well as the theoretical uncertainties in $\pi_S$ and in the functions
$M_K(M_\host)$ and $A_K(\pi_\rel)$.  However, here my goal is to investigate
the robustness of the method as a whole and to identify the regions of
parameter space in which it may be degraded or even fail completely.
Thus, I begin by assuming that all measurements errors and theoretical
uncertainties are zero, and I will subsequently relax these assumptions.

By assumption, the host has yielded a measurable $K$-band flux.
This already implies that the host is luminous in $K$, i.e., not a
brown dwarf (BD), white dwarf (WD), neutron star (NS), or 
black hole (BH), for which the method will fail, at least in its
``basic'' form.  I will also ignore for the moment the possibility
(extremely rare in microlensing) that the host is an evolved
star\footnote{Such hosts would likely become targets for spectroscopic
observations that would not generally require extreme telescope resources.
}.  Then, $M_K(M_\host)$ is a monotonically
declining function, which can thus be inverted to 
$M_\host(\pi_L) = M^{-1}_K(\Delta K(\pi_L))$, where 
$\Delta K(\pi_L) = K + 5\log(\pi_L/\mas) - A_K(\pi_L) - 10$.
Because $M^{-1}_K(\Delta K)$ and $\log(\pi_L/\mas)$ are monotonically 
increasing, while $A_K(\pi_L)$ is monotonically decreasing,
$M_\host(\pi_L)$ is monotonically decreasing.  Given my initial approximation
that $\pi_S$ is known, it follows that $M_\host(\pi_\rel)$ is also
monotonically decreasing.  

Equation~(\ref{eqn:mpirel1}) is easily solved for
$M_\host(\pi_\rel)$ by dividing both sides by $(1+q)\kappa\pi_\rel$.
The naive hope would be that the $M_\host$--$\pi_\rel$ curves resulting
from these two equations would intersect in one and only one place.
Following the assumption that the measurements and physical relations
are exact, they must intersect at at least one point, namely the
$(M_\host,\pi_\rel)$ values of the actual host.  However, inspection of
Equation~(\ref{eqn:mpirel1}) shows that it is not the case that this
intersection is unique.
One sees that $M_\host\rightarrow \infty$ in both of the limits of 
$\pi_\rel\rightarrow \infty$ and $\pi_\rel\rightarrow 0$ (respectively,
$D_L\rightarrow 0$ and $D_L\rightarrow D_S$), while $M_\host$ achieves
a minimum $M_{\host,\min}$ at $\pi_{\rel,\min}$, given by
\begin{equation}
\pi_{\rel,\min} = {\mu_{\rel,\hel}\au\over v_{\oplus,\perp}} =
0.95\,\mas\biggl({\mu_{\rel,\hel} \over 6\,\masyr}\biggr)
\biggl({v_{\oplus,\perp} \over 30\,\kms}\biggr)^{-1},
\label{eqn:pirelmin}
\end{equation}
\begin{equation}
M_{\host,\min} = 2{\mu_{\rel,\hel}v_{\oplus,\perp}\over (1+q)\kappa\au}t_\e^2
(1 - \cos\phi) =
0.063\,M_\odot\biggl({\mu_{\rel,\hel} \over 6\,\masyr}\biggr)
\biggl({v_{\oplus,\perp} \over 30\,\kms}\biggr)
\biggl({t_\e \over 30\,{\rm day}}\biggr)^2 (1 - \cos\phi),
\label{eqn:mmin}
\end{equation}
where 
$\cos\phi\equiv \bmu_{\rel,\hel}\cdot\bv_{\oplus,\perp}/\mu_{\rel,\hel}v_{\oplus,\perp}$
and where I have made the final evaluation assuming $q\ll 1$.

For the general case, there will then be an even number of points of 
intersection between the curves derived from Equation~(\ref{eqn:mpirel1}) 
and (\ref{eqn:mpirel2}), and, as a practical matter, this ``even number''
will almost always be ``two''.  We will see that if these two points
are both on the low-$\pi_\rel$ (i.e., $\pi_\rel<\pi_{\rel,\min}$) side of
Equation~(\ref{eqn:mpirel1}), then this can lead to a 
``continuous degeneracy''.  I will address this case in 
Section~\ref{sec:errors}.  Here, I focus on the great majority of cases,
i.e., those with a discrete two-fold degeneracy for which
one solution has $\pi_\rel>\pi_{\rel,\min}$ and the other has 
$\pi_\rel<\pi_{\rel,\min}$.  If there
were no way to resolve this degeneracy, it would seriously undermine the 
``basic method''.  However, there are several approaches to resolving
this degeneracy.   See Sections~\ref{sec:degen} and \ref{sec:errors}.

\begin{figure}
\plotone{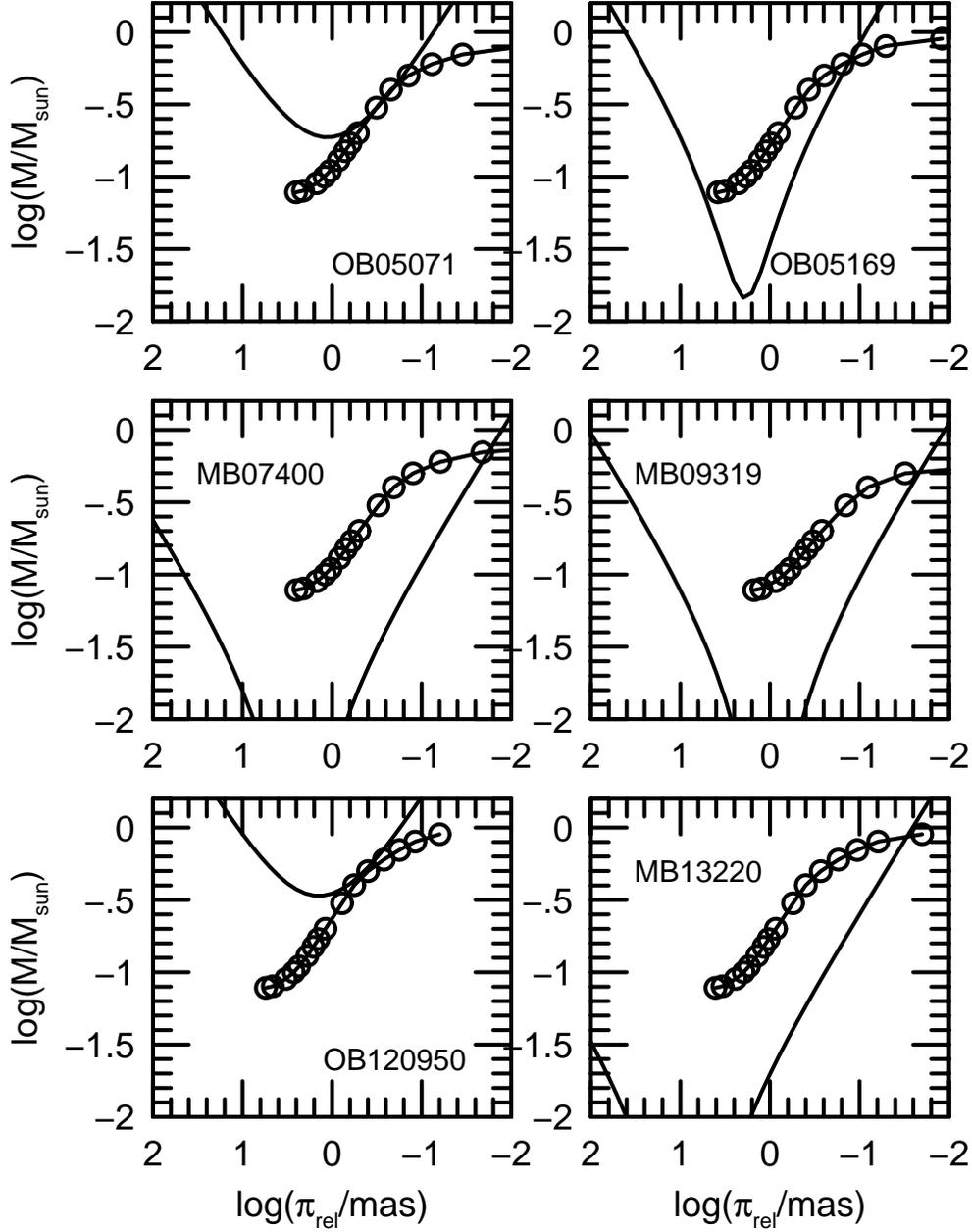}
\caption{Equations~(\ref{eqn:mpirel1}) and (\ref{eqn:mpirel2}) 
are illustrated
for six planetary events (internal labels) for which the source and lens
were separately resolved, enabling the measurements of $K_\host$ and
$\bmu_{\rel,\hel}$. \citet{baraffe15} $K$-band 1-Gyr isochrones are assumed.
For OGLE-2005-BLG-169, I adopted $K_\host = H_\host - 0.11$.  To the right,
Equation~(\ref{eqn:mpirel1}) asymptotically approaches constant
Einstein radius, i.e., 
$\theta_\e\equiv \sqrt{\kappa M\pi_\rel}\rightarrow\mu_{\rel,\hel}t_\e$, while
to the left, it asymptotically approaches constant microlens parallax, i.e.,
$\pi_\e\equiv{\pi_\rel/\kappa M}\rightarrow \au/v_{\oplus,\perp}t_\e$.  For none
of the six cases does Equation~(\ref{eqn:mpirel2}) cross both branches
of Equation~(\ref{eqn:mpirel1}), but it comes close for OGLE-2005-BLG-169.
The two curves are nearly tangent for two cases: OGLE-2005-BLG-071 and
OGLE-2012-BLG-0950.
}
\label{fig:all1}
\end{figure}

\begin{figure}
\plotone{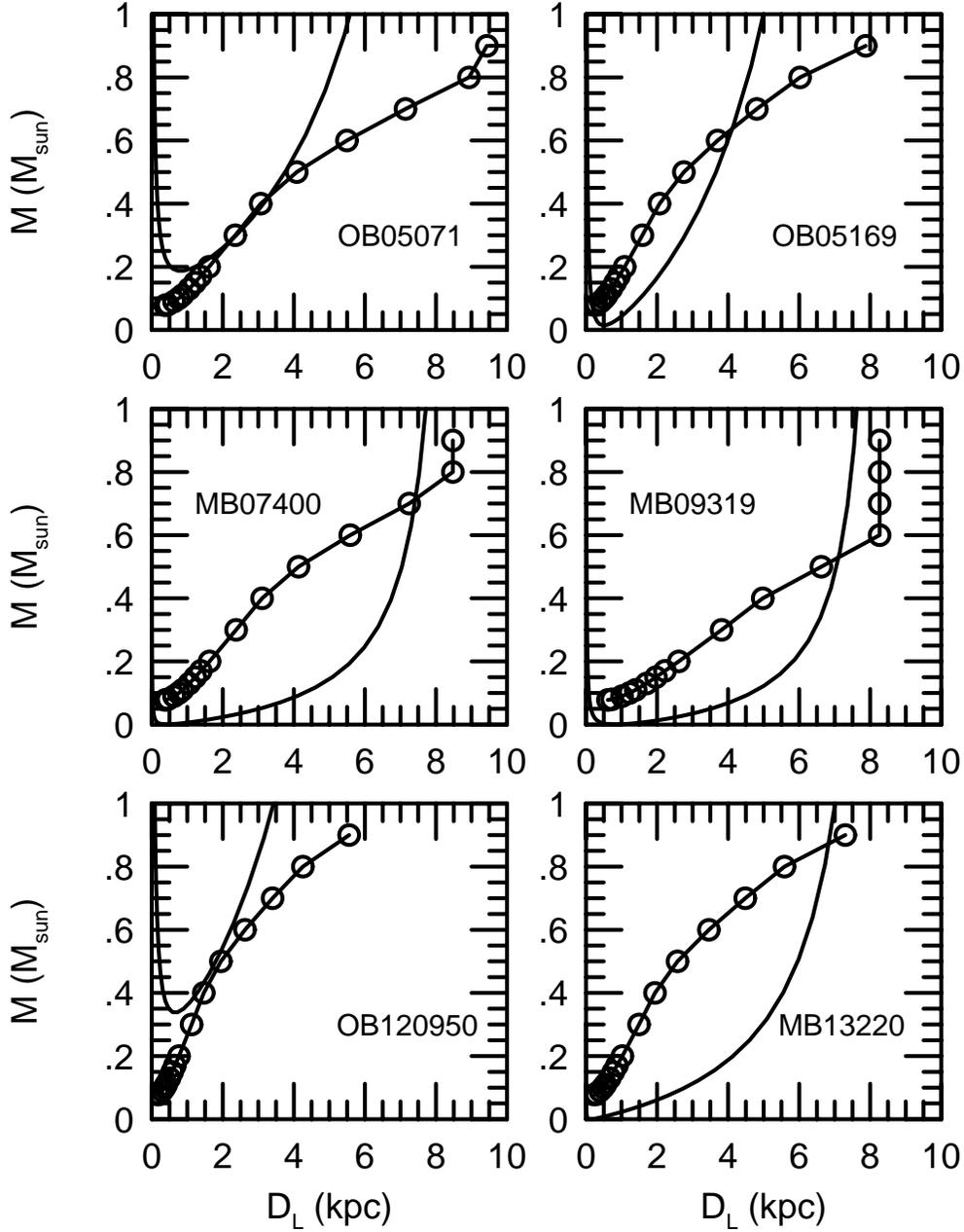}
\caption{Alternate representations of the information in 
Figure~\ref{fig:all1},
which is more similar to what is often displayed in the Keck AO papers.  
Note that the left (low-mass) branch of Equation~(\ref{eqn:mpirel1}) is
clearly visible in this representation in only two of the six cases. It has
not previously appeared in published diagrams.
}
\label{fig:all2}
\end{figure}

\begin{figure}
\plotone{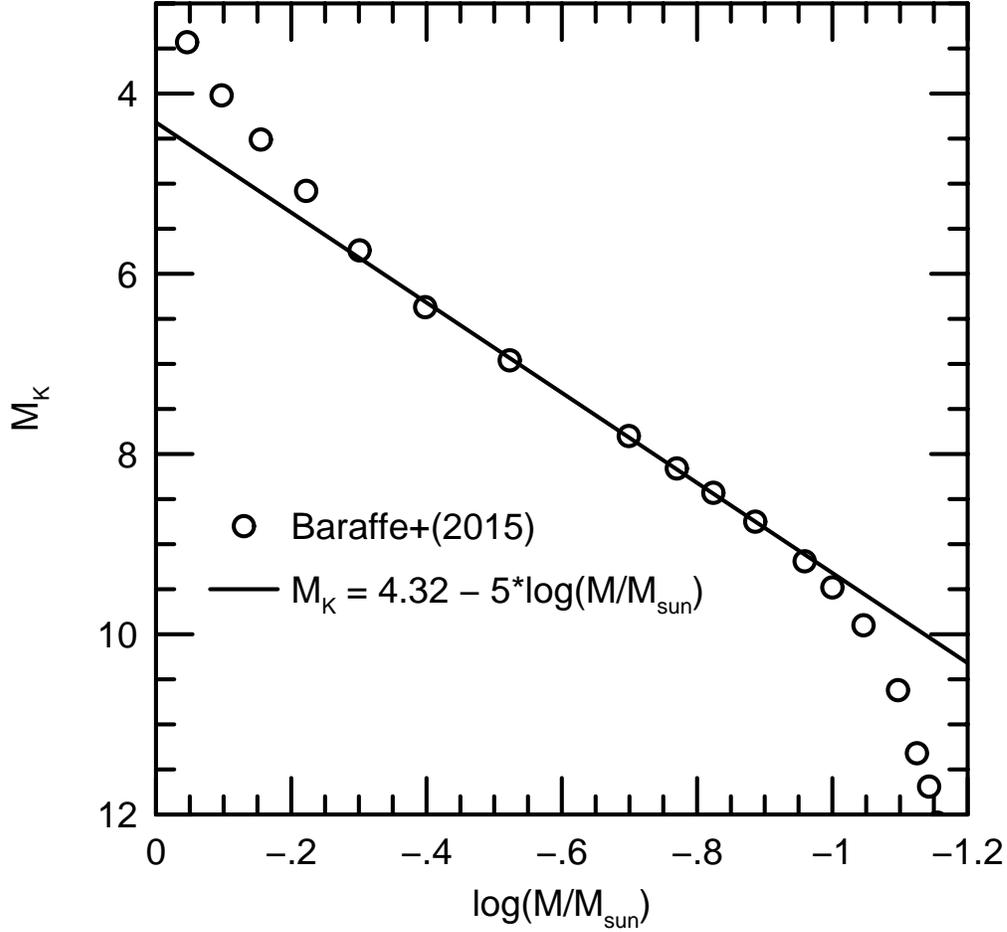}
\caption{$M_K$ versus $\log M$ mass-luminosity relation in the $K$ band.
The points show the 1-Gyr, solar-metallicity models of \citet{baraffe15}.
Over the mass range $-0.4\ga\log(M/M_\odot) \ga -0.9$ (in which the
star is fully convective and supported by the ideal gas law, $P=nkT$),
these are well
approximated by the solid line.  This corresponds to a scaling
$L_K\propto M^2$ of the $K$-band luminosity $L_K$.  Hence, over this
mass range, the function $H(M) \equiv -(5d M_K/d\log M + 1)\rightarrow 0$.
See Figure~\ref{fig:hofm}.}
\label{fig:klogm}
\end{figure}

Figures~\ref{fig:all1} and \ref{fig:all2} illustrate the ``basic method''
using $K_\host$ photometry and $\bmu_{\rel,\hel}$ measurements from the six
planetary events mentioned above. (For OGLE-2005-BLG-169, I adopt
$K_\host = H_\host - 0.11$).  For the mass luminosity relation, I adopt
the \citet{baraffe15} 1 Gyr isochrones, which are shown in 
Figure~\ref{fig:klogm}.  The source distance is held fixed at 
$D_S=1.08\times 10^{(I_{0,\rm clump}- M_{I,\rm clump})/5 -2}\,\kpc$ where
$M_{I,\rm clump}=-0.12$ and $I_{0,\rm clump}$ is given by 
Table~1 from \citet{nataf13}.  I employ a dust model with a 
scale height of 120 pc.
These figures have some interesting
features, some of which I point out now, while others will be noted in
Section~\ref{sec:degen}.

The first point is that while figures with the same axes as 
Figure~\ref{fig:all2} appear in four of the six Keck papers
(corresponding to the four lower panels of Figure~\ref{fig:all2}), none of these
show the high-$\pi_\rel$ branch of Equation~(\ref{eqn:mpirel1}).
For MOA-2009-BLG-319 (Figure~5 of \citealt{mb09319b}) and 
MOA-2009-BLG-220 (Figure~5 of \citealt{mb13220b}), 
this is because the figures are restricted to the region near the solution.
For OGLE-2012-BLG-0950, Figure~6 from \citet{ob120950b}
is not directly comparable to Figure~\ref{fig:all2} because it only shows
the $\bmu_{\rel,\hel}$ constraint after it is combined with a $\pi_\e$ constraint
that is derived from the light curve.
On the other hand, for MOA-2007-BLG-400, Figure~4 of \citet{mb07400b} 
displays the entire parameter 
space, but does not show this feature.  However, it is also barely visible
in Figure~\ref{fig:all2}, where it is mainly superposed on the y-axis,
although it is clearly visible in Figure~\ref{fig:all1}. 
Nevertheless, one can see from Figure~\ref{fig:all2} (and better 
from Figure~\ref{fig:all1}) that there are cases for which 
Equation~(\ref{eqn:mpirel2}) intersects Equation~(\ref{eqn:mpirel1})
on its high-$\pi_\rel$ branch.  I will show in Sections~\ref{sec:degen}
and \ref{sec:errors}
that \citet{ob05169bat} were correct to ignore this second solution, but
this shows that, in general, these solutions must be considered.

Second, it is striking that for two of the 6 events, i.e., OGLE-2005-BLG-071
and OGLE-2012-BLG-0950, the ``intersection'' of the two equations takes the
form of an ``extended tangent''.  At first, one might think that this a
rare coincidence, but it is actually rooted in the peculiar ``inflection''
which is highlighted in Figure~\ref{fig:klogm} and (as I will explain in 
Section~\ref{sec:errors}) is especially critical for relatively nearby lenses.

\begin{figure}
\plotone{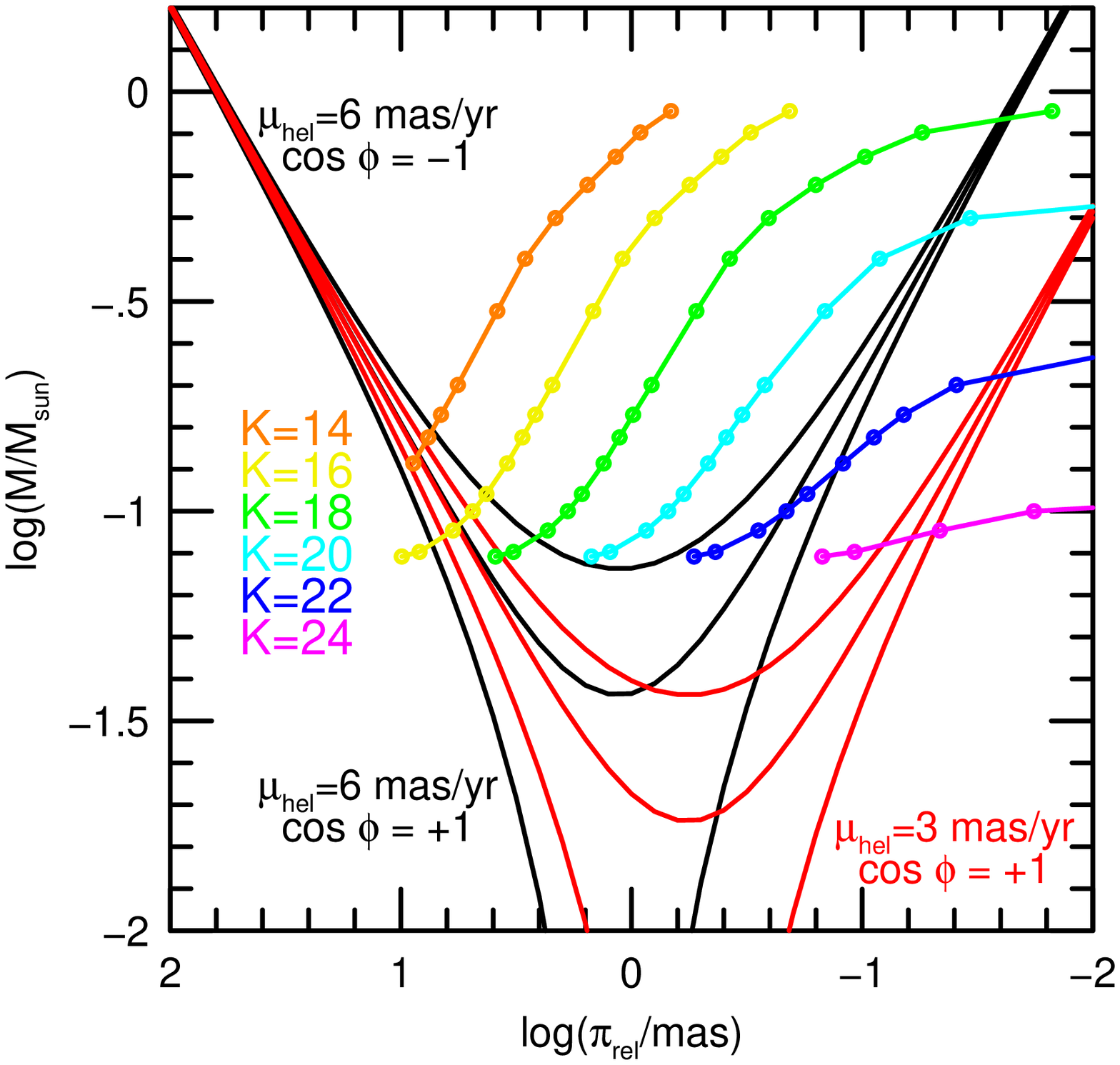}
\caption{Illustration of the variation of Equations~(\ref{eqn:mpirel1})
and (\ref{eqn:mpirel2}) with various input parameters.  Several parameters
are held fixed for all curves: $t_\e= 25\,$day, $\pi_S= 0.115\,\mas$,
$A_K=0.13$, $b=-3.0$, and $v_{\oplus,\perp}=25\,\kms$. Equation~(\ref{eqn:mpirel1})
is illustrated for two values of $\mu_{\rel\hel}$, i.e., 3 (red) and 6 (black) 
$\masyr$, and three values of 
$\cos\phi\equiv \bmu_{\rel\hel}\cdot \bv_{\oplus,\perp}/\mu_{\rel\hel}v_{\oplus,\perp}$
i.e., $-1$ 0, and $+1$ (top to bottom). Equation~(\ref{eqn:mpirel2}) is 
illustrated for 6 values of $K_\host =(14,16,18,20,22,24)$ (top to bottom).
}
\label{fig:6fake_events}
\end{figure}

Figure~\ref{fig:6fake_events} illustrates how Equations~(\ref{eqn:mpirel1})
and (\ref{eqn:mpirel2}) behave as key parameters change.  For all 12 curves
shown in this figure, I hold fixed 
$t_\e= 25\,$day,
$\pi_S= 0.115\,\mas$,
$A_K=0.13$,
$b=-3.0$, and
$v_{\oplus,\perp}=25\,\kms$.  That is, I fix all the quantities that are
known (within errors) in advance of the high-resolution observations.

The three black curves show Equation~(\ref{eqn:mpirel1})
for $\mu_\rel=6\,\masyr$ and three values (from top to bottom) of 
$\cos\phi=(-1,0,+1)$.  Similarly, the red curves show 
Equation~(\ref{eqn:mpirel1}) for $\mu_\rel=3\,\masyr$. The curves with small
circle show Equation~(\ref{eqn:mpirel2}) for 6 different values (top to bottom)
of $K_\host=(14,16,18,20,22,24)$

Before continuing, I note the ``basic method'' for measuring host masses
from resolved host-source pairs is closely related to another method
that measures host masses from unresolved host-source pairs.  Both
methods generally rely on high-resolution imaging 
to exclude (or limit) light from other stars,
and both derive the mass from a combination of measurements of $K_\host$
(or flux in some other band) and $\theta_\e$.  The ``unresolved'' method
has the major advantage that there is no need to wait for the host and
source to separate.  However, it has several major restrictions and
disadvantages.  First, it is restricted to cases for which the host is not too
much fainter than the source: otherwise, the source light cannot be
accurately subtracted from the combined flux.  Second, it is restricted
to the subsample of planets (roughly 2/3, see Section~\ref{sec:geocent}) 
for which there is a $\theta_\e$ measurement from the light curve.
Third, it runs a greater risk of falsely identifying another star as the
host.  See the examples of MOA-2016-BLG-227 \citep{mb16227}
and MOA-2008-BLG-310 \citep{mb08310b} in Sections~\ref{sec:2016.1} and
\ref{sec:ufun}, respectively.  Nevertheless, this method has important
applications, e.g., for 
OGLE-2006-BLG-109 \citep{ob06109b},
OGLE-2007-BLG-349 \citep{ob07349}, and
OGLE-2017-BLG-1434 \citep{ob171434b}.
See Sections~\ref{sec:ufun} and \ref{sec:2016.1}.

\section{Methods to Break the Discrete Degeneracy}
\label{sec:degen}

Even within the context of the ``basic method'', the discrete degeneracy 
that was discussed in Section~\ref{sec:basic}
can be broken definitively in a substantial majority
of cases.  Moreover, for a majority of the cases that it cannot be
broken definitively, the low-$\pi_\rel$ solution will be strongly
favored statistically.  Finally, it can essentially always be broken
by taking one additional observation.

\subsection{Geocentric Proper-Motion Measurements}
\label{sec:geocent}

In most cases (see the paragraph after next), 
the planetary event will have itself yielded an 
independent measurement of $\theta_\e$, via the source-size parameter
$\rho = \theta_*/\theta_\e$, where $\theta_*$ is the angular size of the
source.  This parameter is required to fit the light curve whenever
it is strongly impacted by caustics, and, when $\rho$ is needed, it is usually
measured to better than 10\%.  Then $\theta_*$ can also be measured
using standard techniques \citep{ob03262}, also usually to better than 10\%.
Hence, $\theta_\e=\theta_*/\rho$ and $\mu_\rel = \theta_\e/t_\e$ are usually
known to about 10\%.  For a typical event, the true value of $\pi_\rel$ will
be small, e.g., $\pi_{\rel,\rm true}\la 0.05\,\mas$, whereas in the alternate
solution, it will be $\pi_{\rel,\rm alt}\ga 1\,\mas$.  Using the same fiducial
parameters as above, the two solutions would predict
$\mu_\rel \simeq \mu_{\rel,\hel}$ and $\mu_\rel \ga 1.5\mu_{\rel,\hel}$, respectively.
Hence, they could be distinguished by $\rho$-based measurements of $\mu_\rel$
that are of typical quality.

The historical example of OGLE-2005-BLG-169 illustrates this point.
Although, nominally, Equation~(\ref{eqn:mpirel2}) ``ends'' before it intersects
the high-$\pi_\rel$ branch in Figure~\ref{fig:all1}, it does so only
marginally, and this would not be the case once measurement errors were
taken into consideration.  However, lines with slopes of +1 on this figure
have constant $\theta_\e$.  Thus, one can see by eye that the two solutions
are offset by $\Delta\log \theta_\e = 0.5(\Delta\log M + \Delta\log \pi_\rel)
\simeq 0.42$, i.e., a factor 2.6.  Because $\theta_\e$ was reasonably well
measured, the alternate solution would be ruled out by this argument.
In fact, we will see that it would be ruled out by two other arguments
as well.

However, as predicted by \citet{Zhu:2014} and confirmed by \citet{2018subpr},
of order half of microlensing planets do not have caustic crossings.
In particular, \citet{2018subpr} found that of 33 planets from 2018 that
were suitable for statistical studies, only 16 had caustic crossings.
Nevertheless, a review of the remaining 17 shows that five 
(OGLE-2018-BLG-0298Lb\footnote{This event has two solutions, one in which
the source crosses the caustic and another in which it crosses a ridge
extending from a cusp.  It is classified as ``non-caustic-crossing'' because
the latter solution has lower $\chi^2$.  In any case, the two solutions have
similar values and error bars for $\rho$.},
KMT-2018-BLG-0087Lb,
OGLE-2018-BLG-1185Lb,
OGLE-2018-BLG-1011b, and
OGLE-2018-BLG-1011c)
have good measurements of $\rho$, while three others 
have useful constraints on $\rho$
(OGLE-2018-BLG-0977Lb,
OGLE-2018-BLG-0506Lb, and
OGLE-2018-BLG-1647Lb).

Based on this experience, I estimate that this method can resolve the
degeneracy for $\sim 2/3$ of planets.

\subsection{Microlens Parallax}
\label{sec:parallax}

A substantial fraction, probably a majority, of the alternate,
high-$\pi_\rel$ solutions can be decisively rejected based on
microlensing-parallax constraints that are derived from 
the original light curves.
This statement may seem surprising because only a minority of published
microlensing-planet analyses even report parallax parameters, and, of those
that do, many give only constraints rather than measurements.

However, there are two factors that make microlensing parallax a much more
powerful tool for rejecting high-$\pi_\rel$ alternate solutions than
for making parallax measurements.  First, the alternate solutions
generally have very large values of the microlensing parallax, $\pi_\e$.
Second, these alternate solutions make very precise predictions of the direction
of the parallax vector, $\bpi_\e$.  Hence, the high-$\pi_\rel$ solution
can be ruled out, even when the parallax analysis was not considered
to be sufficiently constraining to warrant being reported in the original
papers.  Here,
\begin{equation}
\bpi_\e = \pi_\e{\bmu_\rel\over\mu_\rel}; \qquad
 \pi_\e\equiv{\pi_\rel\over\theta_\e} = \sqrt{\pi_\rel\over\kappa M}
= 1.57\biggl({\pi_\rel\over 2\,\mas }\biggr)^{1/2} \biggl({M\over 0.1\,M_\odot}
\biggr)^{-1/2}.
\label{eqn:piedef}
\end{equation}

The first point is that the illustrative values used in 
Equation~(\ref{eqn:piedef}) are typical of the real problem under
consideration.  From ~Equation~(\ref{eqn:pirelmin}), 
$\pi_{\rel,\min}\sim 1\,\mas$, while (by definition) the high-$\pi_\rel$ branch has
$\pi_\rel > \pi_{\rel,\min}$.  And because, typically, 
$\pi_{\rel,\rm true}\ll 1\,\mas \sim \pi_{\rel,\min}$ (and 
$M_\host(\pi_L) = M^{-1}_K(\Delta K(\pi_L))$ is monotonically declining),
it is likely that the host mass in the high-$\pi_\rel$ solution is a factor
several lower than the true mass.  Hence, typically, $\pi_\e\ga 1$
for the high-$\pi_\rel$ solution.

However, this fact, in itself, would not be enough to rule out such
solutions for typical planetary events.  To understand why, I examine
some ``typical'' cases of ``marginally interesting'' microlens parallax
constraints.  By this I mean, cases for which the parallax solutions would not
traditionally be published because they were regarded as ``not constraining'',
but which were in fact published due to my own initiative.  I focus on
KMT-2018-BLG-2004 and OGLE-2018-BLG-1367, which are both
illustrated in Figure~3 of \citet{2018prime}.  In both cases, the parallax
amplitude is consistent with all values $0\leq \pi_\e \la 3$. and so appears to 
contain essentially no information.  The parallax constraints were nevertheless
included in the Bayesian analyses of that paper because the effectively
1-dimensional (1-D) contours \citep{gmb94} in the 2-D $\bpi_\e$ plane do 
in fact contain some information.  Note, however, that for the illustrative
example of Equation~(\ref{eqn:piedef}),
the constraint $\pi_\e\la 3$ would not exclude the high-$\pi_\rel$ solution,
i.e., $\pi_\e = 1.57$.

Nevertheless, inspection of Figure~3 from \citet{2018prime} shows that if
$\bpi_\e$ were additionally constrained to lie at least $10^\circ$ from
the long-axis of these contours, then this solution would be excluded.  

Now, if one were to assume that the direction
of $\bpi_\e$ for the (wrong) high-$\pi_\rel$ solution was randomly
distributed relative to this long axis, then one would conclude that
$1 - (4\times 10^\circ/360^\circ)=89\%$ of spurious high-$\pi_\rel$ solutions
could be excluded, provided that the parallax-contour measurements were
of comparable quality to Figure~3 of \citet{2018prime}.  In fact, however,
the situation is substantially more favorable.  The first point
is that for the high-$\pi_\rel$ solution, by definition, 
$\pi_\rel>\pi_{\rel,\min} = \mu_{\rel,\hel}\au/v_{\oplus,\perp}$.  Hence, the direction
of $\bpi_\e$ (same as that of 
$\bmu_\rel = \bmu_{\rel,\hel}- \bv_{\oplus,\perp}\pi_\rel/\au$),
is more closely aligned to the direction of $\bv_{\oplus,\perp}$ than to
$\bmu_{\rel,\hel}$, and usually substantially so.  However, for a substantial
majority of the microlensing season, $\bv_{\oplus,\perp}$ points 
approximately due east (or due west).  

On the other hand, for a substantial majority of the microlensing
season, the major axis of the 1-D parallax contours is roughly aligned
north-south.  This is because, for events that are short compared to
a year, Earth's acceleration can be approximated as constant, and the
parallax vector is much better constrained in the direction of this
acceleration \citep{smp03,gould04}, which is usually close to either due
east or due west.  This ``happy coincidence'' that Earth's projected
acceleration is typically aligned (or anti-aligned) with its projected
velocity is due to the fact that the bulge microlensing fields lie near the
ecliptic.  By contrast, toward the Large Magellanic Cloud, Earth's projected
acceleration is perpendicular to its projected velocity.

The case of OGLE-2005-BLG-169 also illustrates the power of this
parallax argument.  Lines with a slope of $-1$ in Figure~\ref{fig:all1}
have constant parallax, so one sees immediately that the alternate
solution has $\Delta\log\pi_\e=1.18$, i.e., 15 times larger.   In more
detail, $M=0.078\,M_\odot$, $\pi_\rel=6.0\,\mas$,
$\bmu_{\rel,\hel}(N,E) = (+4.87,+5.60)\,\masyr$, and
$\bv_{\oplus,\perp}(N,E) = (+3.1,+18.5 )\,\kms$, 
together with Equation~(\ref{eqn:bmu_hel}), imply
$\bpi_\e = \sqrt{\pi_\rel/\kappa M}\bmu_\rel/\mu_\rel = (+0.2,-3.1 )$.
By contrast, \citet{ob05169} published a 1-D parallax measurement
at the time of the original discovery, $\pi_{\e,\parallel}=-0.086\pm 0.261$,
which is the component of $\bpi_\e$ in the direction of the apparent
(``geocentric'') acceleration of the Sun, i.e., $\psi = -79^\circ$
(north though east). By comparison, the alternate model predicts
$\pi_{\e,\parallel}=0.2\cos\psi + (-3.1)\sin\psi = +3.1$.  
Hence, although the original
parallax ``measurement'' appeared to be almost completely
unconstraining at the time, it rules out the alternate model in this case.

Thus, even when the parallax contours are substantially ``fatter''
than those shown in Figure~3 of \citet{2018prime} (and hence are rarely
published unless they are strongly inconsistent with zero, i.e.,
indicating very large $\pi_\e$), they can still be adequate to exclude the
high-$\pi_\rel$ solution.  This reflects a general principle that seemingly
``useless'' information can suddenly be the key to unlocking a 
puzzle\footnote{``Any man can make use of the useful, but it takes
a wise man to make use of the useless'', - Lao Tzu}
when new techniques become available, such as old clothing in police
storage lockers after the advent of DNA testing.  In particular, this
emphasizes the importance of archiving final-reduction light-curve 
photometry of all microlensing planets.

Of course, there are some cases that the high-$\pi_\rel$ solution is correct.
For reasons that are discussed further below, these are quite rare.
Nonetheless, it is important to point out that this same test will typically
confirm those solutions.  Indeed, it is likely that in those cases, the
high parallax will already have been noticed and the host will already
be recognized as being of exceptionally low mass and/or nearby.

\subsection{Unique Main-Sequence Solutions}
\label{sec:unique}

For a substantial fraction of cases, Equation~(\ref{eqn:mpirel2}) 
(provided that it is restricted to main-sequence hosts) will ``end''
before it can intersect with the $\pi_\rel>\pi_{\rel,\min}$ branch of
Equation~(\ref{eqn:mpirel1}).  I will cover the case of luminous BD hosts 
later in this section.  For the sake of discussion, I adopt
$M_K=11.0$ and $M=0.078\,M_\odot$ for a star at the hydrogen-burning 
threshold\footnote{It may be that the dimmest ``stars'' in Figure~22 of
\citet{benedict16}
are actually relatively young BDs that are still cooling.  Such young BDs
are over-represented in the solar neighborhood relative to typical
microlenses, which are several hundred pc from the plane.  Thus, as a technical
point, non-detections could be consistent with
very slightly more massive BDs, e.g., $0.08\,M_\odot$.  The value adopted
here, $0.078\,M_\odot$, is adequate for the illustrations of this section.}
from Figure~22 of \citet{benedict16}.  And, for simplicity, I assume that
extinction is negligible for the following example, wherein
$\mu_{\rel,\hel}=6\,\masyr$, $v_{\oplus,\perp}=30\,\kms$, $t_\e=30\,$day
(i.e., the fiducial parameters of Equations~(\ref{eqn:pirelmin}) and
(\ref{eqn:mmin})), $K=21$, $\cos\psi=0$, and $\pi_S=0.12\,\mas$.  Then, the 
pair $(M_\host,\pi_L)=(0.078\,M_\odot,1.00\,\mas)$, corresponding to
$(M_\host,\pi_\rel)=(0.078\,M_\odot,0.88\,\mas)$ is the ``end point''
of the mass-luminosity relation: there are no solutions with higher $\pi_\rel$
because the sources would have to be dimmer than the bottom of the main
sequence.  Therefore, because the minimum from 
Equation~(\ref{eqn:pirelmin}), $\pi_{\rel,\min}=0.95\,\mas$, is already
larger than this end-point value, there are no solutions on the
$\pi_\rel>\pi_{\rel,\min}$ branch of Equation~(\ref{eqn:mpirel1}) for this case.

However, this argument will fail in a large fraction of cases.
For example, the fiducial value $\mu_\rel=6\,\masyr$ is quite typical of
microlensing events, and the fiducial value $v_{\oplus,\perp}$, while also
fairly typical is, in addition, the maximum possible value.  Hence, one
broadly expects $\pi_{\rel,\min}\sim 1\,\mas$.  On the other hand, a large
fraction of hosts will be brighter than the above example, i.e., 
$K=21-\delta K$, where $\delta K$ is a few magnitudes.  For example,
an $M_\host = 0.5\,M_\odot$ star at $D_L=5\,\kpc$ would have $K\sim 19.5$.
In such cases, the end point of the main sequence will be at 
$\pi_\rel \sim (10^{\delta K/5} - 0.12)\,\mas\rightarrow 1.88\,\mas$, 
so the high-$\pi_\rel$ branch
will not necessarily pass ``under'' the end point.

Figure~\ref{fig:all1} shows three clear examples of this argument,
namely, MOA-2007-BLG-400, MOA-2009-BLG-319, and MOA-2013-BLG-220.
The argument clearly fails for OGLE-2005-BLG-071 and OGLE-2012-BLG-0950,
for which the ``two solutions'' are both on the low-$\pi_\rel$ branch. In both 
cases, these solutions are very close to each other.  These cases will be
discussed in greater detail further below.  Finally, if one could
really assume zero errors, the argument would apply to OGLE-2005-BLG-169.
However, as discussed above, this application is marginal, and, in practice,
one must consider the alternate solution.

Moreover, even when this argument does successfully exclude the alternate
high-$\pi_\rel$ solution for main-sequence hosts, it will still always
permit a high-$\pi_\rel$ BD host.  The first point is that the entire
``basic method'' depends on the existence of a mass-luminosity relation
that is approximately independent of other parameters, in particular,
age and metallicity.  The $K$-band relation
is approximately independent of metallicity.  And, as I will now discuss,
it is also basically independent of age for (unevolved) main-sequence stars.  
This would not quite be true of M dwarfs, particularly late M dwarfs, which can
take several 100 Myr to reach the zero-age main sequence.  However,
the great majority of lenses are either in the bulge or are several
hundred pc from the Galactic plane, where such young stars are exceedingly
rare.  Thus, this age dependence can usually be ignored.

However, it cannot be ignored for alternate, high-$\pi_\rel$ solutions because
these are close to the Sun and so, given their low Galactic latitude,
also close to the Galactic plane, where young stars are more plentiful.
And this means that one must consider cooling BDs as well.

The main argument against such solutions, assuming that they cannot be
excluded by geocentric proper-motion or microlens-parallax arguments
(Sections~\ref{sec:geocent} and \ref{sec:parallax}), is statistical.
In addition to general statistical arguments favoring distant over
nearby lenses for discrete degeneracies (e.g., \citealt{gould20}),
these BD solutions are possible for only a brief range of ages for
each of a narrow range of masses.

In brief, there are two broad classes of this discrete degeneracy:
one in which the mass-luminosity relation (Equation~(\ref{eqn:mpirel2})
intersects the high-$\pi_\rel$ branch of 
Equation~(\ref{eqn:mpirel1}) while it is still on the main sequence) and
the other where the intersection would indicate a cooling BD.  The 
$\mu_\rel$ (Section~\ref{sec:geocent}) and $\bpi_\e$ (Section~\ref{sec:parallax})
arguments can be applied against either class and will likely rule out
the great majority of the high-$\pi_\rel$ solutions.  For the few remaining
events, those that are in the second class, will be strongly disfavored
by probabilistic arguments.

\subsection{Decisive Resolution by Extra Observations}
\label{sec:extra}

For any particular case, it is straightforward to decisively resolve
the degeneracy between nearby and distant solutions by performing two
epochs of late-time AO observations instead of one.  A total of three epochs
is all that is required to measure the heliocentric lens-source relative proper
motion and parallax, $\bmu_{\rel,\hel}$ and $\pi_\rel$.  These could be $t_0$ 
(when the lens and source are known to have essentially zero offset)
and two epochs, say in April and September, many
years later.  These would be sufficient to distinguish between small $\pi_\rel$
and the values $\pi_\rel \ga 1\,\mas$ that are typical of the high-$\pi_\rel$
solutions.  The decision on whether to make such additional observations
will depend on the required purity of the sample for deriving statistical
conclusions about planets.  At present, it appears that of order 80\% to 90\%
of the planets with large high-$\pi_\rel$ alternate solutions will be
decisively resolved by the methods of Sections~\ref{sec:geocent} and 
\ref{sec:parallax}), while the statistical uncertainty of the remainder
will be small.  However, if greater purity is desired, then this method
remains available.

Another method, which also requires an extra observation but does not
require any special timing, is to observe the source in a second, substantially
bluer band, e.g., $J$-band.  The high-$\pi_\rel$ solution will almost
always have substantially lower mass and hence be substantially redder.
In some cases of high-extinction fields, the lower predicted extinction
at the higher $\pi_\rel$ might effectively cancel this intrinsic difference,
but this will be rare and, moreover, such wasted observations can easily
be avoided by predicting the observed colors of the two solutions in advance.

This color method is, in fact, the third one that can be applied to the
case of OGLE-2005-BLG-169.  I considered only $K$-band observations
when I constructed Figures~\ref{fig:all1} and \ref{fig:all2} because
I wanted to focus on issues related to the ``basic method''.  However,
several of these events were observed in two or more bands, including
OGLE-2005-BLG-169, which was observed in optical bands using {\it HST}
by \citet{ob05169ben}.  Their resulting solution was consistent results with the
one derived by \citet{ob05169bat} based on Keck IR observations,
while these would have been totally inconsistent
if the alternate solution had been correct.

Thus, while \citet{ob05169bat} and \citet{ob05169ben} did not
explicitly consider the alternate solution, they did explicitly remark
on all three of the arguments that objectively rule it out:
agreement with the \citet{ob05169} proper-motion measurement 
(Section~\ref{sec:geocent}),
consistency with the \citet{ob05169} parallax measurement 
(Section~\ref{sec:parallax}), and
consistency between results obtained in red and blue bands
(Section~\ref{sec:extra}).

\subsection{Remaining Problematic Cases}
\label{sec:remain}

The main source of problematic cases will be events with low
$\mu_{\rel,\hel}$, which will have strong overlap with members of the input sample
that have small $\mu_\rel$ (if known).  While low, e.g.,
$\mu_\rel< 2\,\masyr$, events are expected to be rare
($p\sim(2/2.9)^3/6\,\sqrt{\pi}\sim 3\%$, \citealt{gould21}), planetary
events with $\mu_\rel< 2\,\masyr$ appear to occur at a higher rate.
For example, of the 71 planetary events with proper-motion measurements
that are examined in Section~\ref{sec:practical}, 9 have $\mu_\rel<2\,\masyr$,
of which 2 have $\mu_\rel<1\,\masyr$, compared to the 2.0 and 0.25 that would
be expected if planetary events followed the underlying event distribution.
However, there is a significant selection bias toward lower $\mu_\rel$ for events
with detectable planetary anomalies because,
other event characteristics being equal, 
$\Delta\chi^2\equiv\chi^2({\rm 1L1S})-\chi^2({\rm 2L1S})\propto \mu_\rel^{-1}$.
See Section~\ref{sec:distmu}.  Here $n$L$m$S means $n$ lenses and $m$ sources.

Assuming that these events prove to have $\mu_{\rel,\hel}\sim \mu_\rel$, then
from Equation~(\ref{eqn:pirelmin}), they  would have
$\pi_{\rel,\min}\la 0.3\,\mas$.  It might still be that the actual lens
was in or near the bulge, in which case the alternate solution would still have
very large $\pi_\rel$ and so would be susceptible to the various rejection
arguments given above.  However, it is also possible (if less likely)
that the actual lens had $\pi_\rel\sim 0.35\,\mas$ in which case the
alternate solution might have $\pi_\rel\sim 0.25\,\mas$.  In this
case, it might be very difficult to apply either of the methods from
Sections~\ref{sec:geocent} and \ref{sec:parallax}, and the statistical
arguments would likewise fail.  Moreover a reliable measurement of 
the difference in trigonometric parallaxes, i.e., $0.1\,\mas$, would
probably require extraordinary effort.

\section{{False Positives}
\label{sec:false_pos}}

In Section~\ref{sec:degen}, I showed that if two stars are detected
in a single epoch of late-time AO follow-up imaging, and 
{\it if these stars are actually the microlensed source and the host},
then one can make unambiguous host mass and distance determinations in the
great majority of cases and can make strong statistical arguments 
regarding these determinations in most of the remaining cases.

However, the inference that the star identified as the host is in fact 
the host is far from trivial.
Logically, there are four broad outcomes from a first epoch of late-time 
imaging: either zero, one, two, or more than two stars are detected in the 
neighborhood of the anticipated target.  I mention zero detections only
for completeness: it is very difficult to imagine the circumstances
under which a source that was detected (even highly magnified) by a 1.6m
optical telescope, could fail to appear in an 8m (or 30m) $K$-band
image (other than if a brighter star were superposed, in which case there
would still be at least one star detected in the image).

For this reason, if only one star is detected, it is almost certainly the
source (or a brighter companion of the source).  
The main potential reasons for this would be that the host has
not separated sufficiently to be seen due to the glare of the source
(which could be quite severe if it is, e.g., a giant), or the host is
dark (BD, WD, NS, or BH).  The specific circumstances, in particular
whether there was a geocentric proper-motion measurement, would have
to be evaluated to determine whether to wait for another observation
or to conclude that the host was dark.

If more than two stars are detected, then some additional work
(possibly including additional observations) is required to determine 
which among these several stars is the source and which is the host.  
I discuss this in Section~\ref{sec:second}.

In this section, I focus on the case that exactly 2 stars are detected.
The principal operational issue is that, assuming one can confidently
identify which is the source, one can, in principle, proceed to directly
apply the ``basic method'' of Section~\ref{sec:basic} to measure the
host mass and distance.  Then, for the case that the ``host'' is actually
a false positive, these measurements will, in fact, be incorrect.  Thus,
the main questions are: how likely is this to occur, under what conditions
will there be ``warning signs'' that the measurements are incorrect,
is the frequency of undetected ``bad measurements'' at an acceptable level,
and, finally, if not, what can be done to reduce the frequency of
``bad measurements''?

In general, for the case that two stars are detected, a ``false positive''
requires two conditions: first, the true host is not detected, and second,
some other star is detected.  Logically, there are only three choices
for the ``other star'': a companion to the host, a companion to the source,
and an unrelated field star.

When concrete scenarios are needed, I will assume that the late-time
AO observations are made by MICADO in the $K$-band,
on the 39m European Extremely Large Telescope (EELT).  Hence, I will
assume that the images have a FWHM $\theta_{\rm FWHM} = 14\,\mas$.  
The implications would not be qualitatively different if I were to use
the 25m Giant Magellan Telescope (GMT), with $\theta_{\rm FWHM} = 22\,\mas$.  
However, it is important to note that
substantial initial progress can be made toward the goals outlined in this paper
using present-day, 8m class, telescopes, whose imaging characteristics are very
different.  Nevertheless, most questionable cases can only be resolved using
next-generation telescopes, so my assumption is relevant to the final
resolution.  Moreover, I do not want to clutter the exposition with many
different cases, and I therefore rely on the reader to adapt the reasoning
presented here to other particular observational configurations and strategies.

\subsection{Stellar Companion to the Host}
\label{sec:hostcomp}

Among the three possible causes of false positives, the one most likely to
corrupt the ``basic method'' of 
Section~\ref{sec:basic} will be companions to the host that
are heavier and more luminous than the host.  Such configurations can
easily account for {\it both} the presence of a star other than the host
{\it and} the failure of the host star to appear.  

To give a concrete sense of the problem, let us imagine that an
$M_\host =0.9\,M_\odot$ G dwarf
is detected 10 years after the event at 50 mas from the source.   
I will further specify that this star is at $D_L=5\,\kpc$.  The 
first point is that if the G dwarf is the host, then the ``basic method''
will (approximately) yield the correct mass and distance of this star.
However, if this G dwarf is a companion to the true host, which is
an $M_\host = 0.4\,M_\odot$ M dwarf, then the solution generated by
Equations~(\ref{eqn:mpirel1}) and (\ref{eqn:mpirel2}) will be wrong
for two reasons.  First, the inferred $\bmu_{\rel,\hel}$ will be incorrect,
and second, the ``$M_\host$'' entering the two equations will be different.
Thus, there is a good chance that we would not even realize that this
was in fact a ``G dwarf''.

Nevertheless, as an all-knowing outsider to this process, I can ask:
``given that there is a G dwarf at 50 mas from the source, what is the
chance that it gave rise to the observed microlensing event'', which
took place 10 years earlier?  According to \citet{dm91}, about 40\%
of G dwarfs do not have stellar companions.  For this class,
the probability that the G dwarf was the host is 100\%.  The remaining
60\% of G dwarfs have companions at various mass ratios and periods.
However, those with separations 
$\Delta\theta \ga 4\,\theta_{\rm FWHM}\sim 60\,\mas$, corresponding
to projected separations $a_\perp\ga 300\,\au$, and so periods
$\log P/{\rm day}\ga 6.4$, would have been seen regardless of mass
(unless they were BDs, WDs, NSs, or BHs).  This corresponds to about
35\% of those with companions.  Another 25\% have periods 
$\log P/{\rm day} < 4.0$, in which case they would have given rise
to noticeable binary-lens effects in typical planetary microlensing
events (which have higher magnifications than average).  Thus,
only $40\%\times 60\% = 24\%$ of G dwarfs have companions that could
have escaped detection during both the microlensing event and during
the AO follow-up observations.
However, those that do escape detection are very good candidates
for the host.  That is, each component of a stellar binary has a
probability to generate a microlensing event
proportional to its $\theta_\e$, i.e., to the square root
of its mass.  So, for typical mass ratios of a G-dwarf companion, $Q\sim 0.5$,
the probability that the companion was responsible for the microlensing
event is $\sqrt{Q}/(1+\sqrt{Q}) \rightarrow 40\%$.  Hence, before considering
other mitigating factors, the overall probability of a false
positive for this G-dwarf example is about 10\%.

Nevertheless, there are two potential methods for recognizing the
presence of such a false positive.  Moreover, I will show that the
G-dwarf example is unrepresentative of the general case and is more adverse
than typical cases.

First, as discussed in Section~\ref{sec:geocent}, about 2/3 of planetary
events will have good (e.g., 10\%) measurements of $\mu_\rel$.  While
$\mu_\rel\not=\mu_{\rel,\hel}$, a difference of e.g., 25\% would raise
a ``red flag''.  In the above example, the region immune to such red flags
would be an annulus from 37 mas to 63 mas from the source, compared to
the 60 mas radius from the observed G dwarf, within which the true host
could be ``hiding''.  That is, only about 25\% of this region would
fail to trigger a red flag.  Note that if a red flag were triggered, this
would still not yield an accurate mass measurement, but it would trigger
an additional investigation, which I discuss in Section~\ref{sec:second}.

Second, it is possible that the original light curve contains
substantial parallax information.  Recall that this information is
normally highly directional in nature, so it could reveal strong
conflicts with the measured direction of $\bmu_{\rel,\hel}$ even if it,
by itself, provided no constraint on the lens mass.  Unfortunately,
however, the fraction of events with such parallax constraints is 
relatively small, and to be conservative, I ignore this channel in my
assessment of the robustness of the method.

Thus, for single-epoch detections of G dwarfs as the putative host, I conclude
that of order 10\% of events without $\rho$ measurements and of order
3\% of events with $\rho$ measurements will be corrupted by binary companions
and will escape any red flags that would alert us to this corruption.

As I discuss in Section~\ref{sec:second}, 
one can easily vet these alerts by taking
an additional AO image after several years.  Because these alerts will
arise in only a small minority of all events, they can be acted upon
at small fractional cost in the observing time of the overall program.

We do not expect that a major fraction of the detected ``putative hosts''
will be G dwarfs.  Microlensing is intrinsically sensitive to stars
in proportion to their Einstein radii, $\theta_\e=\sqrt{\kappa M \pi_\rel}$, and
hence, $\propto M^{1/2}$.  As there are vastly more M dwarfs than G dwarfs,
there are also vastly more M-dwarf than G-dwarf events.  For a sample
of planetary events (i.e., the current focus), the bias toward M-dwarf
events is expected to be less severe because, it is currently believed (based
on planets that are much closer to their hosts than microlensing planets), that
more massive stars host more planets.  Nevertheless, we still expect that
the majority of hosts will have lower mass than G dwarfs.  And, while
of order half of all G dwarfs have M dwarf companions, only a small fraction
of M dwarfs have G dwarf companions (simply because there are far more
M dwarfs overall).  Thus, the typical detected ``putative hosts'' will
be K and M dwarfs, for which the fraction with companions is lower
than described in the above G-dwarf example.

Overall, I conclude that unrecognized corruption of the sample by
luminous binary-star companions will be a small, though non-negligible
effect.  I discuss how the contamination of the sample can be further
greatly reduced in Section~\ref{sec:second}.


\subsection{Stellar Companion to the Source}
\label{sec:hostsource}

The great majority of microlensed sources from ground-based surveys
are either on the upper main sequence or are evolved stars.  Hence, they
are of order $\sim 1\,M_\odot$ and so have companion distributions similar
to G dwarfs.  Hence, source companions will populate the field in a manner
very similar to the above G-dwarf example.  However, their path to becoming
false positives is substantially different than for host companions.

I first consider the case (about 2/3 of all planets) for which there
is a good $\rho$ measurement.  Again I use the example that $\mu_\rel=5\,\masyr$
with an error of about 10\% and that the observation is taken 10 years after
the event.  Hence, I adopt a ``permitted range'' of $\pm 25\%$, i.e., 
$37\,\mas < \Delta\theta < 63\,\mas$ for an observed source companion to
be mistaken for the host.  This corresponds to periods
$ 6.40 \la \log(P/{\rm day}) \la 6.75$.  Only about 4\% of G dwarfs have
companions in this period range.  In addition, if this source companion is to
be mistaken for the host, the host itself must be absent. This can occur only
if the host is dark or if it is sufficiently close to the source companion
to disappear in its glare.  Because (in contrast to the host-companion
scenario of Section~\ref{sec:hostcomp}) the source companion is drawn
randomly from the mass-ratio distribution, I adopt a 25 mas radius typical
black-out zone.  Thus, in this example, only about 
$(2\times 25)/(2\pi\times 50)\sim 16\%$ of luminous hosts would be successfully
blocked out.  Hence, the source-companion channel of false positives is about
1\% (for the case that $\rho$ is measured during the event).

Next I consider the remaining 1/3 of planetary events for which $\rho$ is not
measured.  Before proceeding with this analysis, I note that a single late-time
observation will yield a very precise measurement of the source position.
This can be combined with a measurement of the source position at the time
of the event (based on difference images), which normally has a precision
of about 10 mas, to yield a source proper motion measurement with precision
$\sigma(\bmu_S)\sim 1\,\masyr$.  This greatly restricts the domain of
plausible $\bmu_{\rel,\hel}$ relative to the case that $\bmu_S$ is unknown.
In practice, one would solve Equations~(\ref{eqn:mpirel1}) and
(\ref{eqn:mpirel2}) under the assumption that the detected star was the host
and ask whether the resulting transverse velocity of the lens was consistent
with known Galactic populations.  For example, suppose that, in the solution,
$\pi_\rel \simeq 10\,\mu$as (implying that the lens was almost certainly in
the Galactic bulge) and $\bmu_S\simeq 0$ in the bulge frame.  Then
observed values of $\mu_{\rel,\hel}>6\,\masyr$, i.e., $\Delta\theta>60\,\mas$
would be regarded as suspicious, while $\Delta\theta<14\,\mas$ would be
unobservable.  Only 9\% of G-dwarfs have companions in the remaining region.
And only about 1/4 of these would succeed in ``blacking out'' the actual
host.  Hence, while contamination by unflagged false positives for the
case of unmeasured $\rho$ is a factor several higher than the measured-$\rho$
case, it is still very small.

\subsection{Random Field Stars}
\label{sec:random}

In principle, if there is one star observed (in addition to the source),
it could be a random field star.  To evaluate this possibility,
I normalize this population to the source binary-companion distribution
just analyzed.  \citet{holtzman98} report that there are 
4.2 stars ${\rm arcsec^{-2}}$ with
$M_I\leq 9$ toward Baade's Window, which corresponds 
to $M\ga 0.25\,M_\odot$, while \citet{dm91}
report that G dwarfs have an average of 0.56 companions in this mass range.
Based on the clump-giant surface densities reported by \citet{nataf13},
I find that microlensing planets typically lie in fields with
2--3 times the surface density of Baade's Window.  Therefore,
the frequency of source companions is comparable to the frequency of
random field that lie within $0.053\,\rm arcsec^2$, i.e., a circle of
radius 130 mas.  Thus, it will be somewhat less common to find field stars
at the separations in the typical range of interest than to find
binary companions to the source.  As I have already shown that the latter
present a relatively minor problem, I do not rehearse the details of
that analysis as applied to random field stars.

\subsection{Confusion Between Source and Host}
\label{sec:confusion}

In all the preceding analyses, I have implicitly assumed that
if exactly two stars are detected, then it will be clear which
is the source.  For the great majority of cases, the source 
$K$-band magnitude can be predicted from the source $I$-band magnitude,
which is usually well measured during the event, and the source $V-I$
color.   This is often well measured, but if not, it can be estimated
from the source magnitude together with the \citet{holtzman98}
color-magnitude diagram (CMD).  Then, in most cases, one of the
two observed stars will have approximately this brightness, while the
other will be very different.  

However, confusion may be possible for two reasons.  First, the source and
host may be of comparable brightness.  Second the source may have a close
companion that is not separately resolved, so that the ``star'' at the
source location is much brighter than the source.  For example, a companion
at $a_\perp\sim 15\,\au$ would be separated by $\sim 2\,\mas$.  The fact that
there were two stars would not leave any detectable trace on the AO imaging,
while (barring rare geometries) the source companion would not leave
any discernible trace on the microlensing event.

I now show that the cases for which this cannot be resolved with good
confidence are rare and that the practical impact of the remaining
cases is small.

The first point is that, as mentioned above, the proper motion
of each star can be measured separately in the bulge frame
from their positional offsets relative to the precise (difference image)
measurement of their position at $t_0$.  For the case that the lens is
in the disk, this measurement will often identify it as a disk star.
Second, there will sometimes be directional information from the microlens
parallax, although this will be relatively rare.  Third, it is actually
mainly for nearby disk lenses (i.e., just those that are most easily
resolved kinematically) that the problem of comparable brightness
arises.  That is, source stars are typically upper-main-sequence or
evolved stars in the bulge.  It would be rare that a bulge host would be
as luminous in $K$ as an upper-main-sequence star and extremely rare
to be as luminous as a sub-giant star.  On the other hand, a host at
$D_L=4\,\kpc$ or $D_L=2\,\kpc$ would gain an advantage of 1.5 or 3 magnitudes
from smaller distance modulus.  Hence, at these two distances, respectively,
early and middle M-dwarf hosts would have comparable brightness to G-dwarf 
sources.  Fourth, as source stars are usually bright, the chance that
they will have still brighter companions is small.  For example,
\citet{dm91} find that only 4.6\% of G dwarfs have companions of
comparable or greater mass.

Finally, we should ask how the estimates of the host mass and system 
distance would be affected if the wrong identification were made.  There would
be two effects.  First, obviously, the source brightness would be attributed
to the host, which would impact Equation~(\ref{eqn:mpirel2}).  Second,
the heliocentric relative proper motion would be assigned opposite sign:
$\bmu_{\rel,\hel} \rightarrow -\bmu_{\rel,\hel}$.  Hence, the quantity
entering Equation~(\ref{eqn:mpirel1}) would transform
$|\bmu_{\rel,\hel} - \bv_{\oplus,\perp}\pi_\rel/\au|
\rightarrow
|\bmu_{\rel,\hel} + \bv_{\oplus,\perp}\pi_\rel/\au|$.  Because the second term
in these expressions is usually small compared to the first, this
usually would be a minor issue.  However, in the great majority of cases,
the whole problem originates from the fact that the host and source have
comparable brightness.  Hence, the fact that one has mistaken the source
for the host does not have a major impact on this estimate.  And,
because the issue is mostly resolved for nearby (i.e., large $\pi_\rel$)
hosts, the error in Equation~(\ref{eqn:mpirel1}) induced by reversing the
sign of $\bmu_{\rel,\hel}$ is small.

\subsection{Second Epoch of Late Time AO Observations}
\label{sec:second}

In my view, the statistical uncertainties induced
by the various effects discussed in 
Sections~\ref{sec:hostcomp}--\ref{sec:confusion} on the final results,
e.g., planetary mass function and 2-D planet+host mass distribution,
are likely to be modest.  Hence, reliable results can be obtained by 
following a general policy of obtaining a single epoch of late-time 
$K$-band imaging.  However, there will be, as I have discussed, a
significant minority of individual cases that will be flagged for further
investigation, and most of these would greatly profit by making an additional 
AO observation after two or so years.

Comparison of the two images would immediately give an independent
measurement of the relative proper motion of the two observed stars,
regardless of their relation to the microlensing event.  If the
``putative host'' were actually a companion to the source, then this
relative proper motion would be essentially zero, which would directly
contradict the tentative conclusion that its displacement from the source
represented host-source relative proper motion.

Similarly, if the ``putative host'' were in fact a random field star,
its {\it vector} proper motion relative to the source would almost
certainly be inconsistent with the result derived from their
first-epoch separation.  That is, with precise astrometric measurements,
the chance for agreement in one dimension is already small, but
it is negligible in two dimensions.

For the great majority of the cases for which the ``putative host'' was
 a companion to the host, the second measurement would immediately make this
clear.  Combining the putative-host/source relative proper motion 
derived from the two epochs with the putative-host position in the first,
one could derive the position of the putative host relative to the source
at $t_0$, likely with errors of order 1 mas.  In almost all cases, this
separation would be much larger than 1 mas because such a close companion
would have given rise to strong microlensing effects during the event.
Thus, the fractional error in the projected angular separation between
the host and its companion would be small.  Unfortunately, this additional
information would not help break the mass-distance degeneracy, although
it would give a measurement of $\bmu_{\rel,\hel}$ and so (via
Equation~(\ref{eqn:mpirel1})) a good estimate of $\theta_\e$,
which would be especially important for events without a $\rho$ measurement.
Note that in most cases, the internal motion of this binary would be
very small compared to its motion relative to the source.

At this point, there would be three choices: drop the event from the sample,
include it as a probabilistic measurement with two constraints ($\theta_\e$
and upper limit on the $K$-band flux), or perform further investigatory work.
This might include taking substantially deeper images in the hope of
detecting the host or a spectrum of the companion to the host.
Such a spectrum (combined with the $K$ magnitude) could give a reasonable
good estimate of the system distance.  Combining this with the $\theta_\e$
measurement would then yield the host mass.

Because each of these cases (companion to host, companion to source, and
random field star) would be resolved individually by a second AO
epoch, it follows immediately that all such possibilities would be
resolved simultaneously for the case that more than two stars were detected.

If the late-time imaging were done in a second band, e.g., $J$,
then it could also aid in resolving the issues of confusion between
the source and host.  Of course, before conducting such observations,
one would have to assess that it would be likely to re-detect the
host, particularly if it were faint and red and/or the field had 
high extinction.  However, if feasible, measurements of the $(J-K)$
colors of the two stars, combined with an empirical $IJK$ color-color diagram,
would allow precise predictions of their $I$-band magnitudes, which could
then be compared with the one derived for the source during the event.
Moreover, such $J$-band second epochs would be generally useful, as they
would allow for an independent check on the mass and distance estimates
from the ``basic method''.

There is another class of events for which such a $(J-K)$ measurement of
the source could be useful.  For some high-magnification events, the product
of the source flux and the Einstein timescale, $f_S t_\e$ is much better
measured than either parameter separately.  For such events, $t_q\equiv q t_\e$
is also usually an invariant \citep{mb11293}.  If $I_S$ were determined
from the $J$ and $K$ measurements (as just described), this would then
both greatly reduce the uncertainty in $t_\e$, which enters the host-mass
determination via Equation~(\ref{eqn:mpirel1}), and reduce the uncertainty
in $q$ (and hence in $M_{\rm planet} = q M_\host$).

Finally, late time images, particularly in a second band, such as $J$,
could clarify the small subset of cases for which the source is 
corrupted by a nearby (i.e., unresolved) companion.  If the companion
is as bright or brighter than the source, then it will substantially
change the measured flux of the (apparent) source.  However, even
if the companion is a magnitude or so fainter than the source, it can still
affect the astrometry.  For example, at a separation of 10 mas and
at 1 magnitude fainter, such a companion would most likely not be separately
resolved and would displace the measured ``source'' position by about 3 mas.
Moreover, the $\sim 0.35\,$mag ``excess'' in $K$-band flux might also
escape notice, depending on the precision of the original light-curve modeling.
Then one might, at least initially, attribute this astrometric offset
to a binary companion to the host.  In fact, such a small offset would
almost certainly be ruled out by the lack of binary-microlensing signatures
in the original light curve, as discussed above.  However, much larger
offsets are possible if the source companion is brighter than the source.

Such large excesses in the (apparent) source flux would, in most cases,
be easily recognized from the prediction of $K_S$ flux (from
$I_S$ and $(V-I)_S$).  However, if $(V-I)_S$ was not measured or was
poorly measured, then a late-time $(J-K)$ color could determine whether
the measured $K$ flux was indeed excessive.

\section{{Analytic Error Estimates}
\label{sec:errors}}

In this section I give analytic error estimates for the
measurements of $M_\host$ and $D_L$ due to measurement uncertainties in $t_\e$,
and theoretical uncertainties in the $\pi_S$ (equivalently, $D_S$), in the
extinction profile $A_K(\pi_L)$, and in the mass-luminosity relation 
$M_K(M_\host)$.
I initially assume (as will usually be the case) that the errors in the 
measurements of the observed flux, $K$, and in $\bmu_{\rel,\hel}$ are small.
However, I subsequently address the impact if they are not.

The purpose of these estimates is to facilitate comprehensive understanding
of the measurement process.  They should not be applied to determine
the error bars of actual measurements.  Rather these should be determined,
as I outline below, by standard Bayesian procedures.  In particular,
I will adopt simplified forms of the analytic equations in order to make
them more intuitively accessible.   Moreover, the approach given here
ignores the constraints that arise from the measurements of the source 
and lens kinematics, which can reduce the size of the error bars in some
cases.  That said, I believe that this approach yields approximately correct
error estimates for the great majority of cases.

I begin by solving Equation~(\ref{eqn:mpirel1}) for $\pi_\rel$ and
substituting it into Equation~(\ref{eqn:mpirel2}), which yields
an equation for $K$ as a function of $M_\host$ (plus several parameters and
functions).  Strictly speaking this would lead to two equations, one
for each branch.  However, I simplify Equation~~(\ref{eqn:mpirel1}) by
setting $\bv_{\perp,\oplus}\rightarrow 0$.  This is not strictly necessary
from a mathematical standpoint, but
it makes the results much more transparent.  It has little practical impact in
most cases because the second term in this equation is usually small.
There is an important subclass of events for which this is not a good
assumption.  However, this subclass is most easily identified by considering
the simplified formalism, and it is best addressed separately.  
See Section~\ref{sec:hofm0}.
I also suppress the subscript ``host'' on $M$
in the interests of readability.  This yields
\begin{equation}
K = M_K(M) - 5\log\biggl({\pi_L\over\mas}\biggr) + 10 + 
A_K(\pi_L); 
\qquad
\pi_L = {(\mu_{\rel,\hel} t_\e)^2\over (1+q)\kappa M} + \pi_S.
\label{eqn:mpirel3}
\end{equation}
For purposes of this section, I adopt the \citet{baraffe15}
1-Gyr, solar-metallicity $K$-band
isochrone, which is illustrated in Figure~\ref{fig:klogm}.  Note that
over the mass range $-0.4\ga\log(M/M_\odot) \ga -0.9$, the $K$-band 
mass-luminosity relation is given by $M_K = 4.32 - 5\times \log(M/M_\odot)$,
i.e., $L_K \propto M^2$.  Because this will play an important role in the
error analysis, it is worthwhile to remark on the origins this 
``episode'' of power-law behavior in the mass luminosity relation.

Because $M\propto R$ for main-sequence stars, this power-law region
can be expressed as $L_K\propto R^2$, i.e., constant surface brightness
(in the $K$ band) as the mass changes.  Physically, this implies
constant temperature.  Indeed, one finds from the \citet{baraffe15}
isochrones that the temperature evolves slowly in this interval,
although it is not strictly constant.  From a stellar interiors point
of view, this nearly constant temperature arises from a fully convective
interior supported by ideal-gas pressure, $P=nkT$.  Starting at about
$M\sim 0.35\,M_\odot$ the radiative zone gradually expands (until convection
is eliminated at about $M\sim 1.3\,M_\odot$).  Below about $M\sim 0.13\,M_\odot$,
stars are increasingly supported by degeneracy pressure (until this becomes
dominant in the BD regime, $M\la 0.075\,M_\odot$).

I begin by finding the change $\delta K$ from a combination of small changes 
$\delta M$, $\delta\pi_S$, and $\delta t_\e$
\begin{equation}
\delta K =
{\partial K\over \partial M}\delta M +
{\partial K\over \partial \pi_S}\delta \pi_S  +
{\partial K\over \partial t_\e}\delta t_\e 
\label{eqn:deltas}
\end{equation}
where
\begin{equation}
{\partial K\over \partial M} = M_K^\prime + Z{\pi_\rel\over M};
\quad
{\partial K\over \partial \pi_S} =  -Z;
\quad
{\partial K\over \partial t_\e} = - 2Z{\pi_\rel\over t_\e};
\qquad
Z\equiv {5\over \pi_L\ln 10} - A^\prime_K.
\label{eqn:deltas_eval}
\end{equation}

\subsection{Error Induced by $\pi_S$}
\label{sec:pis}
I evaluate the effect of an error $\delta\pi_S$ by considering all other
parameters and functions to be correct and finding the required change in
$M$ to enforce $\delta K=0$:
\begin{equation}
{\delta M\over M} = {\delta \pi_S\over \pi_\rel + (M/Z)M^\prime_K}.
\label{eqn:dmdpis}
\end{equation}

As I now show, the role of $A^\prime_K(\pi_L)$ is small, and therefore,
in order to facilitate analytic treatment, I set 
$A^\prime_K \rightarrow 0$ (i.e., $Z\rightarrow 5/(\pi_L\ln 10)$).  
First, a substantial majority of lenses are behind nearly all the dust,
in which case $A^\prime_K$ is negligibly small.  Second, for very nearby lenses
($D_L\la 1\,\kpc$), $dA_K/d D_L \sim 0.1/\kpc$, so that the ratio of
the extinction term to the distance term in $Z$ is 
$\sim (0.1/\kpc)/[5/(\ln(10)D_L)]\simeq 0.046(D_L/\kpc)$.  For typical
lines of slight, the relative effect peaks at about $D_L\sim 3\,\kpc$,
where $dA_K/d D_L \sim 0.05/\kpc$, so the ratio is about 7\%.  While
the effect of dust certainly cannot be excluded from mass and distance
estimates, its impact on the errors in these quantities (the focus
of the present section) is almost always small, 
and so can be ignored.  I note, however,
that the sign of the effect of ignoring the dust is to slightly
underestimate the size of the errors (typically by a few percent).
That is, at fixed $\theta_\e$, the predicted flux of the lens
$F\propto L/D^2$
will increase with distance because its increasing luminosity, $L$, (from
higher mass) dominates over the distance term in the denominator.  This
increase is reduced by the increasing column of dust with distance,
thus reducing the leverage of the $K$-band flux measurement.

\begin{figure}
\plotone{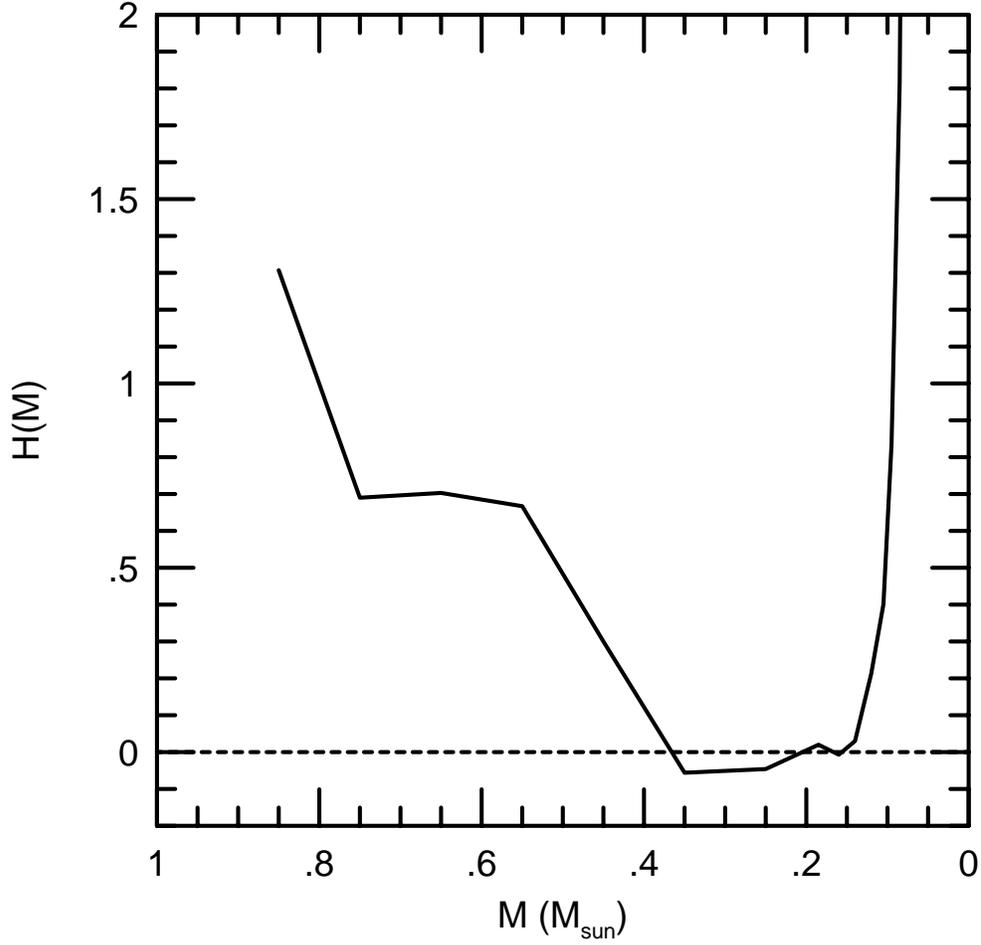}
\caption{The function $H(M)=-(5dM_K/d\log M +1)$, which plays
a critical role in the error propagation.  In particular, $H(M) = 0$
for the mass range $-0.4\ga\log(M/M_\odot) \ga -0.9$, i.e.,
$0.4\ga(M/M_\odot)\ga 0.13$.  See, e.g.,  Equations~(\ref{eqn:dmdpis2}),
(\ref{eqn:dmdte}), and (\ref{eqn:dmdk}).
}
\label{fig:hofm}
\end{figure}

After a small amount of algebra, Equation~(\ref{eqn:dmdpis}) then becomes
\begin{equation}
{\delta M\over M} = {-\delta \pi_S\over \pi_S +  H(M)\pi_L}
= {\delta D_S\over D_S}\biggl[1 + {D_S\over D_L}H(M)\biggr]^{-1};\quad
H(M) \equiv -{1\over 5}{dM_K\over d\log M} - 1.
\label{eqn:dmdpis2}
\end{equation}
The function $H(M)$ is shown in Figure~\ref{fig:hofm}, as calculated
from the online discrete representation of the \citet{baraffe15} isochrones.
I note that these isochrones are in excellent agreement with the 
high-precision M-dwarf mass measurements of \citet{benedict16} over the mass 
range $0.1< M/M_\odot\la 0.45$.  However, \citet{benedict16} show a cluster of
5 stars near $M\sim 0.6\,M_\odot$ that lie about 0.15 mag below the 
\citet{baraffe15} prediction.  Reconciliation of these details, as well as 
precision testing of the \citet{baraffe15} isochrones for 
$0.6\la  M/M_\odot\la 0.9$,
should be carried out before employing them in practice.  However,
from the present perspective, I am only concerned with the general
form of $H(M)$.  The main features are that 
\begin{equation}
H(M)\simeq 0
\qquad
(0.13< M/M_\odot < 0.4),
\label{eqn:hofm}
\end{equation}
and that it rises toward both lower and higher masses.
Adopting $\sigma(D_S)/D_S\sim 12\%$ for typical lines of sight,
one sees that the error induced by this uncertainty in $D_S$ is a 
maximum in the mass range shown in Equation~(\ref{eqn:hofm}), where it is
$\sigma(M)/M \sim \sigma(D_S)/D_S \sim 12\%$, and that it is substantially
smaller away from this mass range.

Using $\delta\pi_\rel = -\pi_\rel(\delta M/M)$ and 
$\delta \pi_L = \delta\pi_\rel + \delta\pi_s$, one obtains
\begin{equation}
{\delta D_L\over D_L} 
= {\delta D_S\over D_S}{1 + H(M)\over 1 + H(M)(D_S/D_L)}.
\label{eqn:dmdpis3}
\end{equation}
Hence, for bulge lenses ($D_L\sim D_S$), the error in $D_L$ is essentially
the same as the error in $D_S$ (as one would naively expect) and, in particular,
is independent of the mass.  For disk lenses, the fractional error in
$D_L$ is the same as that of $D_S$ for the mass range shown in 
Equation~(\ref{eqn:hofm}), while it is lower for other masses (because
$H(M)>0$ and $D_S/D_L > 1$).

\subsection{Error Induced by $t_\e$}
\label{sec:tE}

Before beginning, I note that the fractional errors in $t_\e$ that are
derived from microlensing light curves cover a very wide range, from
$<1\%$ to several tens of percent.  In the former case, their impact
on the errors in $M_\host$ and $D_L$ are negligible compared to those
of uncertainties in other quantities, such as $D_S$.  In the latter,
the present perturbative approach cannot fully capture their impact.
Furthermore, as I detail below, there can be additional information
that is not captured by this formalism and that can substantially reduce
the impact of the $t_\e$ errors.

From Equation~(\ref{eqn:deltas_eval}), 
$\partial K/\partial t_\e = 2(\pi_\rel/t_\e)(\partial K/\partial \pi_S)$.
Hence, we can immediately transform Equation~(\ref{eqn:dmdpis2}) to obtain
\begin{equation}
{\delta M\over M} = {-2(\delta t_\e/t_\e)\pi_\rel\over \pi_S +  H(M)\pi_L}
= -2{\delta t_\e\over t_\e}{1 - (D_L/D_S) \over H(M) + (D_L/D_S)}.
\label{eqn:dmdte}
\end{equation}
For bulge lenses $(D_L\sim D_S)$, the impact of the $t_\e$ errors on the
mass estimate is very small.  The physical reason is clear: if the host
is approximately at the bulge distance then the measurement of $K$ directly
gives the mass, without any significant
input from the $\theta_\e$ determination.  However, for disk lenses, the
impact can be far larger, particularly for those in the mass range of
Equation~(\ref{eqn:hofm}), for which
Equation~(\ref{eqn:dmdte}) yields
$\sigma (M)/M\rightarrow 2[\sigma (t_\e)/t_\e](D_S/D_L-1)$. 
This can be quite problematic for events with large fractional errors in $t_\e$.

The fractional error induced in $D_L$ is 
\begin{equation}
{\delta D_L\over D_L} 
= 2{\delta t_\e\over t_\e}{[1 - (D_L/D_S)]^2 \over H(M) + (D_L/D_S)},
\label{eqn:dmdte2}
\end{equation}
i.e., smaller than Equation~(\ref{eqn:dmdte}) by a factor
$[1-(D_L/D_S)]$.

I note that the late-time AO measurement of $\bmu_{\rel,\hel}$
can substantially improve the determination of $t_\e$ relative
to what was possible based on the microlensing light curve alone.
For high-magnification events of faint sources, $t_*\equiv \rho t_\e$
may be well-measured (an invariant), even if $\rho$ and $t_\e$
have large errors because this ``source self-crossing time''
is directly constrained by the interaction of the source with the
caustic.  If we, for the moment, ignore the difference between
$\bmu_{\rel,\hel}$ and $\bmu_\rel$, then the angular source size
is given by the product $\theta_* = \mu_{\rel,\hel}\times t_*$, both of
which are well-measured.  However, $\theta_*$ is related to
the dereddened source color and magnitude $[(V-I),I]_{S,0}$ 
by $\theta_* \propto 10^{-I_{S,0}/5} G(V-I)_{S,0}$ where $G$ is
a known, empirically determined function.  Thus, $I_{S,0}$ can
be determined.  Then, inverting standard procedures 
(e.g., \citealt{ob03262}), one can determine the instrumental
magnitude $I_S$, using the intrinsic magnitude of the clump
(from Table~1 of \citealt{nataf13}) and the location of the
clump on an instrumental CMD.  This equivalently yields the
source flux, $f_S$ in the instrumental system.  For such
high-magnification events with invariant $t_*$, it is also
the case that $f_S t_\e$ is an invariant \citep{mb11293}.
Hence, because $f_S$ is well-determined, so is $t_\e$.
For an actual event, the correction from heliocentric
to geometric proper motion would have to be included,
but in only rare cases would this significantly undermine the
above logic.

In the cases for which $t_\e$ is poorly measured but $f_S t_\e$ is an
invariant, one can also strongly constrain $t_\e$ by ``directly'' measuring
$f_S$.  In fact, this cannot be done within the context of the ``basic
method'' (i.e., single, late-time $K$-band image) because, in order to
infer $f_S$ in the $I$-band from $K$-band photometry, one also needs
photometry in a second band that is closer to $I$, e.g., $J$.  However,
it is a broadly applicable approach, if it is needed.

This again emphasizes the importance of evaluating the 
mass and distance in the context of a full Bayesian analysis,
including taking account of the correlations in the microlensing
light-curve parameters, including especially those between
$f_S$, $t_\e$. $\rho$, and $q$.  This is particularly important
if the fractional error in $t_\e$ is large.  For these cases,
it is probably fine to use the invariant quantities and their
error bars if these are reported in the original paper.  If not,
they should be re-derived from new light-curve modeling.

Finally, I note that the most difficult cases, i.e., nearby lenses
in the mass range of Equation~(\ref{eqn:hofm}), are also intrinsically
rare (because they occupy little phase space) and are the most
accessible to other constraints and measurements.  For example,
a host at $D_L\sim 2\,\kpc$ and at the bottom of this mass range
would have $K_\host\sim 19$ and a large lens-source relative parallax,
$\pi_\rel \sim 0.4\,\mas$, making it susceptible to a precise
trigonometric parallax measurement.  Indeed, its microlensing
parallax would also be large, $\pi_\e\sim 0.6$, implying that it was
likely to have been measured during the event, a measurement that could
be greatly improved by imposing the directional constraint from the
late-time measurement of $\bmu_{\rel,\hel}$.  Finally, such ``bright''
(by 30m standards) hosts would be good spectroscopic targets.

\subsection{Errors Induced by $M_K(M_\host)$ and $A_K(\pi_L)$}
\label{sec:mofk}

Because $M_K(M_\host)$ and $A_K(\pi_L)$ are functions rather than parameters,
they can in principle differ from their assumed forms in very complex
ways.  However, in order to illustrate the role of these potential 
differences, I will assume that locally, i.e., in the neighborhood
of the mass and distance of the actual lens, they differ by some
offset, $\delta K$, and that the derivatives of these functions are
unaltered.  Then, the induced error in $M$ is simply 
$\delta M = -\delta K/(\partial K/\partial M)$.  Following the same
procedures as in the previous two subsections, I find,
\begin{equation}
{\delta M\over M} = {-\delta K/M\over -M^\prime_K - Z\pi_\rel/M}
\rightarrow {-(\ln 10/5)\delta K\over H(M) + D_L/D_S}.
\label{eqn:dmdk}
\end{equation}

Regarding $A_K(\pi_L)$, a conservative estimate is $|\delta K| \la 0.07$.
First, for typical lines of sight, the full column of dust toward the
bulge is only $A_K\sim 0.3$, and there are few lines of sight that
are substantially above $A_K\sim 0.5$.  Second, the great majority
of the lenses are either in the bulge (i.e., behind essentially
all the dust) or in the disk but well over 100 pc from the Galactic
plane (so behind most of the dust).  The total extinction toward the
bulge is well measured from the position of the clump \citep{gonzalez12}, so 
for these two classes of lenses, $A_K(\pi_L)$ is either the full extinction or
slightly less than this.  Third, given the small total extinction,
a continuous dust model (based, e.g., on a constant dust scale height)
would be adequate to achieve 0.1 mag accuracy, even in the near-disk
regions. Fourth, it is likely that there will be good 3-D dust maps
toward the bulge, based on Gaia data, by the time of first AO light on 30m
class telescopes.

Based on the scatter in Figure~22 from \citet{benedict16}, I estimate
$|\delta K| \la 0.07$ for $M_K$ and thus a quadrature sum
$|\delta K| \la 0.1$ for both effects combined.  This implies that
the numerator in Equation~(\ref{eqn:dmdk}) is limited by
$|(\ln 10/5)\delta K| \la 0.045$.

From the functional form of Equation~(\ref{eqn:dmdk}), the errors
are most severe for very nearby lenses in the mass range of 
Equation~(\ref{eqn:hofm}).  That is, this source of error is qualitatively
similar to the $t_\e$-based error that was just discussed in 
Section~\ref{sec:tE}, except that the numerator of Equation~(\ref{eqn:dmdk})
is more constrained.  Hence, exactly the same remarks apply about the
most difficult cases as in the close of that section.

Finally, I note that the fractional distance error is smaller than
the fractional mass error,
\begin{equation}
{\delta D_L\over D_L} = 
{-(\ln 10/5)\delta K\over H(M) + D_L/D_S}
\biggl(1 - {D_L\over D_S}\biggr).
\label{eqn:dmdk2}
\end{equation}

\subsection{Error Induced by $\bmu_{\rel\hel}$ and $K_\host$}
\label{sec:mu+K}

While the uncertainties in $\bmu_{\rel,\hel}$ and $K_\host$ are generally 
expected to be small, it is straightforward to evaluate their impact using the
above analysis.

Because $\mu_{\rel,\hel}$ and $t_\e$ appear symmetrically in
Equation~(\ref{eqn:mpirel3}), the impact of the
uncertainty in $\mu_{\rel,\hel}$ is directly given by 
Equations~(\ref{eqn:dmdte}) and (\ref{eqn:dmdte2}) in Section~\ref{sec:tE}
(with the substitution $\delta t_\e/t_\e \rightarrow
\delta\mu_{\rel,\hel} /\mu_{\rel,\hel}$).

Given the method I have used to estimate the effects of uncertainties
in the mass-luminosity relation and the dust profile, the impact of the
uncertainty in $K_\host$ is directly given by Equations~(\ref{eqn:dmdk}) and
(\ref{eqn:dmdk2}) in Section~\ref{sec:mofk}.

\subsection{{Character of the $H(M)\sim 0$ Regime}
\label{sec:hofm0}}

All the equations for error propagation that are derived above 
for the mass estimates have the form,
\begin{equation}
{\delta M\over M} \propto {1\over H(M) + D_L/D_S}
\label{eqn:hofm0}
\end{equation}
Hence, in the mass interval of Equation~(\ref{eqn:hofm}), the errors
diverge for $D_L\ll D_S$,  implying that a practical understanding of this
regime is important.  From a mathematical perspective, this divergence
occurs because Equations~(\ref{eqn:mpirel1}) and (\ref{eqn:mpirel2})
intersect at an acute angle, or, stated otherwise, are nearly parallel
at their point of intersection.  This is because, in the approximation 
$\bv_{\oplus,\perp}\rightarrow 0$, $\log M = -\log \pi_\rel + const$, while
in the approximation $A_K\rightarrow 0$, and in the regime of $H(M)\simeq 0$
(i.e., $L_K\propto M^2$), $\log M = -\log \pi_L + const$.  Hence,
for $D_L\ll D_S$ (so $\pi_L\sim \pi_\rel$), the slopes are similar.
This means that the various other terms that affect the slopes cannot
really be ignored in this regime.  In particular, the assumption
$\bv_{\oplus,\perp}\rightarrow 0$ is actually
$\pi_\rel\bv_{\oplus,\perp}\rightarrow 0$, and this is not consistent
with the regime $D_L\ll D_S$, unless $\bv_{\oplus,\perp}$ is identically zero.

Considerable insight into this regime can be gained by examining the subpanels for
OGLE-2005-BLG-071 and OGLE-2012-BLG-0950 in Figure~\ref{fig:all1}.
In both cases, there is an extended region where the two curves
are basically tangent.  These are centered at 
$(M/M_\odot,D_L/\kpc)\sim (0.35,2.7)$ and (0.45, 1.7), respectively, i.e.,
near the upper edge of Equation~(\ref{eqn:hofm}).  In both cases,
particularly for OGLE-2012-BLG-0950, Equation~(\ref{eqn:mpirel1})
curves upward in the near-tangent region, which reflects the
fact that $\pi_\rel$ is not much smaller than $\pi_{\rel,\min}$, i.e.,
the minimum of this curve.  And in both cases, Equation~(\ref{eqn:mpirel2})
curves downward in this region.  Hence, in both cases, had $K_L$ been
somewhat brighter, there would have been a discrete two-fold degeneracy,
with the higher-mass solution well above the regime of 
Equation~(\ref{eqn:hofm}) and the lower-mass solution being at the low-mass
end of this regime.  Although there are no such cases among the six historical
examples, they are likely to be relatively common in a large sample.
These discrete degeneracies can be addressed by the methods discussed
in Section~\ref{sec:degen}.

However, the continuous degeneracies represented by the actual cases
of OGLE-2005-BLG-071 and OGLE-2012-BLG-0950 cannot be properly captured
by the perturbative approach of the present section.  Rather, they require
direct numerical estimates.  Therefore, I will compare
the idealized treatments given here with the numerical solutions for these
two events that were given by \citet{ob05071c} and \citet{ob120950b},
respectively.  Interestingly, as noted in Section~\ref{sec:basic},
neither paper presents a version of Figure~\ref{fig:all2}, although
such figures are common in papers that derive lens properties from
high-resolution imaging.  The reason is that each event had substantial
$\bpi_\e$ information, and the authors of both papers focused
on incorporating this information to make the best estimate of the host
mass and distance, rather than asking what could be learned from
AO measurements alone.  In particular, when there is a 1-D $\bpi_\e$
measurement that is strongly inconsistent with zero, then a precise
measurement of $\bmu_{\rel,\hel}$ can yield very good measurements
of the lens mass and distance without any photometric information
\citep{ob03175,gould14}.

Here, however, my goal is to understand the challenges to applying the
``basic method'', and I evaluate the role of $\bpi_\e$ and other auxiliary
information in that context.  From this perspective, it is useful to derive
the mass and distance estimates from the ``basic method'' and then check to 
see whether (or to what extent) they are confirmed by the $\bpi_\e$ measurement.

\begin{figure}
\plotone{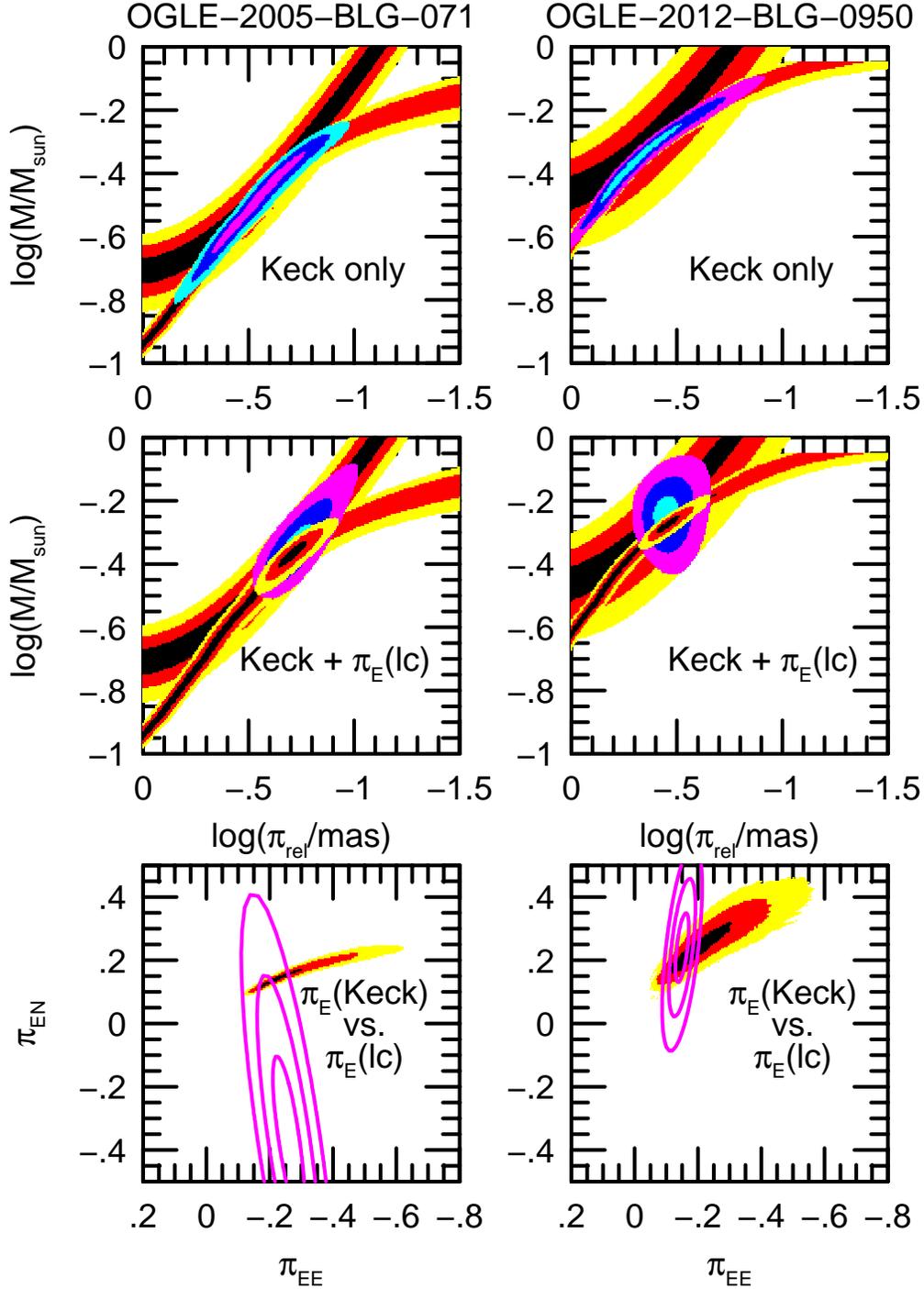}
\caption{Two Keck AO mass measurements with continuous degeneracies
that are broken by light-curve $\bpi_\e$ measurements:
OGLE-2005-BLG-071 (left) and OGLE-2012-BLG-0950 (right).
Top row: Equations~(\ref{eqn:mpirel1}) and (\ref{eqn:mpirel2}), i.e.,
``bottom'' and ``middle'' contours, respectively, are combined to yield
mass measurements (``top'' contours)
with $1\,\sigma$ ranges of $\Delta\log M = 0.4$ and 0.3, respectively.
Bottom row:  Combined solution from top row is projected on the $\bpi_\e$
plane (filled contours), where it can be compared to $\bpi_\e$ contours
derived from the light curve (open contours).  In both cases, the two
sets of contours are roughly orthogonal, so the combination is strongly
constrained.
Middle panel: ``top'' contours are the result of combining 
Equations~(\ref{eqn:mpirel1}) and (\ref{eqn:mpirel2}), i.e., two ``bottom''
sets of contours, with the $\bpi_\e$ constraint from the light curve.
The second level contour is the combination of this constraint
with just Equation~(\ref{eqn:mpirel1}), i.e., the
$\bmu_{\rel,\hel}$ measurement.}
\label{fig:2par}
\end{figure}

The top panels in Figure~\ref{fig:2par} show the tracks of 
Equations~(\ref{eqn:mpirel1}) and (\ref{eqn:mpirel2}) on the 
$\log M$-$\log\pi_\rel$ plane, i.e., the same as in Figure~\ref{fig:all1},
except here showing error contours.  The ``top level'' contours
(displayed in a different color scheme) show the result of combining the
two constraints, i.e., adding the $\chi^2$ values from each set of 
contours.  In both cases, there is a strong 1-D degeneracy at the
$1\,\sigma$ level, which arises because 
Equations~(\ref{eqn:mpirel1}) and (\ref{eqn:mpirel2}) are both slightly
curved and nearly tangent.  For this reason, the 2-$\sigma$ and 3-$\sigma$ 
errors are not proportionately larger than the 1-$\sigma$ errors

The middle panels show the impact of including the light-curve based
$\bpi_\e$ measurements.  The ``top level'' contours show the resulting
error contours: they are dramatically reduced in the direction of
the 1-D degeneracy.  The physical origins of this reduction is
made clear in the bottom panels where the filled contours show
the $\bpi_\e$ contours derived from Equations~(\ref{eqn:mpirel1}) and 
(\ref{eqn:mpirel2}), while the open contours show the $\bpi_\e$
constraints from the light curve.  In both cases, these are nearly
orthogonal, so their combination greatly restricts the modulus of
$\bpi_\e$, which is essentially the direction of the 1-D degeneracies
in the top panel, i.e., $\log\pi_\e \sim (\log \pi_\rel - \log M)/2$.
Returning to the middle panel, the ``second level'' contours (in a different
color scheme) show the result of combining the light-curve $\bpi_\e$
constraint with Equation~(\ref{eqn:mpirel1}) alone, i.e., the method
of \citet{ob03175} and \citet{gould14}.

For both events, I constructed the light-curve $\bpi_\e$ contours
(magenta) by making analytic representations of the parallax figures
from \citet{ob05071b} and \citet{ob120950b}.  In the first case, this
was straightforward and would not warrant particular mention.  However,
for OGLE-2012-BLG-0950, the numerical parallax contours are quite complex,
partly because the two solutions, which arise from the ``ecliptic degeneracy''
\citep{ob03238}, overlap and partly because the solution suffers from 
the ``jerk-parallax degeneracy''
\citep{gould04}, which typically generates two well-separated
minima along the $\pi_{\e,\perp}$ direction.  In the present case, the two minima
are barely resolved, with the southern solution favored by roughly 
$\Delta\chi^2\sim 2$, which is not significant.
As the southern solution is ruled out by the Keck observations, I
made my analytic representation using the northern solution.

Figure~\ref{fig:2par} illustrates the key issues regarding the
most difficult cases of applying the ``basic method''.

First, it shows that these difficult cases are closely associated with
the ``power-law episode'' in the mass-luminosity relation, i.e.,
the $0.13 \la M/M_\odot \la 0.4$ interval of Equation~(\ref{eqn:hofm}), but they
can extend beyond it.  In particular, for OGLE-2012-BLG-0950, the actual
mass, $M\simeq 0.56\,M_\odot$, lies substantially beyond this mass
interval.  However, this mass value is only known via the incorporation
of the light-curve $\bpi_\e$ constraint: using the ``basic method'' alone
(and only using the Keck $K$-band data as I have done and ignoring the 
{\it HST} data), the range of solutions broadly overlaps 
Equation~(\ref{eqn:hofm}) at the $1\,\sigma$ level.  This is because,
while the local slopes of the two equations have begun to diverge at the
true value of $M_\host$, the curves themselves remain consistent within their 
error bars.

Second, these cases illustrate the possibility of discrete degeneracies
for which both solutions have $\pi_\rel <\pi_{\rel,\min}$, even though
neither is an example of such a case.  That is, in both cases, if the
true host mass had been greater (so that the measured $K$-band flux
would have been brighter and thus Equation~(\ref{eqn:mpirel2}) would
have been higher on the plots), then the curves would have intersected
in two locations, i.e., to the left and right of the current 1-$\sigma$
range.  This possibility arises because (for the regime of 
Equation~(\ref{eqn:hofm})), the slope of Equation~(\ref{eqn:mpirel2})
is very nearly $-1$, while the slope of Equation~(\ref{eqn:mpirel1}) 
is substantially less negative than $-1$ (because $\pi_\rel$ is not truly 
small compared to $\pi_{\rel,\min}$).
In the present two cases, this discrete degeneracy
would still have easily been broken by the light-curve $\bpi_\e$ measurement.
In both cases, these good (albeit 1-D) $\bpi_\e$ measurements were
facilitated by the events having relatively bright $(I_{S,0}\sim 18.5)$,
mildly extincted $(A_I\la 1)$, well-magnified $(A_\max\ga 10)$ sources,
and, especially, long ($t_\e\sim 70\,$day) timescales.  Regarding
the first three characteristics, conditions will vary strongly for
other events.  Regarding the long timescales, these were the ratio of
rather large $\theta_\e$ ($\sim 1\,\mas$) with fairly typical $\mu_\rel$.
In turn, the large $\theta_\e$ were primarily due to the fact that these
are both disk lenses (large $\pi_\rel$).  For the case of discrete
degeneracies, in which the true solution is the more massive (and
more distant) lens, this is less likely to be the case.  However,
it will still be the case that the alternate (incorrect) solution
will predict a large $\pi_\e$, so that even if $\bpi_\e$ is
not subject to even a 1-D measurement, the alternate solution can be
ruled out.  Thus, at least for events with relatively high-quality
light curves, it may often be possible to break this discrete degeneracy.


\section{{A Practical Approach}
\label{sec:practical}}

To transform the mass-ratio measurements that are routinely
derived from microlensing light-curve analyses into planet masses
requires late-time imaging for essentially the entire sample.
There are a relative handful of events that have mass measurements
from the microlens parallax effect, but even these should probably
be cross-checked with the imaging method for better understanding
of potential problems of both methods.

This effort can begin with a single late-time image for each event.
This imaging should wait until there is a reasonable expectation that
the host and source will be separately resolved.  In particular, one
must balance the chance that the lens will be bright enough to detect
(given the glare of the source), at whatever separation is expected,
against the cost of failed observations.  For dwarf-star sources,
separation of 1.3 times the diffraction-limited FWHM is a reasonable 
threshold because a large fraction of potential hosts will be visible at 
this separation.  In the $K$ band, this corresponds to
$\Delta\theta = 72\,(D/10{\rm m})^{-1} \,\mas$, where $D$ is the mirror
diameter.

For events with
light-curve based $\rho$ measurements, the decision on the wait time
will rest on the $\mu_\rel$ estimates, in addition to the
characteristics of the observing instrument.  For those events 
without $\rho$ measurements (roughly 1/3 of the final sample),
one must adopt a conservative
lower limit, e.g., $\mu_{\rel,\hel}\ga 2\,\masyr$, in order to minimize the
chance of wasting extremely valuable telescope time.  Thus, for
planets discovered in 2016--2022, the conservative estimates of
separation in 2030 are 28--16 mas.  Hence, applying the
1.3-FWHM criterion, we immediately see that it would be inappropriate
to image any of these events that lack $\rho$ measurements prior
to AO first light on 30m class telescopes.  On the other hand, it will
be appropriate to do so at first AO light on these telescopes, or shortly
thereafter, on essentially all of them.

Nevertheless, it will be feasible to image some fraction
of the events that have $\rho$ measurements using current instruments.
Some of these observations will reveal only the source.  
The most likely explanation
for this outcome is that the proper-motion has
been correctly estimated, but the lens is too faint to be detected at the
inferred separation.  Hence, the indicated response is to wait for 30m AO
to obtain a second epoch of imaging.
Some will have ambiguous implications because more than
two stars are detected.  However, as discussed in Section~\ref{sec:second}, 
these ambiguities can be resolved by a second AO epoch taken after two years
or so.  Others will have ``red flags'', such as an inconsistency between
the heliocentric (from imaging) and geocentric (from the light-curve analysis)
lens-source relative proper motions, or an inconsistency between the
heliocentric proper motion and constraints from the microlens-parallax
analysis.  These, likewise, can be investigated by additional epoch of imaging
two or so years later.

However, it is likely that a substantial fraction of these current-instrument
observations will yield
good host-mass measurements and so (unless the errors in $q$ are very large)
good planet-mass measurements, and these can be the basis for 
preliminary planet-mass and host-mass function studies.  Once 30m-class AO
is available, then first epochs can be obtained for the rest of the sample,
while those requiring 30m-class second epochs can also be observed, which
will lay the basis for comprehensive analyses of the sample as a whole,
including events without $\rho$ measurements.  In a few cases of ``red flags''
or non-detection of hosts with giant-star sources,  there may be reason
for second-epoch 30m observations.

From the pattern of host-mass measurements as a function of $\theta_\e$,
it will then likely be possible to distinguish which non-detections are
due to BDs and which to WDs (or more massive remnants).  For example,
\citet{kb190371} showed that events with $\theta_\e,\la 100\,\mas$ (and
$\mu_\rel \la 10\,\masyr$) are expected to have late M-dwarf or BD
hosts.  If this expectation is mostly confirmed for those with detections
(including those without $\rho$ measurements),
then the remainder of those without detections but with $\rho$ 
measurements can be inferred, with good confidence, to have BD hosts.
For events that lack both $\rho$ measurements and host detections,
a similar analysis can be conducted based on $t_\e$ alone (together
with constraints on $\bpi_\e$, when available), although these
designations will generally be less secure.

Thus, while the full analysis of the sample must await 30m AO observations,
initial progress can be made using present facilities, particularly
during the latter part of the 2020s.  Moreover, by systematically applying
present-day telescopes to events to which these are accessible, the burden
on and duration of the 30m AO observations will be reduced.

With these prospects in mind, I present two comprehensive tables of
planetary events that are both likely to enter the final sample and
are well-analyzed today, i.e., are either published (including on arXiv)
or are in a late state of preparation that I have personally reviewed.
The sample is defined as planets that have been (or will be) found
by the KMT AnomalyFinder \citep{af2} in 2016-2019 and 2021-2022 and
that have ``good'' $\log q$ measurements.  For present purposes, I define these
as having error bars $\sigma(\log q)<0.2$ and discrete degeneracies
(defined as $\Delta\chi^2<10$) of $|\Delta\log q|<0.25$.  I conceptually
include 2022 because the observing season is mostly complete and there
are no major data issues so far, although no planet analyses have yet been
completed for this season.

In addition to the planets that are likely to enter
the statistical sample, which I have just described,
I also include those from the same six seasons
that might enter the sample based on further information derived from AO
observations, as well as a few others that may be of particular interest
for a variety of reasons.  These are indicated by the notes that are 
described below.

Table~\ref{tab:tab1} has 82 events with $\rho$ measurements 
(or strong upper limits), and
Table~\ref{tab:tab2} has 29 events without $\rho$ measurements.  The purpose
of these tables is to support strategic planning of observations of
the sample as a whole.  They are not intended as a substitute for the
original papers, including tables, which contain many more details 
than are summarized in these two tables.  On the other hand, I have endeavored
to provide information directly related to decisions
regarding AO follow-up observations in a compact form, including some
information that is either absent from or not easily extracted from
the discovery papers.

Of all the seasons, only 2018 may
be considered as complete \citep{2018prime,2018subpr}.  As mentioned
above, no planets have yet been properly analyzed from 2022, while
many planets are still being discovered and/or analyzed for the remaining
4 seasons.  I will attempt to provide updated tables and/or supplements
as new planets are analyzed.

Each planet is described by two lines in Table~\ref{tab:tab1}.  
The first column gives the event name in the first row and the
KMT name (if different) in the second row.  The second column gives
$t_0$, i.e., the peak of the event, while the third column gives the
proper motion and its error.  The fourth through sixth columns give the
dereddened magnitude and color of the source and its $K$-band extinction.
The sixth column gives the Galactic coordinates.  The seventh through tenth
give the logarithms of the Einstein timescale $t_\e$ (in days), the impact
parameter $u_0$, the normalized source radius $\rho$, and the mass ratio $q$,
as well as their errors.  All of these quantities are given for the 
lowest-$\chi^2$ solution.  Where there are major differences between solutions
that are within $\Delta\chi^2<10$, these are discussed in the notes.
The next column gives Earth's projected velocity $\bv_{\oplus,\perp}$ at
$t_0$ in (N,E) coordinates and in $\kms$, followed by $M_{\rm cr}$, which
is expressed in units of $0.075\,M_\odot$, i.e., the hydrogen-burning limit,
and which is defined by
\begin{equation}
\label{eqn:mcrit}
M_{\rm cr}\equiv {\theta_\e^2\over \kappa} {v_{\oplus,\perp}\over \au\mu_\rel}
= {\mu_\rel v_{\oplus,\perp}t_\e^2 \over \au\,\kappa}
\end{equation}
The penultimate column
gives 5 codes that are described below, while the final column
gives the discovery reference.

The only quantity in Table~\ref{tab:tab1} that is likely to be unfamiliar
is $M_{\rm cr}$, which
is the lens mass at which $|\bmu_\rel - \bmu_{\rel,\hel}| = \mu_\rel$.
At higher host masses, i.e., $M_\host>M_{\rm cr}$, the correction from 
geocentric to heliocentric will be 
smaller.  Hence, if this quantity is well below the hydrogen-burning
limit (i.e., $\ll 1$ in Table~\ref{tab:tab1}), then one need not be
concerned about this correction because the hosts for which it is significant
cannot be seen anyway.  Otherwise, the correction must be carefully
considered.

The parameters $t_0$, $\mu_\rel$, $\sigma(\mu_\rel)$, $I_{s,0}$, $(V-I)_{s,0}$, 
$A_K$, and the Galactic coordinates $(l,b)$ are
the most important for deciding at what point
the event can profitably be observed.  The first three of these allow
one to estimate the separation and its uncertainty as a function of time,
while the next two allow one to estimate the source type.  In general,
giants will require substantially greater separations than dwarfs.  The
last three of these columns are usually of minor importance, but a large
value of $A_K$ may affect observing strategy, while a small value of $|b|$
may indicate a substantial probability that the source cannot be assumed
to be in the bulge.  Note that the expected source flux in $K$ can be estimated
by combining $[(V-I),I]_{S,0}$, and $A_K$, 
together with tabulated $VIK$ photometry
(e.g., \citealt{bb88}).  However, for purposes of a quick estimate,
one can just approximate $K_S \sim I_{S,0} - (V-I)_{S,0} + A_K$.
Note if the $\mu_\rel$ estimate is a lower limit (rather than a measurement)
it is shown in bold face.

The four parameters $t_\e$, $u_0$, $\rho$, and $q$ (and their errors),
can help understand the potential role of AO observations, as follows.
If the fractional
error in $t_\e$ is small, e.g., $\sigma(\log t_\e)\la 0.02$, then this
error does not undermine the host mass and distance estimates relative
to other sources of error.  See Section~\ref{sec:tE}.  Otherwise, it will
be of interest whether the $t_\e$ error can be reduced via its correlations
with $f_S$ and $\rho$, as described in Sections~\ref{sec:second} and
\ref{sec:tE}.  Then, if the logarithmic errors in $t_\e$ and $u_0$ are
very similar, it is a good indication that $f_S$ and $t_\e$ are strongly
anti-correlated, implying that it may well be possible to reduce the
error in $t_\e$ using this technique.  An analogous logic applies if
the logarithmic errors in $t_\e$ and $\rho$ are similar.  The logarithmic
error in $q$ is of interest for two reasons.  First, a high value of this
parameter would imply that the event is less useful for constraining fine
details of the planet mass function.  Second, if the logarithmic errors
in $q$ are not much bigger than those in $t_\e$, this probably means that
$t_q\equiv q t_\e$ is much better measured than $q$, so that if 
$\sigma(\log t_\e)$ can be reduced by the above-mentioned techniques, then
$\sigma(\log q)$ can be reduced as well.

The projected velocity $\bv_{\oplus,\perp}$ is presented because it is
an input into $M_{\rm cr}$: if the $M_{\rm cr}$ test mentioned above indicates
that the correction from heliocentric to geocentric must be taken into
account, then the vector form of this quantity will be important.

Table~\ref{tab:tab2} is similar to Table~\ref{tab:tab1} except that it omits 
the columns: $\mu_\rel$, $\log\rho$, and $M_{\rm cr}$,
for which there is no information.

Table~\ref{tab:tab3} gives the meanings and distributions of the five
codes in the penultimate columns of Tables~\ref{tab:tab1} and \ref{tab:tab2}.
The first code (0 or 1) indicates whether there is comment in the following
two sections about the event.  The second code gives my evaluation of whether
the event will enter a statistical sample of AnomalyFinder detected
planets with AO imaging.  The third code tells the largest discrete degeneracy
in $\log q$.  My orientation is that if there is no degeneracy, or if
$\Delta\log q < 0.1$, there should be no concern.  Otherwise, I indicate
the magnitude of this parameter. The fourth code gives the lens/source
multiplicity.  Most events are 2L1S.  Those that are 3L1S are divided between
events with two planets and events with planet+binary.  
Here, $n$L$m$S means $n$ lenses and $m$ sources.
The final code tells whether the event has {\it Spitzer} data.
Each classification in Table~\ref{tab:tab3} gives the total number, in 
parentheses, of 
entries in Tables~\ref{tab:tab1} and \ref{tab:tab2} with that classification.
Note, in particular, that there are 4 3L1S 2-planet systems,
which have a total of ``8'' planets.

\subsection{Comments on Events in Table 1}
\label{sec:table1}

\subsubsection{2016}
\label{sec:2016.1}
{\bf MOA-2016-BLG-227}:
\citet{mb16227} conducted Keck AO observations on HJD$^\prime=7613.85$,
i.e., at $\tau = (t-t_0)/t_\e\sim 5.6$ after peak.  They detected excess
flux but concluded that it was not likely to be due to the host.  Regardless
of whether this assessment turns out to be correct, these images form
a valuable first epoch.
{\bf MOA-2016-BLG-319}
has only an upper limit $\rho>0.01$, but this places a significant
constraint on $\mu_\rel$, so it is included in Table~\ref{tab:tab1}.  

{\bf OGLE-2016-BLG-0613} 
is a 3L1S system, i.e., a stellar binary with a planet.  Solution ``C''
is favored over solutions ``B'' and ``D'' by $\Delta\chi^2=10.0$ and
13.2, respectively.  However, its model source flux is about 1.4 mag
too faint, given its color.  The other two solutions are brighter by
0.75 mag and 1.04 mag, respectively, and so they are favored in this sense.
Late-time AO in two bands, one near the $I$ band (e.g., $J$) and the other
at $K$ (to determine the color correction) could easily distinguish
between ``C'' and ``B''/''D''.  Both ``B'' and ``D'' predict proper 
motions that are a factor $\sim 1.5$ larger than ``C'', so this would provide
a second test.  If ``B''/''D'' are preferred, it would be difficult
to distinguish between them.  However, they differ in mass ratio by
only $\Delta \log q_3 = 0.06$, compared to $\Delta \log q_3 = 0.29$, for
``D'' relative to ``C''.  This event entered the AnomalyFinder sample
due to its binary-star component, rather than its planet.  Hence,
the search would have to be expanded to look for planets in all binary
events for this planet to enter a statistical sample.  This may well
occur prior to first AO light on 30m class telescopes.
{\bf OGLE-2016-BLG-0693}:
For this event, the source flux is very poorly measured, so the parameter
combinations $t_\e\times(u_0,\rho,q,f_S) = (5.5,0.15,8.8,2.6)\,$day
are much better measured than $(u_0,\rho,q,f_S)$.  Here, $f_S$ is the flux
on an $I=18$ scale.  In Table~\ref{tab:tab1}, I report the values and errors of
all quantities assuming that $f_{S,\rm OGLE-IV} = 0.0158$, i.e., the 
``fiducial value'' adopted by \citet{ob160693}.  A central goal of AO
observations should be to measure $I_S$, which can be done by observing
in $J$ (a close proxy for $I$) and $K$ (to determine the color correction).
Note that in a free fit, $q\sim 0.02$ which is within the range of study
of current statistical analyses of KMT data, while the fiducial $I_S$
yields $q\sim 0.06$ (which is not).  These issues can only be resolved
by AO observations.
%
%
{\bf OGLE-2016-BLG-1067} 
has a {\it Spitzer} parallax measurement \citep{yee15}.  However,
as there is only an upper limit on $\rho$, this did not yield mass and 
distance estimates, Because the $\rho$ upper limit places a significant 
constraint on $\mu_\rel$, I have included this event in Table~\ref{tab:tab1}.
{\bf OGLE-2016-BLG-1093} 
has a {\it Spitzer} parallax measurement \citep{yee15}, which
yields a host mass, planet mass, and system distance of 
$M_\host \sim 0.46\,M_\odot$, $M_{\rm planet} \sim 0.71\,M_{\rm Jup}$ 
and $D_L\sim 8.1\,\kpc$.   
{\bf OGLE-2016-BLG-1190} 
has a {\it Spitzer} parallax measurement \citep{yee15}, which
yields a host mass, planet mass, and system distance of 
$M_\host \sim 0.91\,M_\odot$, $M_{\rm planet} \sim 13.4\,M_{\rm Jup}$ 
and $D_L\sim 6.8\,\kpc$.   There is substantial orbital motion information
from the fit, and the predicted period and semi-major axis are 
$P\sim 3\,$yr and $a\sim 2\,\au$.  The host is
expected to be $K_\host\sim 18.5$.  Hence, RV observations could confirm
and greatly refine the orbital parameters presented by  \citet{ob161190}
in their Figure~10 and Table~7.
{\bf OGLE-2016-BLG-1195} 
has a {\it Spitzer} parallax measurement \citep{yee15}, which
yields a host mass, planet mass, and system distance of 
$M_\host \sim 0.08\,M_\odot$, $M_{\rm planet} \sim 1.4\,M_\oplus$ 
and $D_L\sim 3.9\,\kpc$ \citep{ob161195b}.  If correct, this 
would imply $K_\host\sim 23.7$, i.e., $\Delta K\sim 6.3\,$mag fainter than
the source.  Moreover, within errors, the host could be below the
hydrogen-burning limit and thus much fainter.  However, the {\it Spitzer}
flux variation is only about 2.5 units, i.e., just a few times larger
than the level of systematic effects seen in other events.  Moreover,
the full solution would imply a counter-rotating object in the Galactic disk.
A Bayesian analysis carried out without the {\it Spitzer} data yields
$M_\host \sim 0.37\,M_\odot$, $M_{\rm planet} \sim 5.1\,M_\oplus$ 
and $D_L\sim 7.2\,\kpc$, and thus
would imply $K_\host\sim 21.0$, i.e., $\Delta K\sim 3.6\,$mag fainter than
the source \citep{ob161195a}.  This would certainly be observable at
first AO light of 30m class telescopes and possibly before that using
8m-class telescopes.  Note that the parameters in Table~\ref{tab:tab1} are the
weighted average of those reported by \citet{ob161195b} and \citet{ob161195a},
who analyzed completely independent data sets.
{\bf OGLE-2016-BLG-1227}:
Because the source is a bright giant and the host is likely to be
a late M dwarf in or near the bulge, a separation of 
$\Delta\theta\sim 5\,$FWHM$\sim 70\,\mas$ for EELT is likely required to 
resolve the source and host.  Given the extremely low proper motion, 
$\mu_\rel=0.8\,\masyr$, this would require almost a century wait time
for EELT, but perhaps ``only'' 35 years for VLTI GRAVITY \citep{kojima1b}, 
which is effectively a 100m-class telescope.   Note that $M_{\rm cr}$ is
small, so the heliocentric and geocentric proper motions will not differ 
greatly.

{\bf KMT-2016-BLG-0212}:
Late-time AO observations can definitely distinguish between two
classes of degenerate solutions with $\Delta\log q=2.88$ \citep{kb160212}.
``Class I'' (which is favored by $\Delta\chi^2=6.6$) predicts
a source flux that is fainter by $\Delta I_S=1.1\,$mag.  Thus, combining
a nearby band (e.g., $J$) to approximate $I$ with a more distant one
(e.g., $K$) to make the color correction, will distinguish between these.
If the smaller-$q$ (higher $\chi^2$) ``Class II'' solutions are confirmed, 
then among these, the ``wide 2b'' solution has a larger $q$ by a 
factor 1.7 compared to ``wide 2a'' and ``wide 3''.  However,  
``wide 2a'' predicts a 20\% larger proper motion than ``wide 2b'', so that 
if the former were confirmed by AO observations, then this would also confirm 
the lower mass ratio, $q=5\times 10^{-5}$.  Unfortunately, the reported errors 
in  the $\mu_\rel$ predictions are larger than their difference, and these
are basically rooted in the relatively large error in $\rho$, so they
cannot be substantially ameliorated by late time observations.  Note that
if ``Class I'' is confirmed, the mass ratio would be beyond the
$q<0.03$ limit for current systematic searches.
{\bf KMT-2016-BLG-1107}
has a mass ratio that is beyond the $q<0.03$ limit for current 
systematic searches.  In addition, it is likely to be extraordinarily
difficult to resolve because the source is extremely bright and the
lens is expected to be a late M dwarf or BD in or near the bulge
\citep{kb161107}, while $\mu_\rel$ is very low.
{\bf KMT-2016-BLG-2605}
is not in the current AnomalyFinder statistical sample because the
underlying event was not discovered by AlertFinder or EventFinder, but
rather in a special supplementary search that targeted giant-source events.
Hence, it was not searched for planets by AnomalyFinder.  However, it is
interesting because $t_\e$ is the shortest for any microlensing planet, and its
$\theta_\e=0.116\,\mas$ implies that the host lies near the star/BD boundary 
\citep{kb162605}.  Nevertheless, even if it is a star, the contrast
ratio will correspond to $\Delta K\sim 7.5$.  Hence, despite its high
proper motion, $\mu_\rel\sim 12\,\masyr$, it probably cannot be resolved
until the advent of 30m-class AO.  However, as noted by \citet{kb162605}, the 
highly uncertain source color could already be measured by high resolution
(AO or {\it HST}) observations 
in two bands, which would improve the $\theta_\e$ and $\mu_\rel$ estimates
that are critical to its interpretation, whether or not the host is luminous.
It would be straightforward to
extend the AnomalyFinder search to the 2016-2019 special giant-source 
sample \citep{gould22}, in which case this planet would likely
become part of the statistical sample.

\subsubsection{2017}
\label{sec:2017.1}
{\bf OGLE-2017-BLG-0173}
has a $\Delta \log q = 0.41$ discrete degeneracy at $\Delta\chi^2=3.5$, which
favors the lower mass ratio.  The two solutions have
$\Delta\log \rho = 0.039 \pm 0.29$, which fundamentally limits how
well a proper-motion measurement could distinguish between these solutions.
In fact, the source color was not measured during the event, so this
would also have to be precisely measured by AO followup in order to
derive a precise light-curve based proper-motion prediction.
Even if perfectly successful, this could only add at most 
$\Delta\chi^2= 1.8$ to the light-curve preference.  Hence, the two
solutions can be only marginally distinguished, even in principle.
Nevertheless, because both solutions have low-$q$, it would be of interest
to measure the host mass.
{\bf OGLE-2017-BLG-0373}
has a $\Delta \log q = 0.38$ discrete degeneracy.  The two solutions have
nearly identical $\mu_\rel$, so they cannot be distinguished by a late-time
proper-motion measurement.  They have a $\Delta I_S = 0.17\pm 0.07$.
See Table~2 of \citet{ob170373}.  This could, in principle be distinguished, 
but would require two photometric bands, including one near the $I$ band.
{\bf OGLE-2017-BLG-0406}
has a {\it Spitzer} parallax measurement \citep{yee15}, which
yields a host mass, planet mass, and system distance of 
$M_\host \sim 0.56\,M_\odot$, $M_{\rm planet} \sim 0.41\,M_{\rm Jup}$ 
and $D_L\sim 5.2\,\kpc$.  These would imply $K_L\sim 19.3$, i.e.,
$\Delta K\sim 4.5\,$mag fainter than the source.  Hence, the imaging of 
this event should not be attempted before 30m-class AO is available.
I note that the $\bpi_\e$ measurement is derived from the intersection
of two 1-D parallax measurements, one from {\it Spitzer} and one from
the ground, as originally suggested by \citet{gould99}.  AO imaging would
provide an important test of this approach.
{\bf OGLE-2017-BLG-1140}
has a {\it Spitzer} parallax measurement \citep{yee15}, which
yields a host mass, planet mass, and system distance of 
$M_\host \sim 0.21\,M_\odot$, $M_{\rm planet} \sim 1.62\,M_{\rm Jup}$ 
and $D_L\sim 7.4\,\kpc$.  These would imply $K_L\sim 22.3$, i.e.,
$\Delta K\sim 8\,$mag fainter than the source.  Hence, the imaging of 
this event should not be attempted before 30m-class AO is available.
{\bf OGLE-2017-BLG-1434}
has excellent measurements of both $\bpi_\e$ and $\theta_\e$, and these
yield $M_\host = 0.234\pm 0.026\,M_\odot$,
$M_{\rm planet} = 4.4\pm 0.5\,M_\oplus$, and $D_L = 0.86\pm 0.09\,\kpc$.
\citet{ob171434b} confirmed these measurements by measuring the excess
light due to the lens (while it was still superposed on the source in 2018)
using Keck AO. 
{\bf OGLE-2017-BLG-1691}
has a 1L2S solution that is disfavored by $\Delta\chi^2=13.9$, with
both KMTC and KMTS points directly on the anomaly contributing significantly.
Thus, it can be confidently excluded.  Nevertheless, I note that
$\rho_{2,\rm 1L2S}/\rho_{\rm 2L1S}\simeq 1$, while $q_F=0.006$, which together
imply a ratio of predicted proper motions of
$\mu_{\rel,\rm 1L2S}/\mu_{\rel,\rm 2L1S}\sim 0.2$.  Thus,
the late-time AO proper-motion measurement can provide an additional,
and very strong, argument against 1L2S.

{\bf KMT-2017-BLG-0165}:
\citet{kb170165} argue from several lines of evidence that the lens
dominates the blended light.  Because the source and blend are of comparable
brightness and color and are relatively isolated (see their Figure~7),
this implies that they could be resolved at separations 
$\Delta\theta\ga 40\,\mas$, which will occur beginning about 2024.
In principle, this might create ambiguity between identifying the source
versus the host.  However, their Figure~4 shows that the host lies
will lie to the east of the source.
{\bf KMT-2017-BLG-1003}
is listed as having $\rho=5.2\pm 1.2\times 10^{-3}$ 
measurement because this is the result
for the ``outer'' solution, which is very slightly favored by $\chi^2$.  
However, the other (``inner'') solution has a very similar $3\,\sigma$ upper
limit, $\rho<6.7\times 10^{-3}$, so the listed $\mu_\rel$ is a good guide
to determining the wait time.
{\bf KMT-2017-BLG-1038}:
\citet{kb171038} do not report $((V-I),I)_{S,0}$, so I have estimated
$I_{S,0} = 18 - 2.5\log(f_{S,KMTC}) - A_I$, and I have estimated 
$(V-I)_{S,0}$ from the colors of
stars with the same $I$-band offset from the clump (as determined from
Table~1 of \citealt{nataf13}) as those in Baade's Window (as derived 
from {\it HST} photometry from \citealt{holtzman98}).  While there is
a best estimate of $\rho=0.0012$, the finite-source signature is weak.  Hence,
Table~\ref{tab:tab1} values are based on the upper limit $\rho<0.004$.  This means
that the best estimate of the proper motion is roughly 3.3 times larger
than the lower limit in Table~1.
{\bf KMT-2017-BLG-1194}
has only an upper limit $\rho>0.0026$, but this places a significant
constraint on $\mu_\rel$, so it is included in Table~\ref{tab:tab1}.

\subsubsection{2018}
\label{sec:2018.1}
{\bf OGLE-2018-BLG-0506}
has only an upper limit on $\rho$ but this leads to a strong constraint
on the proper motion, $\mu_\rel>6.5\,\masyr$, which can be used to predict
when the lens and source are adequately separated for observations.
{\bf OGLE-2018-BLG-0532} 
has a blend that is about $\Delta K\sim 3$ mag
brighter than the source and $\la 50\,$mas from it.  In principle,
this could be the host, but it is more likely to be a companion to the
host or a random field star.  In any case, its close proximity, together with
the low proper motion, $\mu_\rel = 3.3\,\masyr$, imply that observations
prior to 30m-class AO would be challenging.
{\bf OGLE-2018-BLG-0596}
has a {\it Spitzer} parallax measurement \citep{yee15}, which yields 
a host mass $M_\host \sim 0.23\,M_\odot$ at $D_L\sim 5.6\,\kpc$, which would
be $\Delta K\sim 7\,$mag fainter than the source.  Hence, the imaging of 
this event should not be attempted before 30m-class AO is available.
{\bf OGLE-2018-BLG-0677}
is not in the AnomalyFinder statistical sample because it failed the
$\Delta\chi^2$ criterion in the automated search.  Although $\rho$ is not
measured, it is included in Table~\ref{tab:tab1} because the limit on $\rho$ is 
significant.
{\bf OGLE-2018-BLG-0740} 
has a bright, blue blend that is
almost certainly the host or a companion to the host.  \citet{ob180740}
obtained a spectrum and showed that there is only a 5\% probability
for the host-companion scenario.  The light curve is consistent with $\rho=0$.
The values of $\rho$ and $\mu_\rel$ in Table~\ref{tab:tab1} are derived under the
assumption that the blend (with spectroscopic mass, $M_B = 1.0\pm 0.1\,M_\odot$)
is the host.  If the proper motion from late-time AO were inconsistent,
it would imply that the blend is a companion to the host.  This could
be confirmed by a second epoch, which would demonstrate that the
blend trajectory does not ``point back'' to the source.  Hence, this
is a very interesting case.  Unfortunately, the source is $\Delta K\sim 5$ mag
fainter than the blend.  Therefore, imaging should not be attempted
before 30m-class AO is available.
{\bf OGLE-2018-BLG-0799}
has a {\it Spitzer} parallax measurement \citep{yee15}.  However,
a combination of a weak signal and low-level systematics made it
difficult to reach unambiguous conclusions.  The preferred solution
had $M_\host\sim 0.1\,M_\odot$ at $D_L\sim 4\,\kpc$.  Given that the
source is a lower giant-branch star, this would imply a contrast
offset of $\Delta K\sim 8\,$mag, which would require a lens-source
offset of order 5 FWHM for a detection.  However, this solution predicts
$\mu_{\rel,\hel}\sim 4\,\masyr$, roughly double the value of 
$\mu_\rel\sim 1.8\,\masyr$ shown in Table~\ref{tab:tab1}.  These are compatible
in part because the $\rho$ measurement is consistent with zero at $3\,\sigma$,
and partly because $\pi_\rel v_{\oplus,\perp}/\au\sim 0.76\,\masyr$ at the
preferred distance.  Even with this higher proper motion, it would
be very difficult to detect the host prior to 2030, using current instruments.
Hence, it seems prudent to await 30m-class AO before imaging this event.
Finally, note that based on a systematic analysis of the role of {\it Gaia}
in the interpretation of microlensing planets, \citet{kb210712} concluded
that the {\it Gaia} proper motion of the source should not have been
incorporated into the analysis.
{\bf OGLE-2018-BLG-0932} has {\it Spitzer} data \citep{yee15}, which
show a strong signal, although these have not yet been analyzed.  However,
it is likely that they will be analyzed prior to AO imaging.
{\bf OGLE-2018-BLG-0977} 
has a well-defined $\chi^2(\rho)$ minimum of
$\rho=1.9^{+0.3}_{-0.6} \times 10^{-3}$.  While it is consistent with $\rho=0$ at
$3\,\sigma$, significantly smaller $\rho$ would imply an improbably high
$\mu_\rel$.  In any case, our present concern is that the light curve
robustly predicts that the proper motion is relatively high.  \citet{kb190253}
made a Bayesian estimate, $\mu_\rel = 6.0^{+3.0}_{-1.8}\,\masyr$.  Thus, this
is a plausible target for current instruments in the late 2020s, but it
would be safer to wait for 30m-class AO.
{\bf OGLE-2018-BLG-1185} 
has a {\it Spitzer} parallax measurement \citep{yee15}.  While the flux
only declines about 1 {\it Spitzer} flux unit (i.e., of order the systematic
errors seen in other events), the fact that the decline is not steeper
places strong constraints on the mass of the host.  Hence, this is likely
to be a mid-to-late M dwarf.  If the former, then the host would be
$\Delta K\sim 3.5$ mag fainter than the source, while if the latter,
it would be $\Delta K\sim 6$ mag.  Given the relatively low proper motion,
$\mu_\rel\sim 5\,\masyr$, it would seem prudent to wait for 30m-class AO.
{\bf OGLE-2018-BLG-1269} 
has a bright, $I\sim 15.8$, blend that is very likely to be the lens.
It is $\Delta K\sim 3\,$mag brighter than the source.  Given the relatively
high proper motion, $\mu_\rel \sim 8\,\masyr$, it probably can be imaged
using current instrumentation in the late 2020s.
{\bf OGLE-2018-BLG-1647} 
has a $\mu_\rel$ estimate that is so low as to
invite suspicion, in particular because the $\rho$ measurement is derived
from a ridge crossing, rather than a caustic crossing.  However,
\citet{2018subpr} investigated the constraints on $\rho$ in detail
and found that $\rho<2.3\times 10^{-3}$ was excluded at $2.5\,\sigma$,
which places a limit $\mu_\rel<1.4\,\masyr$.  Hence, this event should
await 30m-class AO before imaging.
{\bf KMT-2018-BLG-0029}
has a {\it Spitzer} parallax measurement \citep{yee15}, 
from which $M_\host\sim 1.2\,M_\host$
and $D_L\sim 3.4\,\kpc$.  If correct, the host would be of order 3.5 mag
brighter than the source.
{\bf KMT-2018-BLG-0087}
has only an upper limit on $\rho$, but this leads to a strong constraint
on the proper motion, $\mu_\rel>7\,\masyr$, which can be used to predict
when the lens and source are adequately separated for observations.
Note, however, that the source is a giant, $K_{S,0}\sim 12.3$, whereas
the Bayesian analysis predicts $M_\host\sim M\sim 0.1\,M_\odot$ and
$D_L\sim 7\,\kpc$, i.e., $K_{\host,0}\sim 23.5$.  Hence, imaging should not
be attempted before 30m-class AO is available.  Possibly, it is a
candidate for VLTI GRAVITY \citep{kojima1b}, 
which is effectively a 100m-class telescope.
{\bf KMT-2018-BLG-0748} has $\theta_\e=0.11\,\mas$, and therefore is
likely to have $M_\host\sim 0.1\,M_\odot$ and to lie in or near the bulge.  
This would imply that it would be $\Delta K\sim 7\,$ mag fainter than the 
source and so, challenging to detect prior to 30m-class AO.
{\bf KMT-2018-BLG-1025} 
is not currently in the AnomalyFinder statistical sample because
it has a discrete degeneracy $\Delta\log q = 0.29$.  However, as noted
by \citet{kb181025} this will likely be resolved by AO observations
because the two solutions predict $\mu_\rel$ values that differ by
significantly more than their error bars.  Note that $M_{\rm cr}$ is small,
so the conversion from heliocentric to geocentric
should not interfere with breaking this degeneracy.  The best solution
has only an upper limit on $\rho$, but it strongly constrains $\mu_\rel$,
so the event is included in Table~\ref{tab:tab1}.
{\bf KMT-2018-BLG-1292}:
The host is likely to be the origin of the blended light, in which case it
would probably be an F or G dwarf at $D_L\sim 3.3\,\kpc$.  To verify this
would require separately resolving them, which is probably best done in
$J$ band because the blend and the source have similar brightness in
$I$ band, while the source is $\sim 4.5$ mag brighter in the $K$ band.
However, because $b=-0.28^\circ$ and $A_I\sim 5.2$, it would be challenging
to apply the ``basic method'' to this event.  Rather, the host mass should
be measured by obtaining a spectrum in the $V/I$ range of the energy
distribution.  Note that $V_B\sim 20.8$, while $V_S\sim 25$, so there
would be very little source contamination in the $V$ band.
{\bf KMT-2018-BLG-1743} 
is not currently in the AnomalyFinder statistical sample because
it has a discrete degeneracy $\Delta\log q = 0.48$.  However, as noted
by \citet{kb181743} this may possibly be resolved by AO observations
because the two solutions predict $\mu_\rel$ values that differ substantially.
Nominally, the predictions of the two solutions are consistent at $1\,\sigma$,
primarily because $\rho$ is poorly measured in the alternate (higher $\chi^2$,
higher $q$ solution).  However, if the measured proper motion is near the
prediction of the favored solution, which has smaller errors, the combined
likelihood of the light-curve fit and the proper-motion prediction may
be sufficient to clearly choose the favored solution.  If, on the other
hand, the proper motion is substantially higher than that of the favored 
solution, the other solution will be clearly selected.
Note that $M_{\rm cr}$ is small, so the conversion from heliocentric to geocentric
should not interfere with breaking this degeneracy.

\subsubsection{2019}
\label{sec:2019.1}

{\bf OGLE-2019-BLG-0299} 
has only an upper limit on $\rho$.  
However, it is included in Table~\ref{tab:tab1} because the resulting 
lower limit $\mu_\rel>1.74\,\masyr$ is significant.  
{\bf OGLE-2019-BLG-0960} 
has a {\it Spitzer} parallax measurement \citep{yee15}, as well as 
a strong ground-based parallax signal, with which it is in moderate
tension.  It is likely that the host is responsible for the blended
light, in which case they could be separately resolved by 2025 \citep{ob190960}.
It is likely that the planet has mass $M_{\rm planet}\sim 2\,M_\oplus$.

{\bf KMT-2019-BLG-0371}
has $q=0.08$, so it lies well beyond the current completeness limit $q<0.03$
of the AnomalyFinder samples.  It is nonetheless an interesting target
because it is a short event ($t_\e=6.5\,$day) with a small 
$\theta_\e=140\,\muas$, and so Bayesian mass estimate $M_\host \sim 0.09\,M_\odot$.
If correct, then $M_{\rm planet}\sim 7.5\,M_{\rm Jup}$, i.e., inside the nominal
planetary range.   Moreover, it would not be difficult to extend the 
current complete sample to short-$t_\e$, low-$q$ events, e.g.,
$t_q \equiv q t_\e < 0.6\,$days, in order to study the whole class of
these objects.
{\bf KMT-2019-BLG-0842}:  
The blended light is $\Delta I\sim 2\,$mag
and $\Delta K\sim 1.5\,$mag brighter than the source.  \citet{kb190842}
argue that it is plausibly the host.  Hence, given the estimated
$\mu_\rel = 8.0\pm 1.8\, \masyr$ proper motion, this could be a good target
with present instruments starting about 2029.
{\bf KMT-2019-BLG-1715}
is a 3L2S event, with a planet in a binary-star system that microlenses
a binary source.  However, because the principal anomaly is due to the planet
magnifying the primary source, its discovery by AnomalyFinder followed
the standard path, and so it should be included in statistical samples.

\subsubsection{2021}
\label{sec:2021.1}

{\bf KMT-2021-BLG-0119}
has only an upper limit $\rho>0.0018$, but this places a significant
constraint on $\mu_\rel$, so it is included in Table~\ref{tab:tab1}.
It has a significant 1-D $\pi_\e$ measurement, which constrains the lens
to lie in the relatively near disk, $D_L\sim 3\,\kpc$.  \citet{kb210119}
note that the blended light is consistent with being the lens, and
because it contains $\sim 40\%$ of the baseline light, this can be checked
with AO observations taken immediately.
{\bf KMT-2021-BLG-0171} 
was discovered as part of a part of a follow-up program by which
the anomaly was monitored intensively from the LCO facility at SSO
\citep{kb210171}.
However, AnomalyFinder recovered the anomaly based on KMTA data.
That is, the planet detection was not influenced by the follow-up data.
Rather, these served to improve the characterization.  Note that there
are alternate solutions, which are disfavored by only $\Delta\chi^2=6.4$
and have smaller $q$ by $\Delta\log q = -0.34$
However, \citet{kb210171} argue that these are heavily disfavored
by phase-space considerations.
{\bf KMT-2021-BLG-0240} 
has two classes of solutions, with 2L2S favored over 3L1S by 
$\Delta\chi^2=10.5$.  \citet{kb210240} argue that neither is decisively
favored because the 2L2S solution would predict violent wiggles unless
the binary-source is seen in deep projection, and these are not present
in the light curve.  In Table~\ref{tab:tab1}, I have listed the 3L1S solution.
To be included in statistical studies, the 3L1S/2L2S degeneracy must
be broken: otherwise the planet (or first planet for 3L1S) has three
possible mass ratios, $q\sim (3.5,6.5,9.5)\times 10^{-4}$, which is
probably too broad a range to be useful.  The two classes predict
different $\mu_\rel$, i.e., $\mu_{\rel,\rm 3L1S}= 3.1\pm 0.4\,\masyr$ and
$\mu_{\rel,\rm 2L2S}= 3.9\pm 0.8\,\masyr$.  Because these overlap, measurement
of $\mu_{\rel,\hel}$ will not decisively distinguish between them.
However, unless the binary source is in extreme projection, the RV
variations should be easily visible from spectroscopy if 2L2S is correct.
That is, for this case, $\sin i\sim 1$, while 
$(M_{S,1},M_{S,2})\sim (1.0,0.6)M_\odot$, so the semi-amplitude of RV
variations will be $v_{\rm semi} \sim 14(a_S/\au)^{-1/2}\,\kms =
110(a_S/a_{\perp,S})^{-1/2}\,\kms$, where $a_{\perp,S} = 0.016\,\au=3.4\,R_\odot$ 
is the projected separation observed at $t_0$.  In the 2L2S scenario, the 
primary source has $I_{S,1}\simeq 21.1$ and $K_{S,1}\simeq 16.6$.  There is
relatively little blended light, so detection of such large RV signals
is probably feasible today, without waiting for the source and lens to
separate.  If the 3L1S solution is confirmed, then the first planet
has a discrete degeneracy $\Delta\log q = 0.17$, which is acceptable,
while the second has $\Delta\log q = 0.32$, which may not be.
{\bf KMT-2021-BLG-0748} 
has an ``alternate solution'', which is disfavored
by $\Delta\chi^2\sim 10$ and which \citet{kb211391} consider to be 
a ``satellite solution''.  The satellite solutions differ somewhat
in their physical implications because $\Delta\log q\sim 0.30$.  They
differ in $I_S$ by 1 magnitude and so could be distinguished using $J$-band 
and $K$-band measurements together with $IJK$ color-color relations.
Because $\mu_\rel\sim 3\,\masyr$, these measurements should wait for 30m AO.
{\bf KMT-2021-BLG-0912} 
has three solutions with very different $q$, i.e.,
$(\Delta\log q,\Delta\chi^2)_{\rm Close\ I}= (+0.42,+9.5)$ and
$(\Delta\log q,\Delta\chi^2)_{\rm Wide}= (-0.79,+19.2)$ relative to the
``Close II'' solution that is listed in Table~\ref{tab:tab1}.  
Once $\mu_\rel$ is measured, it can easily distinguish between the 
heavily disfavored Wide solution and the two Close solutions because 
the former predicts $\mu_{\rel,\rm Wide}=10\pm 1,\masyr$.
The two Close solutions will be more difficult to distinguish because
$\mu_{\rel,\rm Close\ II}=3.8\pm 1.0\,\masyr$, while
$\mu_{\rel,\rm Close\ I} =2.6\pm 0.5\,\masyr$, and hence their predicted 
distributions overlap.  The relatively large errors are dominated by the errors 
in $\rho$, and the much smaller fractional errors in $t_\e$ and $u_0$ in 
Table~\ref{tab:tab1}
imply that an improved measurement of $I_S$ from AO will not substantially
improve the error in $\rho$.  Hence, it is unclear whether the Close I/II
degeneracy will be decisively resolved by AO measurements, although Close II
is substantially preferred by $\chi^2$.
{\bf KMT-2021-BLG-1253} 
will almost certainly {\bf not} be part of an objective sample because
(with $\Delta\chi^2=105$)(with $\Delta\chi^2=105$)
it failed the $\Delta\chi^2>120$
criterion for selection by AnomalyFinder.  Nevertheless,
if this event is pursued, it should be noted that it
has an alternate solution, which is disfavored
by $\Delta\chi^2\sim 1.6$ and has $\mu_\rel \sim 9\,\masyr$.  
Both proper-motion estimates would allow the host to be detected
with current instruments for observations beginning about 2029,
provided that it is sufficiently bright.  However, if the Bayesian estimates
are approximately correct, the host will be $\Delta K\sim 3.3\,$mag
fainter, which could be challenging prior to 30m-class AO.
{\bf KMT-2021-BLG-1391} 
has an alternate solution, which is disfavored
by $\Delta\chi^2\sim 5$ and has $\mu_\rel \sim 2.6\,\masyr$.  Hence, 
imaging should probably not be attempted until 30m-class AO is available.
The solutions can be distinguished by the $\mu_{\rel,\hel}$ measurement,
although the physical implications of the different solutions are very
similar, in any case.
{\bf KMT-2021-BLG-1554} 
has a 1L2S solution that is disfavored by $\Delta\chi^2=10.4$.  The 2L1S
solution looks substantially better by eye, but the 1L2S solution cannot
be absolutely excluded based on $\Delta\chi^2$ alone.  The best-fit 1L2S
solution is unphysical in the sense that the offset between
the sources is $\Delta u_S\sim 0.003$, while the primary source has
$\rho_{1,\rm 1L2S} = 0.098\pm 0.024$.  That is, either the secondary would
be enveloped in the primary (if the physical separation were similar
to the projected separation) or there would be strong eclipses (if the
true separation were much larger).  This conflict could be ameliorated
by reducing $\rho_{1,\rm 1L2S}$, but only at a substantial cost in $\chi^2$.
Thus, the 1L2S solution can be decisively excluded based on a combination
of arguments.  These can be further tightened by a late-time AO proper-motion
measurement.
{\bf KMT-2021-BLG-1689} 
was discovered as part of a part of a follow-up program by which
the anomaly was monitored intensively from the Auckland Observatory
\citep{kb210171}.  There were no KMT data over the anomaly, and
hence it was not recovered by AnomalyFinder.  Therefore, it
does not enter the KMT-based statistical sample.  It may still
enter a statistical sample based on follow-up protocols.  See, e.g.,
\citet{gould10}.  Note that there
are alternate solutions, which are disfavored by only $\Delta\chi^2=2.4$
and have smaller $q$ by $\Delta\log q = -0.11$
However, \citet{kb210171} argue that these are heavily disfavored
by phase-space considerations.  Moreover, they can easily
be vetted by late-time AO observations because they predict proper
motions that are twice as fast.
{\bf KMT-2021-BLG-1898}:
The interpretation will be somewhat complicated by the fact that
there are two sources as well as a bright blend.  At $t_0$, the sources
were separated in the Einstein ring by $\Delta u_S\sim 7.8\times 10^{-3}$,
corresponding to $\Delta\theta_S\sim 2.5\,\muas$ and projected
separation $a_{\perp,S}\sim 4\,R_\odot$.  Hence, they are almost certainly
in a tightly bound orbit and will appear as a ``single star'' in late-time
AO imaging.  Because their separation will be small compared to the
lens-source separation, this will not cause any real difficulty, but it
should be kept in mind.  There is a blend whose brightness is consistent
with a bulge star at the base of the giant branch, but whose color is
not measured.  \citet{kb211898} do not report its offset from the
lens.  If it is indeed a bulge giant, then it will be clearly brighter
than the combined light of the sources in $K$ band.  However, if it lies
the foreground, in front of a substantial fraction of the dust, then
it could be of comparable brightness to the combined source light, 
in which case it could cause some confusion.
{\bf KMT-2021-BLG-2294}
was not found by AnomalyFinder because it had $\Delta\chi^2=59$ (far below
the threshold of 120).  Hence, unless the AnomalyFinder search is modified
(and this event is recovered), it will not enter the statistical sample.
Note that the planetary signal is very obvious by eye in the online pySIS
reductions.  Nevertheless, this failure mode is quite rare and so probably
does not warrant modifying the search.

\subsection{Comments on Events in Table 2}
\label{sec:table2}

\subsubsection{2016}
\label{sec:2016.2}
{\bf OGLE-2016-BLG-0263}
is not in the AnomalyFinder statistical sample because its best-fit solution is
2\% above the formal limit $q=0.03$.  However, because the timescale is
short, $t_\e=16\,$day, its host is plausibly a late M dwarf \citep{ob160263},
in which case the planet is a few Jupiter masses.

\subsubsection{2017}
\label{sec:2017.2}
{\bf KMT-2017-BLG-1146}:
\citet{kb171038} do not report $((V-I),I)_{S,0}$, so I have estimated
$I_{S,0} = 18 - 2.5\log(f_{S,KMTC}) - A_I$, and I have estimated 
$(V-I)_{S,0}$ from the colors of
stars with the same $I$-band offset from the clump (as determined from
Table~1 of \citealt{nataf13}) as those in Baade's Window (as derived 
from {\it HST} photometry from \citealt{holtzman98}).  

\subsubsection{2018}
\label{sec:2018.2}

{\bf OGLE-2018-BLG-1212} 
has a large $\pi_\e=0.77$, which implies a proper motion
$\mu_\rel = \kappa M\pi_\e/t_\e \rightarrow 45(M/M_\odot)\,\masyr$.
Although the Bayesian analysis disfavors masses $M>0.3\,M_\odot$,
the lens-source separation could be quite high in 2030 if this inference proves
to be incorrect.

{\bf KMT-2018-BLG-2602}:  
The source is blended with a clump giant
at $\Delta\theta\la 10\,\mas$, which is $\sim 1.4\,$ mag brighter
than the source and very likely its companion.  This may make it difficult
to measure the source flux, even with EELT.  However, if it can be measured,
then using color-color relations and the invariants $f_S t_\e$ and $q t_\e$,
it may be possible to substantially reduce the errors in $q$ and $t_\e$
relative to those shown in Table~\ref{tab:tab2}.
{\bf KMT-2018-BLG-2718}
has a binary-lens solution that is disfavored by $\Delta\chi^2=12.7$.
In my view, this is sufficient to be included in a ``planetary'' sample
for statistical study, but this is a decision that must finally be made
by those carrying out such studies.  If the source color (in, e.g., 
$(J-K)$), is measured at late times, then it may be possible to
considerably reduce the errors in $t_\e$ and $q$ using the color-color
relations and the invariants $f_S t_\e$ and $t_q = q t_\e$.

\subsubsection{2019}
\label{sec:2019.2}

{\bf KMT-2019-BLG-0253}:  
The blended light is $\Delta I\sim 0.5\,$mag
and $\Delta K\sim 1\,$mag brighter than the source.  \citet{kb190253}
argue that it is likely to be either the host or a companion to the host.
Given that the color and magnitude of the source are well-measured in $V/I$,
it is plausible that the blend could be precisely characterized by 
{\it HST} imaging in similar bands, even before the source and lens
have separated.  Additional work would still by required to distinguish
between the host and companion-to-host scenarios.
{\bf KMT-2019-BLG-0414}
cannot be considered a ``secure planet'' at the present time because
it has an alternate (xallarap) solution that is disfavored by only 
$\Delta\chi^2=4.2$.  Nevertheless, if late-time AO imaging showed that
the proper-motion were sufficiently large, e.g., $\mu_\rel\ga 6\,\masyr$,
the xallarap solution would require a companion mass $M_{\rm comp}\ga 1\,M_\odot$,
which would be inconsistent with light curve (unless the companion were
a NS or BH).  Hence, it is possible that such a
measurement would effectively confirm the planetary solution.  However,
even if confirmed, this planet would not necessarily enter a statistical
sample because of the large discrete ($\Delta\log q = 0.47$) and
continuous ($\sigma(\log q) = 0.26$) degeneracies.

\subsubsection{2021}
\label{sec:2021.2}

{\bf KMT-2021-BLG-2478}:  
As discussed by \citet{kb210712} in their Section~5.3, 
the apparent constraint on $\rho$ is likely to be spurious.
Hence, there are no reliable constraints on the proper motion.

\subsection{Distribution of Planetary-event $\mu_\rel$}
\label{sec:distmu}

As I mentioned in Section~\ref{sec:remain}, there appears to be
an excess of low-$\mu_\rel$ planetary events relative to the expectation
for microlensing events as a whole.  That is, as argued by \citet{gould21},
microlensing is dominated by bulge-bulge lensing, for which the
sources and lenses have both similar mean proper motions and similar
dispersions $\sigma$.  Hence, the $\bmu_\rel$ distribution has approximately
zero mean and $\sqrt{2}\sigma$ dispersion.  While $\sigma$ varies somewhat
according to the line of sight, \citet{gould21} adopted $\sigma=2.9\,\masyr$ as 
representative.  Taking account of the factor $\mu_\rel$ in the rate equation,
this leads to a $\mu_\rel$ distribution
\begin{equation}
p(\mu_\rel)d\mu_\rel ={\mu_\rel^\nu d\mu_\rel \exp[-(\mu_\rel/2\sigma)^2]\over
0.5[(\nu-1)/2]!(2\sigma)^{\nu+1}};
\qquad \nu=2.
\label{eqn:qmudist}
\end{equation}

The primary concern of \citet{gould21} was to predict the fraction of
events with proper motions below some small value of $\mu_\rel\ll 2\sigma$,
which is then 
$\simeq \mu_\rel^{\nu+1}/(\nu+1) p(0) = (\mu_\rel/2\sigma)^{\nu+1}/[(\nu+1)/2]!
\rightarrow (\mu_\rel/\sigma)^3/6\sqrt{\pi}$.
While Equation~(\ref{eqn:qmudist}) does not apply to disk stars,
one does not expect these to greatly alter the low-$\mu_\rel$ form
of the distribution.  First, disk lenses constitute a minority.
Second, because the mean $|\langle \bmu_\rel\rangle |$ is typically
of order $\sim 6\,\masyr$, the height $p(\bmu_\rel=0)$ for disk lenses
is substantially suppressed compared to bulge lenses.  Third,
for small $\mu_\rel$, the functional form $\propto \mu_\rel^{\nu+1}$
remains the same. Fourth, while the disk-lens proper-motion dispersion
is typically smaller than the bulge dispersion, the difference is not
dramatic.  In brief, the contribution of disk lenses to the low-$\mu_\rel$
population is small, and its character is not dramatically different
from that of the bulge lenses.  Hence, we expect that the bulge-bulge
formula will be approximately correct, at least for low-$\mu_\rel$.

However, as noted, this appears not to be the case for the sample
of 71 planetary events with proper-motion measurements that are listed
in Table~\ref{tab:tab1}.  To address this issue quantitatively, I fit
the planetary events with proper-motion measurements to two-parameter
models of the form of Equation~(\ref{eqn:qmudist}).  For this purpose
I remove the two events with $\mu_\rel\sim 15\,\masyr$, which both
have large errors and which are not accounted for by any models in
this class.  Most likely, they are due to incorrect measurements, but
may be part of, e.g., a near-disk population.  In any case, they
are not directly relevant to the problem of an excess of low-$\mu_\rel$
planetary events.

\begin{figure}
\plotone{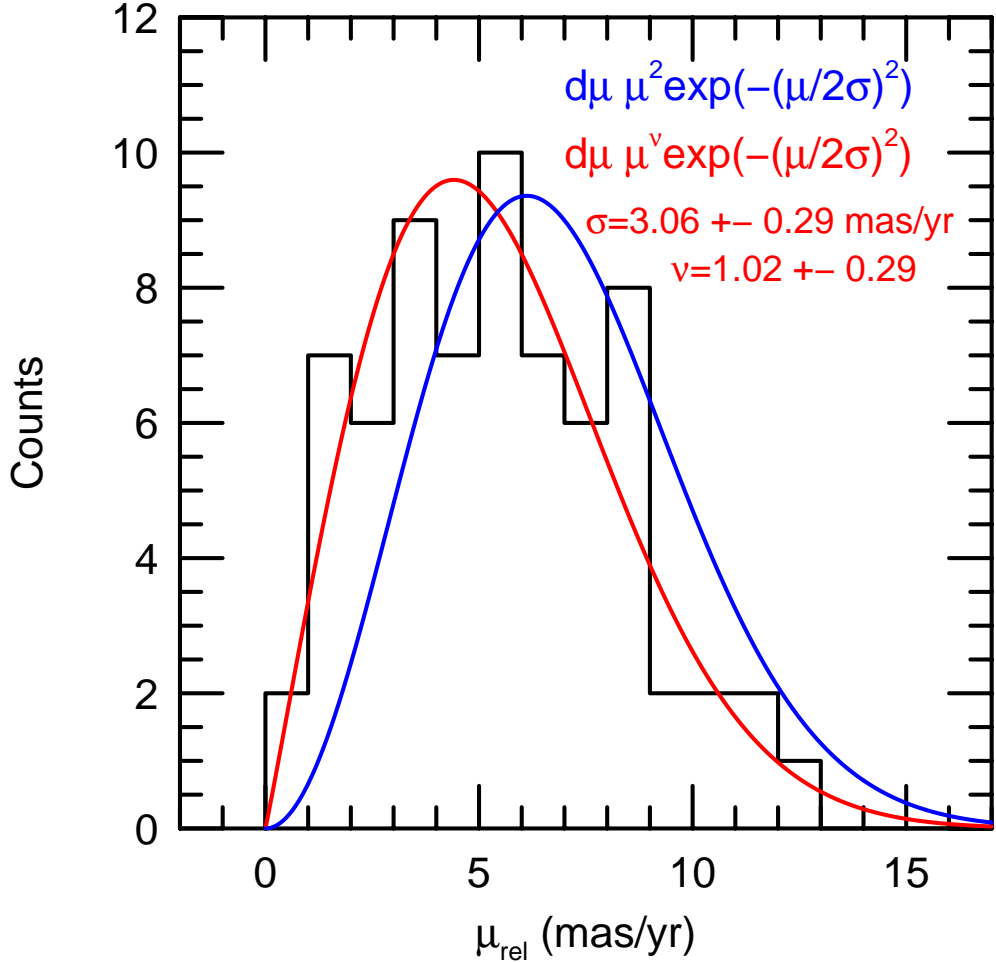}
\caption{Distribution of observed $\mu_\rel$ for 69 planetary events
with proper-motion measurements with $\mu_\rel<15\,\masyr$,
together with two models of
the form given by Equation~(\ref{eqn:qmudist}).  The blue model is an analytic
representation of the distribution expected for microlensing events
as a whole, while the red model is the best two-parameter
fit of Equation~(\ref{eqn:qmudist}) to the
data.  The red model is preferred by $\Delta\chi^2=15$, and it is plausibly
explained by the fact that it is easier to detect planets in lower $\mu_\rel$
events.}
\label{fig:qmu}
\end{figure}

The result is shown in Figure~\ref{fig:qmu}.  The best fit is given by
\begin{equation}
\nu = 1.02\pm 0.29 
\qquad 
\sigma = 3.06 \pm 0.29\,\masyr.
\label{eqn:qmuvals}
\end{equation}
Models with $\nu=2$ and $\sigma\simeq 3.0\,\masyr$ are disfavored by
$\Delta\ln L=7.5$ corresponding to $\Delta\chi^2=15$.  Given that
there is a plausible physical explanation for an excess of
low-$\mu_\rel$ planetary events relative to the underlying population
of events (i.e., $\Delta\chi^2\propto \mu_\rel^{-1}$), the balance of evidence
is that the red curve in Figure~\ref{fig:qmu}, together with its corresponding
analytic representation, is the best predictor that we currently 
have at present for the frequency of low-$\mu_\rel$ planetary events.
Note, further, that the free fit for $(\nu,\sigma)$ recovers the independently
known dispersion of bulge lenses $\sigma=2.9\,\masyr$.  If we impose this
value as a prior, then the constraint on $\nu$ is even tighter:
$\nu=1.16\pm 0.20$.

\subsection{Availability of Targets}
\label{sec:avail}

What are the prospects for making host mass measurements by AO imaging
with current instruments before AO first light on 30m class telescopes
(assumed here to be 2030), and what are the prospects using 30m class
telescopes after first AO light?

\begin{figure}
\plotone{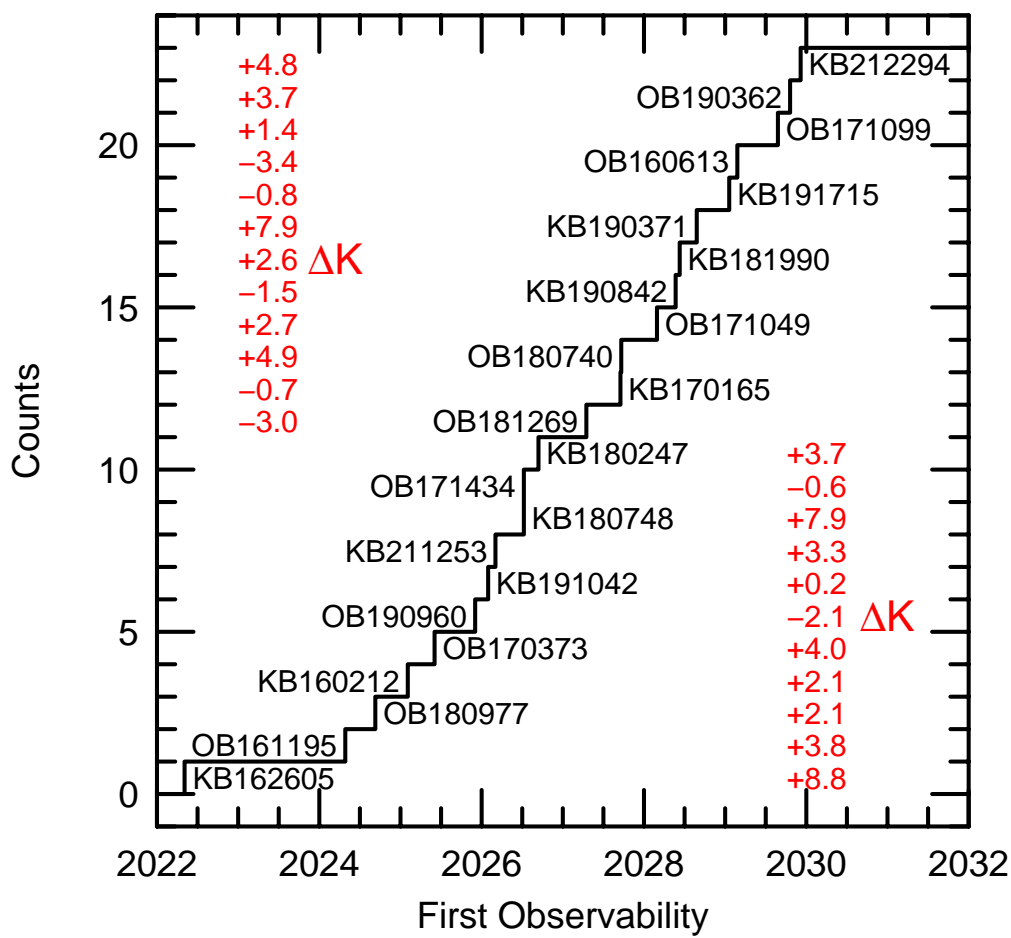}
\caption{Estimated cumulative distribution of planetary events that
can be observed with Keck AO as a function of time.  The estimate is
based on planetary events from Table~\ref{tab:tab1}, which
are labeled in the figure.  However, the actual choice of targets
must be based on much more detailed assessments than were used to make
this figure.  In particular, the vertical red columns at either side
show the estimated $K$-band contrast ratio (expressed as a 
magnitude difference), which are derived from Bayesian
estimates and other information: only about half of the 23 events will
actually be resolvable prior to 30m-class AO.}
\label{fig:cum}
\end{figure}

To address the first question, I first restrict attention to dwarf sources,
which I define at $I_0 - (V-I)_0 < 16$.  Systems with giant sources are
much more likely to have high contrast ratio, which generally requires
larger separations.  I then restrict consideration to the 82 hosts (of 86 
planets) in
Table~\ref{tab:tab1}, i.e., those that have either $\mu_\rel$ measurements,
or $\mu_\rel$ lower limits.  For the first, I adopt the reported value, and
for the second I adopt the lower limit.  I then calculate 
(adopting $\mu_{\rel,\hel} = \mu_\rel$) the date on which
the separation will reach 72 mas, i.e., 1.3 FWHM in $K$ band on the Keck
telescope.  I plot the cumulative distribution in Figure~\ref{fig:cum}.

This is a relatively crude proxy for the true target population as a function
of time.  For example, if there is a reasonable expectation that the
source and host will have comparable brightness (based on Bayesian or other
arguments), then they might be chosen as targets despite having separations
that are only half of this threshold.  On the other hand, if similar arguments
lead to an expectation of a high contrast ratio, they might not be chosen
despite being separated by more than my nominal threshold.  

To test this naive reasoning, I estimate the contrast ratio, expressed as
a $\Delta K$ offset, with negative signs meaning that the host is brighter,
for each of the 23 hosts in Figure~\ref{fig:cum}.  
In most cases, the host brightness is taken from Bayesian host-mass and
distance estimates in the discovery papers.  In a few cases, it is
based on mass measurements, usually from parallax, but sometimes from
excess light.  These are shown in vertical columns in Figure~\ref{fig:cum}.
These estimates show that 9 of the 23 hosts are expected to have
$|\Delta K|\leq 2.5$, meaning that they probably can be resolved at
1.3 FWMH, or perhaps slightly bigger separations.  Another 3 have
$2.5<|\Delta K|\leq 3.0$, and hence may be resolved at somewhat larger
separations, particularly because for the largest of these three $|\Delta K|$,
it is the host that is brighter.  However, three have $\Delta K>7$ and so would
be completely hopeless unless the Bayesian estimates were radically
incorrect, while the remaining 8 would be quite difficult.  Hence, I estimate
that roughly 10 of total sample of 111 planetary systems in 
Tables~\ref{tab:tab1} and \ref{tab:tab2} can be successfully imaged prior
to 30m-class AO.

\begin{figure}
\plotone{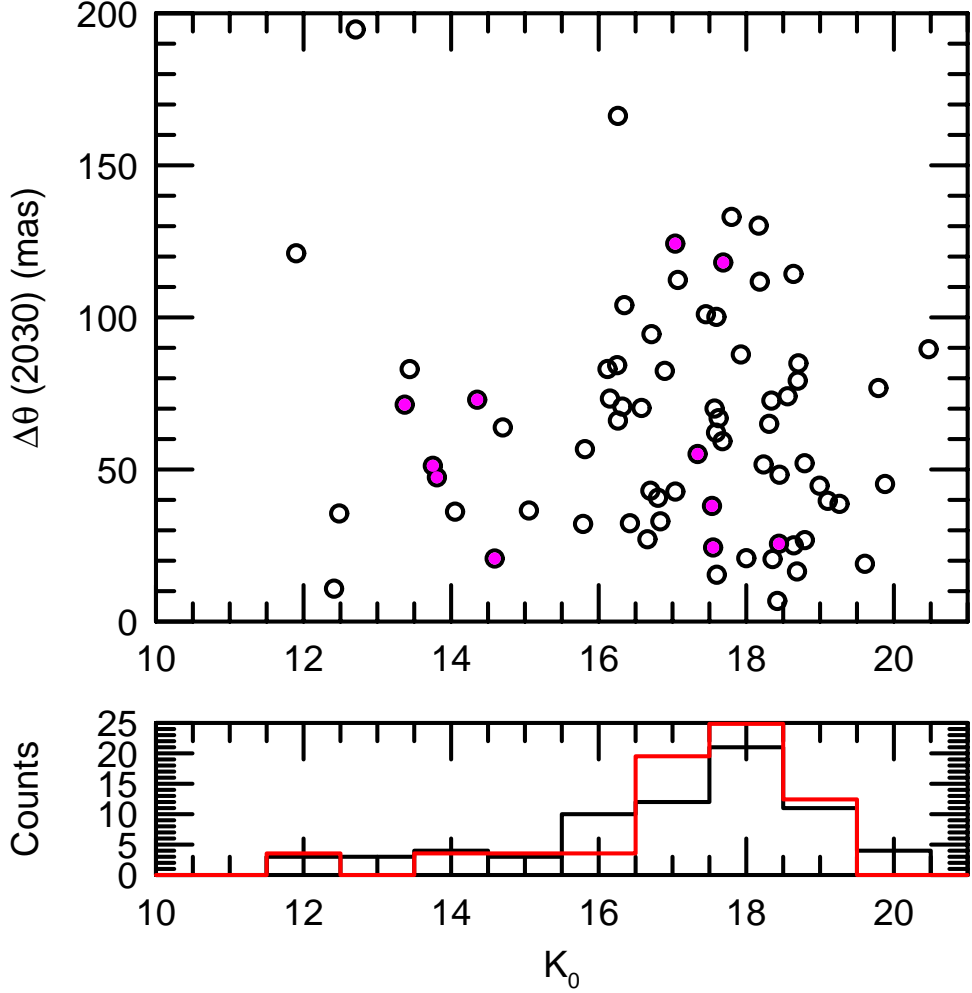}
\caption{Upper panel: Scatter plot of expected lens-source separation in 2030
(i.e., nominal first AO light on 30m class telescopes) versus $K_0$
for 71 potential targets with $\mu_\rel$ measurements.  Events with {\it Spitzer}
measurements are shown in magenta.
Intrinsically brighter sources are likely to have greater source-lens
contrast ratios and so require greater separations.  Only (2, 5) planets have
$\Delta\theta<$(10, 20)$\,\mas$, and four of these five have relatively faint 
sources.  Hence, a substantial majority of the sample should be accessible in 
2030.  Lower panel: Histograms of the same 71 potential targets with 
$\mu_\rel$ measurements (black) and of 40 other potential targets without
$\mu_\rel$ measurements (red).  The latter are scaled to have the same total
area.  The two distributions are similar.
}
\label{fig:ksep}
\end{figure}

To address the second question, I plot the expected separation,
$\Delta\theta$, in 2030
versus $K_0$ in Figure~\ref{fig:ksep} for the 71 planetary events
with $\mu_\rel$ measurements.  The $K_0$ magnitude is important because
intrinsically brighter sources will generally have larger source-lens
contrast ratios, which may require larger $\Delta\theta$ to resolve.
There are only two cases (OGLE-2018-BLG-1647 and OGLE-2016-BLG-1227)
with $\Delta\theta < 15\,\mas$, and only three others 
(KMT-2021-BLG-1077, KMT-2017-BLG-2509, and KMT-2021-BLG-0712)
with $\Delta\theta < 20\,\mas$.  Moreover, four of these five are relatively
faint.  Sources with high contrast ratio probably require separations
$\Delta\theta\ga 5\times$FWHM, which would be 70 mas for EELT in the $K$-band.
Note that there are 10 potential targets with $K_0<15.5$ and 
$\Delta\theta<80\,\mas$, which may be difficult.  In this context, it is
notable that 5 of these 10 have {\it Spitzer}-based parallax, 
which may be helpful in extracting masses in the case on non-detections.

The lower panel of Figure~\ref{fig:ksep} shows a histogram of the same
71 sources with $\mu_\rel$ measurements in black.  It also shows, in red,
a histogram of the 40 sources without $\mu_\rel$ measurements, which
is scaled to have the same total area as the black histogram.  The
two distributions look qualitatively similar.  This is mildly surprising
because one might have guessed that brighter (so bigger) sources
would be more likely to intersect caustics and, more generally, to be
more likely to yield measurable finite-source effects.  However,
if this is the case, the effect is not sufficiently strong to show up
in a sample of this size.  On the other hand, there is a countervailing
effect that, other things being equal, the featureless structures in
non-caustic-crossing events have lower $\chi^2$, so they will more
easily meet a given $\chi^2$ threshold for bright sources.

\section{Application to Other Statistical Samples}
\label{sec:other_stat-sample}

The approach outlined in this paper to transform a statistical sample of
$q$ measurements into a statistical sample of $M_{\rm planet}$ measurements
could be applied to other samples.  Here I discuss four samples, including
three completed and one prospective.

\subsection{High-Magnification Follow-Up Survey (2005-2008)}
\label{sec:ufun}
\citet{gould10} constructed a planet sample from discoveries made by the
Microlensing Follow Up Network ($\mu$FUN) by intensive followup observations
of high-magnification events that were discovered by OGLE and MOA
from 2005-2008.  The process of organizing this
followup was somewhat chaotic, but \citet{gould10} argued that the resulting
sample of $A_\max>200$ events could be considered well-defined because they
were uniformly selected from all $A_\max > 200$ events.  They detected
6 planets in 5 events, out of a total of 13 events in their full sample.
The 13 events are listed in their Table~1.

Of the 5 planetary events (containing 6 planets), 4 events (5 planets)
already have mass measurements, either from $M=\theta_\e/\kappa \pi_\e$
light-curve analyses or from high resolution follow-up observations.
{\bf OGLE-2005-BLG-169} 
\citep{ob05169} was subsequently resolved by 
\citet{ob05169bat} 
and \citet{ob05169ben}, using Keck and {\it HST} respectively, and
was determined to have 
$M_\host = 0.65\,\pm 0.05\,M_\odot$,
$M_{\rm planet}= 13.2\pm 1.2\,M_\oplus$, and
$D_L = 4.0\pm 0.4\,\kpc$.
{\bf OGLE-2006-BLG-109} 
was already known to have 
$M_\host = 0.51\,\pm 0.05\,M_\odot$,
$M_{\rm planet}= 231\pm 19\,M_\oplus$, 
$M_{\rm planet}= 86\pm 7\,M_\oplus$, and
$D_L = 1.51\pm 0.12\,\kpc$
based on the previous analyses of \citet{ob06109} and \citet{ob06109b},
which included both light-curve based $M=\theta_\e/\kappa \pi_\e$ and
unresolved Keck AO follow-up observations.
{\bf OGLE-2007-BLG-349} 
was subsequently analyzed by \citet{ob07349}
based on a combination of the $M=\theta_\e/\kappa \pi_\e$ light-curve method
and already existing {\it HST} data and found to have
$M_\host = 0.41\,\pm 0.07\,M_\odot$,
$M_{\rm planet}= 80\pm 13\,M_\oplus$, and
$D_L = 2.7\pm 0.4\,\kpc$.
Note that this is a circumbinary planet.  The mass of the stellar
companion is $M_{\rm companion} = 0.30\pm 0.07\,M_\odot$.
{\bf MOA-2007-BLG-400} 
\citep{mb07400} was subsequently resolved via 
Keck AO follow-up observations by \citet{mb07400b} and found to have
$M_\host = 0.69\,\pm 0.04\,M_\odot$,
$M_{\rm planet}= 544\pm 86\,M_\oplus$, and
$D_L = 6.9\pm 0.8\,\kpc$.

Of the 6 planets, only MOA-2008-BLG-310 \citep{mb08310} lacks a mass
measurement. While \citet{gould10} list a mass measurement based on
excess light detected when the lens and source were superposed,
\citet{mb08310b} subsequently showed that this excess light was inconsistent
with being the lens.  We note that because 
$\mu_\rel = 4.85\pm 0.15\,\masyr$ \citep{mb08310b}, the lens and source
will be separated by $\Delta\theta= 73\pm 2\,\mas$ in 2023.  Hence,
it is possible in principle that the lens and source could be separately 
resolved ``now'' using Keck AO.  However, based on the original 
$H$-band light-curve measurements,
$H_S=17.69\pm 0.03$ (corresponding to $K_S=17.8$), while
\citet{mb08310b} did a Bayesian analysis that included
the flux constraints from their {\it HST} measurements and
derived $M_\host = 0.21\pm 0.14$ and $D_L=7.7\pm 1.1\,\kpc$.  At the mean
values, the expected host magnitude is $K_\host \sim 22.3$, i.e., 
$\Delta K=4.5$, which would require much greater separations, probably
implying that the host and source will only be resolved at first 
AO light on 30m class telescopes.

\begin{figure}
\plotone{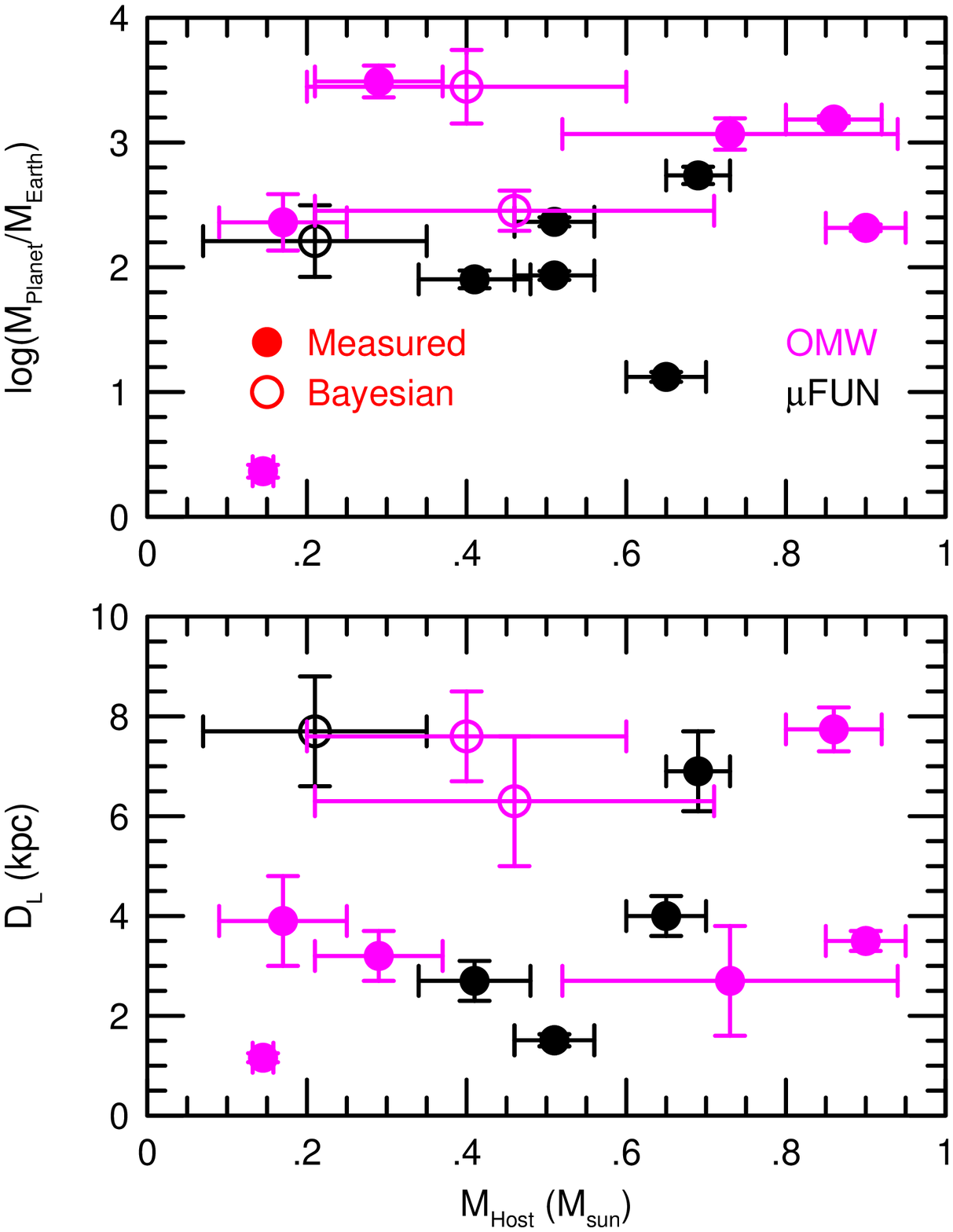}
\caption{Upper panel: Logarithm of planet mass versus host mass for the
samples of 6 planets (from 5 events) from the $\mu$FUN study of \citet{gould10}
(black) and of the 8 planets from the OGLE-MOA-Wise study of 
\citet{shvartzvald16}
(magenta). Lower panel: Lens distance versus host mass for the same sample.  The
open symbols are based on a Bayesian estimate while the filled symbols
are based on direct measurements.  The lower panel suggests that planet
frequency may be higher for more nearby hosts, but no strong conclusion can
be drawn due to small number statistics.}
\label{fig:o+u}
\end{figure}

Nevertheless, as nearly all of this sample has mass measurements, it is 
instructive to plot their values, which I do in Figure~\ref{fig:o+u} (black)..
For this purpose, I plot MOA-2008-BLG-310 at its Bayesian estimates:
$M_\host = 0.21\pm 014$, $M_{\rm planet} = 162\pm 108\,M_\oplus$ and
$D_L=7.7\pm 1.1\,\kpc$.

A naive conclusion from Figure~\ref{fig:o+u} would be that 2/3 of
all microlensing planets are in the near disk, $D_L \la 4\,\kpc$, and that
2/3 have host masses $M_\host \ga 0.5\,M_\odot$.  Of course, these interpretations
would be ignoring the large Poisson uncertainties.  However, they would
also be ignoring a strong sample bias, which must be accounted for
prior to making a statistical analysis.

Note from Table~1 of \citet{gould10} that the underlying sample
has a median and 68\% timescale range of $t_\e = 26^{+95}_{-16}\,$day.
The median and $1\,\sigma$ lower limit are rather typical of microlensing
survey samples, but there is a long tail toward long and very long events.
It is notable that three of the six planets were discovered in the
two longest events.  While the underlying population of $A_\max>200$
events is certainly directly proportional to the number of events
at all timescales, there is likely a bias toward long events 
with $A_\max > 200$ being detected in the surveys and then being recognized
in time to conduct follow-up observations over the peak.  This could
be taken into account by directly using the distribution of
the observed timescales combined with the planet sensitivity functions
presented by \citet{gould10} for the events of various timescales.
Note, for example, that the two least sensitive events (their Figure~4)
are among the shortest,
OGLE-2005-BLG-188 ($t_\e = 14\,$day) and
MOA-2008-BLG-105 ($t_\e = 10\,$day).  If, as an exercise, I remove these
two events, as ``effectively non-sensitive'', then the median rises
to $t_\e = 43\,$days, which is not at all typical of microlensing
samples.  Because $t_\e\equiv \theta_\e/\mu_\rel$, a systematic
bias toward long events favors slow proper motions and large Einstein
radii, and the latter requires large masses and/or nearby lenses, i.e., just
the ``naive results'' mentioned above.

While the selection bias is only directly caused by $t_\e$ (and not
the physical parameters that enter it), it would be of some interest
to know the lens masses and distances for the 8 non-planetary events.
As noted by \citet{gould10}, three of these events already have
such measurements:
OGLE-2007-BLG-050 \citep{ob07050},
OGLE-2007-BLG-224 \citep{ob07224}, and
OGLE-2008-BLG-279 \citep{ob08279}.
Of the remaining 5 events, only one has a proper motion measurement,
OGLE-2005-BLG-188 ($\mu_\rel\sim 4.5\pm 0.5\,\masyr$).


\subsection{OGLE-MOA-Wise Survey (2011-2014)}
\label{sec:omw}
\citet{shvartzvald16} monitored the richest $8\,{\rm deg}^2$ of microlensing
fields from the Wise observatory 1m telescope in Israel during 
the central portions of four seasons, 
2011-2014, and then analyzed the 224 events that were discovered by
OGLE and MOA and that occurred in this area.  The cadence of
the Wise observations was $\Gamma\sim 2\,{\rm hr}^{-1}$, compared to
$\Gamma\sim 4\,{\rm hr}^{-1}$ for MOA and 
$\Gamma\sim 4\,{\rm hr}^{-1}$ and $\Gamma\sim 1.3\,{\rm hr}^{-1}$
for OGLE.  The goal was to carry out an objective survey with
(weather permitting) roughly continuous coverage, i.e., similar to
the subsequent goal of KMTNet but on a smaller scale.  While the total
number of planets found in this survey (8) was more than an order of magnitude
smaller than will come from the 6-year KMT survey described above,
it has, in the present context, the significant advantage that its
median $t_0$ is about 6 years earlier, which should provide a similar
advantage as to when the lenses and sources can first be resolved
using current instruments.

Moreover, this sample has the important advantage that there are
first-epoch AO images for 5 of these 8 events
when the lenses and sources were still superposed, 3 using the 6.5-m Magellan 
adaptive optics system (MagAO; \citealt{magao1,magao2,magao3})
in a follow-up project to systematically characterize the planet hosts of this 
sample \citep{ob140676b,xie21} and two others using Keck
\citep{mb11293b,ob140124b}. [All 3 of the unobserved targets have good 
$\bpi_\e$  measurements, with one of these having a giant-star source, i.e.,
not susceptible to the excess-flux method.]\ \ 
These images may be useful to understand various issues when they
are compared to late-time AO imaging.  Thus, in addition to being an independent
study, it also may provide more general insights that could not be
obtained from events that lack such imaging.

\citet{shvartzvald16} carried out only semi-automated planet detection
for this sample and so did not vet these candidates at the same level
as did the references to Tables~\ref{tab:tab1} and \ref{tab:tab2}.
Nevertheless, other workers did carry out such analyses, including
the exclusion of systems that were not actually planetary.  I should
note that \citet{shvartzvald16} also presented all the non-planetary
(``binary'') candidates in their sample.  These could be the subject
of an additional study, but I do not include them here.
Below I present notes on these 8 events.

{\bf MOA-2011-BLG-293}:
\citet{mb11293b} observed this target on 13 May 2012 using Keck AO
and detected about twice as much $H$-band light as was expected from
the source.  They show that there is only a few percent probability
that this excess is due to a companion to the source or host, and
negligible probability that it is due to an ambient star.  They obtained
$M_\host = 0.86\pm 0.06\,M_\odot$,
$M_{\rm planet} = 1526\pm 95\,M_\oplus$, and
$D_L =  7.72\pm 0.44\,\kpc$.
Still it
would be of interest to probe the first two possibilities.  Because
the blend and source have equal brightness, they can be resolved
when separated by about 35 mas \citep{ob120950b}.  Hence, if the
blend is the host, they were already resolvable beginning 2019.  If the
blend is a companion to the host, this would also be true
unless the original separation were about 35 mas and opposite to the
direction of $\bmu_{\rel,\hel}$.  However, in this case, the blend would
have almost certainly been resolved by \citet{mb11293b}.
Thus, if Keck AO observations done ``today'' did not resolve the
source and blend, they would almost certainly be companions.  Such
an observation might rigorously exclude that the blend is a companion to the 
host, but if not, this could be done by taking another observation a few years
later.  If the proper motion derived from these two observations
``points back'' sufficiently close to the source, then a putative 
host companion would have had to have been so close to the host that
it would have created an additional bump in the light curve.
{\bf MOA-2011-BLG-322}:
\citet{mb11322} did not measure the offset between the blend and the source.
Because the blend is a very rare foreground star (with subsequent {\it Gaia}
parallax $\pi =0.53\pm 0.14\,\mas$), they considered that it might be the
host or a companion to the host, although they did not consider these
scenarios to be likely.
However, \citet{xie21} did measure this offset, finding
$\sim 750\,\mas$, which rules out the ``host = blend'' hypothesis.  If the 
blend is a companion to the host, it lies $1400\,\au$ from it in projection, 
i.e., with a period $\log (P/{\rm day}) \sim 7.5$.  Roughly 9\% of G dwarfs have
companions with such periods or larger \citep{dm91}, so this would not be 
unusual. \citet{xie21} also found that the target was consistent with
no excess $K$-band flux relative to that expected from the source, and
thereby placed an upper limit $M_\host<0.6\,M_\odot$, consistent with the
Bayesian expectations of Figure~5 from \citet{mb11322}.  It would be reasonable
to attempt Keck AO observations at $\Delta\theta\sim 72\,\mas$, but given
the crude $\rho$ measurements, one cannot say very precisely when that will
be.  Nevertheless, adopting the best fit $\mu_\rel=4.7\,\masyr$ would imply
2027.
{\bf OGLE-2011-BLG-0265}
has very precise parallax measurements, but these differ 
substantially between the two solutions, leading to different
host masses, $0.21\,M_\odot$ or $0.14\,M_\odot$, and distances
$4.4\,\kpc$ or $3.5\,\kpc$ \citep{ob110265}.  Hence, the system characteristics 
are approximately known, but could be improved by late-time AO observations.
However, because the host is a giant, $K_{S,0}\sim 14$ and the proper motion
is low ($\mu_\rel = 2.9\pm 0.3\,\masyr$), 
this would not be feasible prior to 30m class AO.

{\bf OGLE-2013-BLG-0341} 
has an excellent parallax measurement, which yields
a host mass, planet mass and system distance of
$M_\host = 0.145\pm 0.013\,M_\odot$, $M_{\rm planet} = 2.32\pm 0.27\,M_\oplus$, 
and $D_L=1.16\,0.09\kpc$ \citep{ob130341}.
The companion to the host, which (in 2013) lay projected at about 15 mas,
would be brighter than the host: $K_{\rm companion}= 18.1 $ and $K_\host= 18.6$.  
These should be compared to $K_S\sim 15.2$ for the source.  Thus, there
would be a net offset of $\Delta K = 2.7$.  Hence, the lens and (binary) source
(with $\mu_\rel =10.2\pm 0.8\,\masyr$)
could be separately resolved by Keck AO, either now or in a few years.
Assuming (as argued by \citealt{ob130341}) that the wide binary solution
is correct, the two components could probably be imaged by EELT AO.
{\bf OGLE-2013-BLG-0911} 
has a good parallax measurement, yielding
a host mass, planet mass and system distance of
$M_\host = 0.29\pm 0.08,M_\odot$, $M_{\rm planet} = 3080\pm 900\,M_\oplus$, and 
$D_L=3.2\pm 0.5\,\kpc$
\citep{ob130911}.  If correct, these would imply $K_\host= 19.9$,  
compared to $K_S\sim 17.5$ for the source, yielding
$\Delta K = 2.4$.  Thus, it would be difficult 
to resolve the host and blend with current instruments even in 2030, 
when they will be separated by $\Delta\theta\sim 40\,\mas$.  Hence,
it would appear best to wait for 30m class AO.
{\bf OGLE-2013-BLG-1721}:
\citet{xie21} measured the baseline object (source + lens) to be
$K=18.04 \pm 0.07$. She did not report this because the position
seemed to be displaced from the OGLE position by $\sim 89\,\mas$.
However, first, reanalysis by Subo Dong (2022, private communication)
shows that the offset is substantially smaller ($58\,\mas$), although
this is still larger than would be expected.  Second, my prediction
for the source flux, based on the event parameters reported by \citet{ob131721}
is $K_S= 18.62\pm 0.23$.  One possible explanation is that there is a fairly
large astrometric error of unknown origin, in which case there is 
$\sim 3\,\sigma$ detection of excess flux, i.e., 
$K_{\rm blend}= 19.00\pm 0.37$.  This would be
consistent with my estimate for the lens flux
$K_L = 20.3$, based on the median Bayesian prediction 
(also with large uncertainty) of \citet{ob131721}, i.e., 
$M_\host=0.46\,M_\odot$, $D_L=6.3\,\kpc$, and $A_K=0.29$. Another explanation
is that the excess light is due to a companion to the lens or source
that lies at a separation of $140\pm 45\,\mas$ and therefore would not be
resolved (for a substantial fraction of this $1\,\sigma$ range)
in the FWHM $\sim 165\,\mas$ seeing of the MagAO images.  Because
of this ambiguity, all we can really says is that
the flux measurement by \citet{xie21} is consistent with the 
Bayesian estimates, but does not further modify them.  The proper motion
is $\mu_\rel = 5.5\pm 1.1\,\masyr$, while the median predicted contrast
ratio in $K$ is 4.7, i.e., $\Delta K=1.7$.  Hence, it is plausible
that the source and lens might be separately resolved when they 
are separated by $\sim 70\,\mas$, i.e., in 2026, although this is
far from guaranteed.

{\bf OGLE-2014-BLG-0124} 
has an excellent parallax measurement from {\it Spitzer} \citep{yee15}, 
yielding a host mass, planet mass and system distance of
$M_\host = 0.71\pm 0.22\,M_\odot$, $M_{\rm planet} = 162\pm 51\,M_\oplus$, 
and $D_L=4.1\pm 0.6\,\kpc$.  Keck AO imaging by \citet{ob140124b}
detected excess light from the superposed source and host, and thereby
refined these measurements to
$M_\host = 0.90\pm 0.05\,M_\odot$, $M_{\rm planet} = 207\pm 13\,M_\oplus$, 
and $D_L=3.5\pm 0.2\,\kpc$.  As discussed in Section~\ref{sec:hostcomp},
single-epoch photometric mass measurements are subject to ambiguity because
the blend light could come from a more massive companion to the host.
This applies more strongly to excess-light mass measurements because
there is no simultaneous measurement of $\theta_\e$.  See \citet{mb08310b}
for an example.  In the present case,
the light-curve based $\theta_\e$ measurement has a 31\% error, which
is the primary source of error in the original $M_\host$ and $M_{\rm planet}$
measurements.  Hence, it remains possible, in principle, that the 
$0.9\,M_\odot$ star detected by Keck is a companion to the host, while the
actual host is, e.g., $0.5\,M_\odot$ and so does not contribute substantially
to the blend light.  Thus, it would still be useful to separately resolve
the source and the blend.  Based on the highly anti-correlated
measurements of $\pi_\e=0.146\pm 0.004$ and the projected velocity 
$\tilde v_\hel = 107\pm 3\,\kms$ \citep{ob140124}, we can expect
$\mu_{\rel,\hel} = \kappa M \pi_\e^2\tilde v_\hel/\au
= (3.8\pm 0.1)(M/M_\odot)\,\masyr$.  Thus, if the star
seen by Keck is truly the host, then $\mu_{\rel,\hel} = 3.4\pm 0.2\,\masyr$.
Because the contrast ratio inferred from \citet{ob140124b} is about 2:1,
the host and blend should be resolvable with Keck AO at 
$\Delta\theta\sim 40\,\mas$, i.e., in 2026.  If they are resolved,
it will lead to an even more precise mass measurement.  If they are not,
it will show that the blend is a companion to the host and thus requires
an additional observation with 30m class AO.
{\bf OGLE-2014-BLG-0676}:
\citet{ob140676b} measured the combined lens+blend light to be 
$K_{\rm base}=16.72\pm 0.04$ at Magellan and in $J_{\rm base}=17.73\pm 0.03$ at Keck.
Because the source is extremely faint ($K_S=19.7\pm0.2$), 
most of this light is due to the blend:
$K_B=16.79\pm 0.04$ and $J_B=17.76\pm 0.03$.  Their analysis showed
that the ``blend=host'' hypothesis is favored over other scenarios by 
more than 100:1 and derived a host mass, planet mass and system distance of
$M_\host = 0.73^{+0.14}_{-0.29}\,M_\odot$, $M_{\rm planet} = 1170^{+220}_{-458},M_\oplus$, 
and $D_L=2.7^{+0.8}_{-1.4}\,\kpc$.  They note that the large errors are due to 
the large error in $\theta_\e$ from the light curve analysis.  This could
be greatly improved by separately resolving the host and blend.  However,
given the high contrast ratio and relatively low proper motion,
$\mu_\rel = 4.3\pm 1.3\,\masyr$, this
may not be possible until 30m class AO is available.

In brief, the OGLE-MOA-Wise survey planet-mass measurements are mostly 
complete.  Of the 8 planet/host systems, 6 have mass determinations
that are based either on good parallax measurements or measurement of
excess of flux (or both).  The only two that lack such measurements
are MOA-2011-BLG-322, for which the imaging provided only upper limits
on the host mass $M_\host <0.6\,M_\oplus$ (corresponding to 
$M_{\rm planet} < 18\,M_{\rm Jup}$) and OGLE-2013-BLG-1721, for which 
the \citet{xie21} measurement was consistent with the Bayesian estimates
but did not further modify them.
Some of the other planets have
relatively large errors, many of which could be improved by additional
imaging in the future.  However, these errors are still small compared
to those of typical microlensing planets, which usually only have 
Bayesian estimates.  I show the masses of the hosts and planets in 
Figure~\ref{fig:o+u} (magenta).  For MOA-2011-BLG-322, I adopt
$M_\host = 0.4\pm 0.2\,M_\odot$,
$M_{\rm planet} = 3800\pm 1900\,M_\oplus$, and
$D_L =  7.6\pm 0.9\,\kpc$,
after considering the Bayesian analysis of \citet{mb11322} and the
flux limits of \citet{xie21}.  For OGLE-2013-BLG-1721, I adopt
$M_\host = 0.46^{+0.26}_{-0.23}\,M_\odot$,
$M_{\rm planet} = 204^{+111}_{-99} \,M_\oplus$, and
$D_L = 6.3^{+1.1}_{-1.6} \,\kpc$
from the Bayesian analysis of \citet{ob131721}.  In all cases, I symmetrize
the error bars for the plot.

It is striking that 5 of the 8 planets have hosts in the near disk,
$D_L\la 4\,\kpc$, a very similar proportion to the 4 out of 6 planets from 
the high-magnification sample,  However, while I pointed out that
the high-magnification sample could be subject to selection bias that
favors such nearby planets, I know of no such argument that could be
applied to the OGLE-MOA-Wise survey.  Moreover, none of the distances
of the 9 ``nearby'' $(D_L\la 4\,\kpc)$ planets from either survey
rely on Bayesian estimates.

Thus, Figure~\ref{fig:o+u} can be regarded as ``suggestive evidence''
that planets may be more common in the nearby disk than in the bulge
and the more distant disk.  Unfortunately, however, the sample is
too small to draw strong conclusions on this issue.

\subsection{MOA Survey (2007-2012)}
\label{sec:moa}

\citet{suzuki16} analyzed 22 ``clear planets'' that were objectively
found from the MOA survey during the six year period 2007-2012.
These are listed in their Table~2.  In addition, they found one companion,
in the event OGLE-2011-BLG-0950, that was ambiguous between a planetary
and binary solution, which was subsequently confirmed to be a binary
by Keck AO resolution \citep{ob110950}.
Thus, the sample contains 22 planets in total.  The main goal of that paper
was to estimate the distributions of $q$ and $s$.  Here I focus
on the prospects of measuring masses for individual events.

As discussed by \citet{suzuki16}, 7 events already had mass measurements
at the time of publication.
One of these, {\bf OGLE-2007-BLG-349}, overlaps the \citet{gould10} sample.
It already has a mass measurement, as discussed in Section~\ref{sec:ufun}.  
Another event, {\bf OGLE-2011-BLG-0265}, overlaps the \citet{shvartzvald16} 
sample. It also has a mass measurement, as discussed in Section~\ref{sec:omw}.

I recapitulate the remaining five.
{\bf MOA-2007-BLG-192} \citep{mb07192} has
$M_\host = 0.060^{+0.028}_{-0.021} \,M_\odot$,
$M_{\rm planet} = 3.3^{4.9}_{-1.6} \,M_\oplus$,
$D_L = 1.0\pm 0.4\,\kpc$,
from the light-curve based $M=\theta_\e/\kappa\pi_\e$ method.
{\bf MOA-2009-BLG-266} \citep{mb09266} has
$M_\host = 0.56\pm 0.09\,M_\odot$,
$M_{\rm planet} = 10.4\pm 1.7 \,M_\oplus$,
$D_L = 3.04\pm0.33 \,\kpc$,
from the light-curve based $M=\theta_\e/\kappa\pi_\e$ method.
{\bf MOA-2010-BLG-117} \citep{mb10117} has
$M_\host = 0.58\pm 0.11\,M_\odot$,
$M_{\rm planet} =172 \pm 32\,M_\oplus$,
$D_L = 6.9\pm 0.7\,\kpc$,
from the light-curve based $M=\theta_\e/\kappa\pi_\e$ method.
{\bf MOA-2010-BLG-328} \citep{mb10328} has 
$M_\host = 0.11\pm 0.01\,M_\odot$,
$M_{\rm planet} =9.2 \pm 2.2\,M_\oplus$,
$D_L = 0.81\pm 0.10\,\kpc$,
from the light-curve based $M=\theta_\e/\kappa\pi_\e$ method.
{\bf OGLE-2012-BLG-0950}:  In this case, the original excess-light based
determinations by \citet{ob120950}, which were summarized by
\citet{suzuki16}, have been superseded by \citet{ob120950b}, who
resolved the source and lens using {\it HST} and Keck AO.  
They found
$M_\host = 0.58\pm 0.04\,M_\odot$,
$M_{\rm planet} =39 \pm 8\,M_\oplus$,
$D_L = 2.19\pm 0.23\,\kpc$.

Note that the mass measurement of MOA-2007-BLG-192 has large uncertainties.
The host is plausibly a BD at 3 kpc.  If so, further refinement will have to
await 30m class AO.  For MOA-2010-BLG-328,
\citet{mb10328} caution that the unusually low mass and
distance could be due to xallarap being misconstrued as parallax, and they
point out that this concern could be addressed by future AO follow-up 
observations.
They report $\mu_\rel = 5.7\pm 0.7\,\masyr$ and $K_{S,0} = 16.89$
(implying $K_S=17.13$).  For their parallax-based solution, they predict
$K_\host = 19.42\pm 0.47$, i.e., $\Delta K = 2.29\pm 0.47$.  If the low
estimates of the mass and distance were due to xallarap, the host would
be brighter, so the contrast ratio would be more favorable.  
Hence, Keck AO observations in 2023, when $\Delta\theta= 74\pm 9\,\mas$ would
resolve the source and lens if the xallarap model were correct, thus
yielding revised mass and distance measurements.  If such observations failed
to resolve the source and lens separately, they would confirm the parallax
model.  Thus, this ambiguity can already be resolved.

Of the remaining 15 planetary events, only one has a  published
mass measurement:
{\bf MOA-2009-BLG-319} \citep{mb09319} was resolved with Keck AO by 
\citet{mb09319b}, who found
$M_\host = 0.52\pm 0.04\,M_\odot$,
$M_{\rm planet} =67 \pm 6\,M_\oplus$,
$D_L = 7.1\pm 0.7\,\kpc$.

Another event, {\bf MOA-2013-BLG-322} was observed using Magellan
by \citet{xie21}, who found only upper limits on the host flux.
As I discussed in Section~\ref{sec:omw}, it may be possible to
resolve the source and lens beginning about 2027, but this is far from
certain.

Three events are very unlikely to be resolved before 30m class AO is
available because they are too bright.
{\bf MOA-2010-BLG-028} and {\bf MOA-2012-BLG-006} both have giant-star
sources, while {\bf MOA-2010-BLG-353} has an unusually red sub-giant source,
which likely has $K_0\simeq 13.6$, i.e., nearly as bright as a clump
giant.

This leaves $22 - 8 - 1 - 3 = 10$ events whose prospects should be evaluated
more closely.  One of these 10, {\bf MOA-2012-BLG-355} appears from 
\citet{suzuki16} to have a proper-motion measurement, but this
measurement is not reported in that work, and there is no published analysis
of this event.  Hence, I ignore it here.  For each of the remaining 9 events,
I evaluate their predicted separation, $\Delta\theta$, in 2030 by adopting
(from the listed reference)
the best estimate of $\mu_\rel$ and assuming $\mu_{\rel,\hel} =\mu_\rel$,
and I evaluate the magnitude offset, $\Delta K$ between the host
and the source by adopting 
the most likely mass and distance according
to its published Bayesian analysis.  These 9 events are listed in
order of my rough estimate of how easy they would be to resolve before 2030.
If the Bayesian estimates are approximately correct, then
Keck AO observations prior to 2030 of the first two of these events should be 
straightforward, the last four will be difficult, and the
remaining three may be feasible.  Thus of the 22 ``clear planets''
from \citet{suzuki16}, about 12 will plausibly have mass and distance
measurements before the advent of 30m class AO, i.e., 6 with $\pi_\e$
measurements, two that were resolved by Keck AO, and perhaps 4 more
that will be accessible to 8m-class AO before 2030.
As anticipated, this is a far larger fraction than can be expected from the 
KMT sample by that date (see Section~\ref{sec:practical}).  However, given that
the parallax subsample is heavily biased toward nearby and/or low-mass
lenses (because $\pi_\e^2\propto \pi_\rel/M$), it would be difficult to
draw firm conclusions from such a partial list of planet mass and distance
measurements.  

Hence, apart from the tentative conclusions that might be drawn from the
relatively small high-magnification (6) and OGLE-MOA-Wise (8) surveys, 
which already
have mostly complete mass measurements, comprehensive catalogs of microlensing
planets with masses and distances will have to await 30m class AO.

\subsection{{\it Roman} Microlensing Survey (2027-2032)?}
\label{sec:roman}
The {\it Nancy Grace Roman Telescope (Roman)}  is currently scheduled for 
launch ``no earlier'' than 2026.  During its nominal 5 year mission,
it will carry out a series of dedicated microlensing campaigns, primarily
using a broad $H$-band filter, with a fraction of observations taken
in a different band to measure source colors.
Currently, the plan is for 6 such campaigns, each lasting 72 days.
See \citet{roman1} and \citet{roman2} for predictions of its yield
in bound planets and free-floating planets, respectively.

{\it Roman} is expected to measure host masses and distance for a much
larger fraction of the bound planets that it detects than ground-based
surveys.  There are three principal reasons for this enhanced anticipated
performance.  

First, the source stars will be fainter (a natural consequence
of a much deeper survey over a much smaller area, $\sim 2\,{\rm deg}^2$),
while the population of host-star lenses will be similar.  This means
that the excess flux due to the host will be measurable in a substantially
larger fraction of cases.  Under the assumption that the host accounts
for this excess flux, its color and magnitude will, by themselves,
significantly constrain the host mass and distance.  In many cases (see below),
there will also be $\theta_\e$ measurements, which (under the same assumption)
will substantially improve these mass and distance determinations.

Second, during the lifetime of the mission, which plausibly could
be extended to 10 years, the image from the combined source-host light
will become extended as they gradually separate (or, for planets found
near the end of the mission, will become less extended).  For cases that
this can be measured, it will give at least the direction of $\bmu_\rel$
(possibly up to a sign ambiguity).
Provided that the lens and host colors differ sufficiently, it could
give the magnitude of $\bmu_\rel$ as well (in which case the ambiguity would
also be resolved).  See \citet{ob03235b}.  Then, one would obtain 
$\theta_\e = \mu_\rel t_\e$ even for the cases for which this quantity
was not measured from the event.  

Third, {\it Roman}'s high photometric stability and long-duration observing
campaigns will enable 1-D parallax measurements ($\pi_{\e,\parallel}$).  
When these are combined
with the direction of $\bmu_\rel$ (same as direction of $\bpi_\e$),
these yield $\bpi_\e$, which can be combined with $\theta_\e$ to yield
$M=\theta_\e/\kappa\pi_\e$ and $\pi_\rel = \theta_\e\pi_\e$.  See \citet{gould14}.

However, there will be many bound planets for which the host mass and
distance cannot be determined from {\it Roman} observations alone, and
ambiguities will remain about a substantial fraction of those that are
``basically measured''.  

First, $\theta_\e$ will not be measured from the light curve for a 
substantial fraction planetary events.  As mentioned in 
Section~\ref{sec:geocent}, \citet{Zhu:2014}  predicted that half of
all planetary events from a ``KMT-like'' survey would lack caustic
crossings, and this estimate was confirmed by \citet{2018subpr} for the
only complete-year sample (i.e., 2018) of KMT planets published to date.  
It was further confirmed by \citet{kb171194} for the 4-year complete sample
of low-$q$ KMT planets.  Some events
without caustic crossings nevertheless yield $\theta_\e$ measurements,
primarily when the source crosses a ridge that extends from a cusp.
Nevertheless, for the KMT sample, about 1/3 still lacked $\theta_\e$
measurements.  This provides our best estimate at present for the fraction
of {\it Roman} bound planets that will lack $\theta_\e$ measurements
from the light curve.

Second, from Figure~12 of \citet{roman1}, $\sim 1/3$ of all planets will have
flux ratios $f_B/f_S<0.1$.  Unambiguous interpretation of such signals
will be extremely difficult, in part because the source-flux parameter
may not be measured with substantially greater precision and in part
because the surface density of ambient stars increases at faint magnitudes.

Third, while many events will have 1-D parallax measurements, this
is far from universal.  See \citet{gouldzhu}.

Fourth, all mass determinations that are based on excess flux are subject
to spurious host identifications due to brighter stellar companions to the
lens and/or companions to the source.  In individual cases, various
arguments can be made to exclude, or at least greatly restrict, these
possibilities, but in many other cases, the only counter-arguments are
statistical.  For example, if a G dwarf is superposed on the source,
the chance that it has some fainter companion is about 70\%.  The
chance that this companion is what gave rise to the microlensing event
scales as the square root of the mass ratio $Q$, so perhaps
$\sqrt{Q}/(1+\sqrt{Q}\sim 40\%$ on average.
If the projected separation is too high, the astrometric offset between
the source and baseline object can be detected.  This limit will vary
by event, but I use 10 mas here for illustration.  If the companion is
too close, then it would have given rise to microlensing effects during the 
event.  Again this will vary but I use $2\theta_\e\sim 1\,\mas$ for 
illustration.  This leaves 2 decades in separation, so 3 decades in period,
or about half of the period distribution for G dwarfs.  In this example,
there would be a $\sim 0.6\times 0.4\times 0.5= 12\%$
chance that the event was actually due to an unseen
companion.  There could be additional arguments based on measurements
of $\theta_\e$ and $\pi_{\e,\parallel}$.  Nevertheless, incorrect identifications
do occur, as demonstrated by the case of MOA-2008-BLG-310 \citep{mb08310b}.

In brief, of order 1/3 of hosts will not generate sufficient excess flux to 
enable mass measurements, and these will be overwhelmingly concentrated among
low-mass hosts.  In addition, a substantial fraction of mass determinations
that rely on excess flux will have a residual uncertainty (although relatively
small) that the excess light is not due to the host.  Late time 
30m class AO observations can clarify both classes of events.

For simplicity, I identify $f_B/f_S<0.1$ as the limit beyond which the mass
cannot be determined from an excess-flux measurement.  Then, from
Figure~12 of \citet{roman1}, about half of these cases will have host-source
flux offsets in the range $2.5<\Delta H < 3.75$ and almost all the rest will
have $3.75<\Delta H < 5$.  As mentioned above, for both classes, about
2/3 will have $\mu_\rel$ measurements, and I will initially restrict 
attention to these.  The first (brighter) class of lenses probably will
require separations of about 2 FWHM, i.e.,  $\sim 30\,\mas$ for $K$-band
observations with EELT.  Hence, they require wait times of
$\Delta t \sim 5\,{\rm yr}(\mu_\rel/6\,\masyr)^{-1}$, where $\mu_\rel$ is the
measured proper motion.  That is, some measurements could begin about
4 years after the start of the mission, 
and a large fraction could be completed within
8 years of the end of the mission.  The second (fainter) class of lenses
might require separations of up to 5 FWHM.  At present, it is difficult
to estimate this limit in the absence of empirical evidence about instrument
performance.  However, based on this estimate, the wait times would
be up to 2.5 times longer.

For the events without $\theta_\e$ measurements, one might adopt
a moderately conservative lower limit, $\mu_\rel\ga 3\,\masyr$, in which
case the wait time would be about 10 years.

The cases with well-detected excess flux will generally have more
favorable contrast ratios, and can mostly be observed at separations 
of 0.5--1.3 FWHM.  Probably, it will be sufficient to spot check these
events, to determine the scale of the problem of false host identifications.
Because the wait times will typically be short, it should be possible
to gain an understanding of the scale of this problem quickly.

\section{Discussion: Pont du Gard}
\label{sec:discuss}

With the advent of 30m-class telescopes, it will ``suddenly''
become possible to make mass measurements of the great majority of
microlensing planets only 5--10 years after their discovery.
Coincidentally, first AO light on these telescopes is expected
about a decade after the discovery of about 150 planets from the
first six full seasons of the KMTNet experiment.  By contrast, using
the $\sim 4$ times smaller current-generation telescopes, one would have
to wait 20-40 years, i.e., until roughly 2050 to obtain a relatively
complete set of planet-mass measurements.

The bulk of the present paper is motivated by and built around the
prospect of exploiting this coincidence to ``rapidly'' obtain a measurement
of the planet mass function over this large parameter space.  A striking
feature of this prospect is that it rests on two completely different
types of scientific initiative, i.e., the massive 8-yr observational
and data-analysis effort of KMTNet (including commissioning year and
Covid-19-induced 2020 semi-hiatus) 
and the construction of massive next-generation
telescopes with advanced instrumentation.  
In terms of capital costs, ELTs are larger by 100:1,
i.e., \$1 billion versus \$10 million.  However, in terms
of amortized capital plus operations, the sign of the imbalance is reversed:
1:10.  That is,
estimating that 100 hours of observations are required (out of 40,000
hours of an initial 20-year telescope ``lifetime'') and approximating
the operational and amortization costs as equal, the comparison is
\$50 million for the KMT project versus \$5 million for the AO followup.

The very high capital costs and the modest amortization/operations costs 
of the AO follow-up (compared to the those of the KMTNet project) inspires
me to ask: how, and under what conditions, can this planet mass function
measurement be made if ELTs are not built.  It is always useful to 
consider such questions, if only to help understand the different
possible paths to achieving a given scientific goal.  However, under
present conditions, it is quite possible that public support
for science, and astronomy in particular, may radically decline, perhaps
for an extended period.  This may seem unthinkable to scientists immersed
in our work, but in my own locale, which is undertaking the largest and
nearest-to-completion of these ELTs, there are active plans
(as of August 2022) to ration heating
and to close factories, and there is wide discussion of possible food 
shortages.  Under such conditions, 30m-class telescopes may come to 
appear as an unaffordable luxury.  Neither does the overall trajectory
of events necessarily portend improvement.

The reason that this issue is particularly worthy of investigation in the
present context is that, as outlined above, 90\% of the scientific
effort (measured in dollars) has already been carried out, and would
be difficult to duplicate in the future if support for science drops
dramatically.  Hence, the conditions under which the ``other 10\%''
can be completed in such adverse circumstances should be considered.

The first-level answer to this question has already been given above:
if ELTs are seriously delayed or canceled, one can use present-day 
telescopes and instruments to carry out the necessary observations,
which can mostly be completed by 2050.  This would be disappointing
to me personally, but it would not raise any scientific questions
that needed to be addressed in the current paper.

However, the same adverse circumstances that led to ELT cancellations
might also lead to abandonment of current 8m-class telescopes,
or at least to failure to maintain their AO capabilities, which
are relatively expensive.

Hence, the real question is, at what point would the fruit of the
original KMTNet investment be ``lost'', so that the experiment would
have to repeated, which, as I have indicated, would be a daunting
prospect.

The main ``loss mechanism'' would be if the lenses and sources
separated sufficiently that they dissolved into the field, so that
they could not be reliably identified relative to random field stars.
For example, after a century, their typical separations would be
about 600 mas, an area that in typical field contains a dozen or
more faint dwarf stars.  At first sight, it seems ``simple'' to
identify the lens and source: they will be the only pair of stars
whose relative proper motion points back to zero, within $\la 1\,$mas, at $t_0$.
Depending on the quality of instruments, such relative proper motions
could be measured by a second epoch taken 10\% later, e.g., a decade
in this illustrative example.

However, in the great majority of cases, either the source or the host
(or both) will have a binary companion, mostly undetected, which
means that the instantaneous proper motion will not point exactly back
to zero.  In most cases, this effect will not be critically important
after only a century.  For example, for a host in a face-on orbit, the
internal proper motion will induce an offset after time $\Delta t$,
\begin{equation}
\Delta\theta_{\rm int}= \mu_{\rm int}\Delta t = 8.2\,\mas\,
\bigg({Q^3 M/(1+Q)^2\over M_\odot/36}\biggr)^{1/3}
\bigg({P\over 100\,\rm yr}\biggr)^{-1/3}
\bigg({D_L\over 5\,\kpc}\biggr)^{-1}
\bigg({\Delta t\over 100\,\rm yr}\biggr),
\label{eqn:muint}
\end{equation}
where I have normalized to a $Q\equiv M_{\rm companion}/M_\host=0.5$ companion
of an $M_\host = 0.5\,M_\odot$ host star.  Hence, in the fiducial case,
the proper motion would still point back to ``zero'' within 10 mas, which
would likely be quite adequate.

The fiducial parameters in Equation~(\ref{eqn:muint}) give a good indication
of the general scale of the problem.  Naively, it would seem that
8 times shorter periods would increase $\Delta\theta_{\rm int}$ by a factor
of 2.  However, two measurements separated by the illustrative 10-year offset
would differ by far less than would be inferred from the instantaneous
internal velocity used in the formula.  The formula shows that the
problem declines slowly with increasing period, so that quite
a broad range of periods, for companions of either the host or the source,
would cause problems of a similar scale.

However, for much larger wait times, e.g., $\Delta t\sim 1000\,$yr,
the problems of identifying the sources and lenses would become more
challenging.  Thus, I conclude that late-time AO could be comfortably
delayed by up to a century or so.

At that point, the sources and hosts would have separated sufficiently
that 3m-class AO would be adequate to carry out the observations.
This reduced requirement could be important if the resources available 
to astronomy were still strapped.

Another issue related to such a possible long hiatus of astronomical
work is data archiving.  While we have no difficulty reading Mayan
or Sumerian tablets, or even parchment and papyrus texts of the 
{\it Dead Sea Scrolls}, 
it is already challenging to read magnetic tapes of the 1980s.
Up until a few years ago, astronomy papers were routinely archived in
paper form.  However, according to my understanding, this is no longer
the case.  While the general problems of scientific archiving are
well beyond the scope of this paper, I suggest that it would be prudent to at 
least print out and carefully store all microlensing planet-discovery papers,
i.e., match the prudence of the archivists of the {\it Scrolls}.


\acknowledgments 
I thank S.\ Dong, J. Kollmeier, M.\ Pinsonneault, Y.\ Shvartzvald, and 
J.C.\ Yee for valuable discussions.
%
%
%
%
%

%






\begin{deluxetable}{lrrrrrrrrrrrrrl}                                  
\tablecolumns{15} \rotate \tablewidth{0pc}                            
\tablecaption{\textsc {Events with $\mu_\rel$ Measurements}}          
\tablehead{\colhead{Name} &                                           
\colhead{$t_0$} &                                                     
\colhead{$\mu_\rel$} &                                                
\colhead{$I_0$} &                                                     
\colhead{(V-I)$_0$} &                                                 
\colhead{$A_K$} &                                                     
\colhead{$l/b$} &                                                     
\colhead{$\log t_\e$} &                                               
\colhead{$\log u_0$} &                                                
\colhead{$\log\rho$} &                                                
\colhead{$\log q$} &                                                  
\colhead{$v_{\oplus,\perp}$} &                                        
\colhead{$M_{\rm cr}$} &                                              
\colhead{codes}  &                                                    
\colhead{Reference} \cr                                               
}                                                                     
\startdata                                                            
MB16227 &7518&{\rm  4.9}&18.54& 0.78& 0.12&$+3.30$& 1.23&$-1.08$&$-2.52$&$-2.03$&$ 1.7$& 0.08&11000&\citet{mb16227}\\
KB160622& &0.2& & & & $-3.24$& 0.00& 0.00& 0.01& 0.01&$ 21.4$\\
MB16319 &7553&{\bf  4.5}&17.49& 0.77& 0.20&$+0.35$& 0.93&$-0.57$&$-2.00$&$-2.41$&$ 0.1$& 0.02&11000&\citet{mb16319}\\
KB161816& &0.0& & & & $-2.17$& 0.01& 0.02& 0.00& 0.01&$ 29.1$\\
OB160596&7487&{\rm  5.1}&18.55& 0.83& 0.41&$-1.01$& 1.91&$-1.95$&$-3.22$&$-1.93$&$ 3.5$& 0.92&01000&\citet{ob160596}\\
KB161677& &0.8& & & & $-2.03$& 0.01& 0.01& 0.06& 0.01&$ 10.0$\\
OB160613&7494&{\rm  5.6}&20.81& 0.85& 0.27&$+1.99$& 1.87&$-1.68$&$-3.66$&$-2.49$&$ 2.1$& 1.04&14020&\citet{ob160613}\\
KB160017& &1.1& & & & $-1.74$& 0.01& 0.02& 0.04& 0.03&$ 12.9$\\
OB160693&7498&{\rm  1.5}&19.99& 1.35& 0.34&$+5.55$& 2.20&$-1.49$&$-3.02$&$-1.21$&$-0.4$& 1.47&13000&\citet{ob160693}\\
KB161248& &0.2& & & & $+2.22$& 0.05& 0.08& 0.07& 0.07&$ 15.1$\\
OB161067&7564&{\bf  2.1}&17.92& 0.74& 0.11&$+4.66$& 1.42&$-0.33$&$-2.26$&$-2.84$&$ 0.6$& 0.11&11001&\citet{ob161067}\\
KB161453& &0.0& & & & $-4.25$& 0.02& 0.03& 0.00& 0.02&$ 29.3$\\
OB161093&7560&{\rm  1.8}&18.55& 0.84& 0.21&$-2.11$& 1.75&$-1.68$&$-2.73$&$-2.84$&$-0.2$& 0.43&11001&\citet{ob161093}\\
KB161345& &0.3& & & & $-3.85$& 0.01& 0.02& 0.07& 0.03&$ 29.3$\\
OB161190&7582&{\rm  1.9}&19.35& 0.77& 0.29&$+2.63$& 1.98&$-1.76$&$-3.04$&$-1.85$&$-0.8$& 1.23&11001&\citet{ob161190}\\
KB160113& &0.2& & & & $-1.84$& 0.00& 0.01& 0.02& 0.01&$ 27.3$\\
OB161195&7569&{\rm  9.2}&17.83& 0.69& 0.24&$ 0.00$& 1.00&$-1.28$&$-2.49$&$-4.32$&$-0.8$& 0.07&11001&\citet{ob161195a}\\
KB160372& &0.8& & & & $-2.47$& 0.00& 0.01& 0.03& 0.05&$ 28.9$& & & \citet{ob161195b}\\
OB161227&7562&{\rm  0.8}&14.07& 1.38& 0.40&$-4.47$& 1.66&$-1.18$&$-1.04$&$-2.10$&$-1.4$& 0.13&11200&\citet{ob161227}\\
KB161089& &0.1& & & & $-1.94$& 0.08& 0.08& 0.08& 0.16&$ 29.2$\\
KB160212&7463&{\rm  8.1}&19.09& 0.77& 0.27&$+0.79$& 1.42&$-0.48$&$-2.92$&$-1.43$&$ 3.0$& 0.05&13800&\citet{kb160212}\\
        & &2.5& & & & $-1.60$& 0.01& 0.02& 0.08& 0.07&$ -2.2$\\
KB161107&7509&{\rm  2.6}&14.14& 1.38& 0.37&$+2.49$& 1.31&$-0.03$&$-1.23$&$-1.44$&$ 0.5$& 0.06&14000&\citet{kb161107}\\
        & &0.4& & & & $+1.50$& 0.02& 0.03& 0.05& 0.07&$ 20.9$\\
KB162364&7601&{\bf  3.2}&19.26& 0.91& 0.52&$+0.96$& 1.31&$-1.55$&$-2.52$&$-2.12$&$-2.0$& 0.08&01000&\citet{kb162397}\\
        & &0.0& & & & $+1.30$& 0.02& 0.03& 0.00& 0.04&$ 21.8$\\
KB162605&7566&{\rm 12.3}&17.57& 1.00& 0.39&$+3.22$& 0.53&$-1.31$&$-1.92$&$-1.92$&$ 1.6$& 0.01&12000&\citet{kb162605}\\
        & &1.0& & & & $-1.60$& 0.02& 0.04& 0.03& 0.04&$ 13.8$\\
OB170173&7838&{\rm  6.5}&14.98& 1.19& 0.43&$+0.41$& 1.48&$-0.06$&$-2.00$&$-4.61$&$ 3.0$& 0.07&14800&\citet{ob170173}\\
KB171707& &0.4& & & & $-1.35$& 0.01& 0.02& 0.02& 0.04&$  3.0$\\
OB170373&7841&{\rm  8.8}&17.97& 0.76& 0.22&$-1.31$& 1.08&$-0.38$&$-2.53$&$-2.81$&$ 4.4$& 0.02&13800&\citet{ob170373}\\
KB171529& &1.0& & & & $-3.71$& 0.01& 0.01& 0.03& 0.10&$  3.7$\\
OB170406&7909&{\rm  5.8}&15.69& 1.02& 0.25&$+0.36$& 1.57&$-2.03$&$-2.23$&$-3.16$&$ 0.7$& 0.59&11001&\citet{ob170406}\\
KB170243& &0.1& & & & $-2.42$& 0.00& 0.00& 0.00& 0.01&$ 28.2$\\
OB171049&7907&{\rm  6.7}&17.14& 0.75& 0.45&$+2.95$& 1.46&$-0.74$&$-2.62$&$-2.02$&$ 0.5$& 0.40&01000&\citet{ob171049}\\
KB170370& &1.4& & & & $-1.46$& 0.00& 0.00& 0.01& 0.02&$ 27.8$\\
OB171099&7917&{\rm  5.9}&19.91& 1.03& 0.44&$-1.68$& 1.28&$-2.40$&$-2.81$&$-2.19$&$-1.1$& 0.16&01000&\citet{ob190299}\\
KB172336& &2.5& & & & $+1.46$& 0.04& 0.11& 0.05& 0.05&$ 29.3$\\
OB171140&7941&{\rm  4.1}&15.09& 1.02& 0.36&$+4.00$& 1.17&$-0.62$&$-1.57$&$-2.14$&$-0.8$& 0.07&11001&\citet{ob171140}\\
KB171018& &0.6& & & & $-1.93$& 0.00& 0.01& 0.04& 0.04&$ 28.0$\\
OB171375&7970&{\rm  3.6}&19.80& 0.70& 0.29&$+0.04$& 2.01&$-1.55$&$-1.44$&$-1.83$&$-2.5$& 2.10&01000&\citet{kb162397}\\
KB170078& &0.5& & & & $-2.76$& 0.02& 0.03& 0.04& 0.02&$ 21.4$\\
OB171434&7985&{\rm  8.1}&18.45& 0.73& 0.23&$-0.28$& 1.80&$-1.37$&$-3.33$&$-4.24$&$-3.2$& 1.30&11000&\citet{ob171434}\\
KB170016& &0.5& & & & $-2.07$& 0.01& 0.01& 0.01& 0.01&$ 15.2$\\
OB171522&7972&{\rm  3.2}&20.61& 1.15& 0.20&$+2.51$& 0.88&$-1.27$&$-2.22$&$-1.80$&$-1.8$& 0.01&01000&\citet{ob171522}\\
KB170460& &0.5& & & & $-2.18$& 0.02& 0.03& 0.05& 0.04&$ 21.1$\\
OB171691&8003&{\rm  5.7}&17.40& 0.71& 0.43&$-1.60$& 1.28&$-1.31$&$-2.45$&$-4.01$&$-3.3$& 0.03&12300&\citet{4subjovi}\\
KB170752& &1.0& & & & $+1.90$& 0.02& 0.02& 0.06& 0.15&$  5.0$\\
OB171806&8024&{\rm  1.7}&18.98& 0.83& 0.30&$+4.09$& 1.82&$-1.59$&$-2.74$&$-4.35$&$-0.4$& 0.09&01200&\citet{kb171194}\\
KB171021& &0.6& & & & $+2.66$& 0.03& 0.03& 0.15& 0.17&$  4.6$\\
KB170165&7854&{\rm  6.9}&19.25& 1.01& 0.32&$+2.14$& 1.62&$-1.46$&$-3.11$&$-2.87$&$ 2.3$& 0.32&11000&\citet{kb170165}\\
        & &1.1& & & & $-2.04$& 0.01& 0.01& 0.06& 0.03&$ 10.1$\\
KB170673&7973&{\rm 15.7}&14.15& 1.10& 0.45&$-4.88$& 1.35&$-0.92$&$-2.12$&$-2.25$&$-4.8$& 0.39&01000&\citet{kb190414}\\
        & &5.1& & & & $-1.73$& 0.01& 0.02& 0.13& 0.10&$ 18.6$\\
KB171003&7873&{\rm  2.6}&17.60& 0.67& 0.24&$+3.42$& 1.41&$-0.75$&$-2.28$&$-4.37$&$-0.2$& 0.09&11200&\citet{kb171194}\\
        & &0.6& & & & $+3.15$& 0.01& 0.01& 0.10& 0.14&$ 19.8$\\
KB171038&7993&{\bf  2.1}&19.00& 0.79& 0.35&$+3.13$& 1.34&$-0.76$&$-2.40$&$-2.28$&$-1.1$& 0.03&11000&\citet{kb171038}\\
        & &0.0& & & & $+2.06$& 0.01& 0.01& 0.00& 0.02&$ 10.9$\\
KB171194&7943&{\bf  1.3}&19.29& 0.71& 0.14&$+6.63$& 1.67&$-0.59$&$-2.59$&$-4.58$&$ 0.6$& 0.21&11000&\citet{kb171194}\\
        & &0.0& & & & $-4.34$& 0.02& 0.03& 0.00& 0.06&$ 28.6$\\
KB172509&7872&{\rm  1.3}&20.18& 1.14& 0.44&$+1.85$& 1.83&$-1.18$&$-2.72$&$-2.36$&$ 0.5$& 0.30&01000&\citet{ob190299}\\
        & &0.6& & & & $+1.99$& 0.03& 0.05& 0.07& 0.05&$ 19.7$\\
OB180298&8189&{\rm  4.8}&16.72& 0.77& 0.57&$-1.63$& 1.51&$-1.67$&$-2.45$&$-3.71$&$ 2.1$& 0.04&01400&\citet{2018subpr}\\
KB181354& &1.0& & & & $+1.16$& 0.01& 0.04& 0.09& 0.10&$ -2.7$\\
OB180383&8199&{\rm  3.1}&16.56& 1.15& 0.27&$+1.19$& 1.05&$-1.15$&$-1.62$&$-3.67$&$ 2.8$& 0.00&01000&\citet{ob180383}\\
KB180900& &0.3& & & & $-1.61$& 0.01& 0.01& 0.04& 0.07&$  0.6$\\
OB180506&8224&{\bf  6.5}&15.10& 1.18& 0.58&$-2.01$& 1.38&$-1.06$&$-1.92$&$-4.09$&$ 3.7$& 0.13&11000&\citet{kb190253}\\
KB180835& &0.0& & & & $-2.45$& 0.01& 0.01& 0.00& 0.12&$ 13.4$\\
OB180532&8220&{\rm  3.3}&21.30& 1.73& 0.13&$+1.54$& 2.14&$-2.08$&$-3.55$&$-4.01$&$ 2.7$& 1.72&11300&\citet{ob180532}\\
KB181161& &0.2& & & & $-2.73$& 0.03& 0.03& 0.04& 0.05&$ 10.2$\\
OB180567&8245&{\rm  3.1}&15.33& 0.98& 0.38&$+1.99$& 1.39&$-0.13$&$-1.75$&$-2.91$&$ 1.5$& 0.10&01000&\citet{ob180567}\\
KB180890& &0.5& & & & $-1.49$& 0.02& 0.02& 0.02& 0.02&$ 21.1$\\
OB180596&8277&{\rm  4.1}&15.30& 1.14& 0.29&$+0.96$& 1.46&$-0.55$&$-1.87$&$-3.74$&$ 0.5$& 0.25&11301&\citet{ob180596}\\
KB180945& &0.4& & & & $-2.13$& 0.00& 0.01& 0.02& 0.03&$ 28.6$\\
OB180677&8230&{\bf  3.4}&17.73& 0.59& 0.22&$-1.61$& 0.69&$-0.99$&$-1.31$&$-4.05$&$ 3.2$& 0.00&15000&\citet{ob180677}\\
KB180816& &0.0& & & & $-3.31$& 0.01& 0.01& 0.00& 0.19&$ 15.3$\\
OB180740&8254&{\rm  7.7}&21.53& 0.87& 0.11&$+1.74$& 1.77&$-1.40$&$-3.65$&$-2.34$&$ 2.4$& 1.61&11200&\citet{ob180740}\\
KB181822& &0.9& & & & $-4.80$& 0.03& 0.03& 0.05& 0.04&$ 23.2$\\
OB180799&8295&{\rm  1.8}&15.90& 1.00& 0.20&$+6.12$& 1.44&$-0.38$&$-1.71$&$-2.58$&$ 0.7$& 0.10&11001&\citet{ob180799}\\
KB181741& &0.2& & & & $-3.73$& 0.01& 0.01& 0.05& 0.03&$ 29.3$\\
OB180932&8301&{\rm  6.2}&14.75& 1.05& 0.29&$+0.81$& 1.43&$-0.07$&$-1.93$&$-2.92$&$-0.9$& 0.33&11001&\citet{2018prime}\\
KB182087& &0.4& & & & $-1.50$& 0.00& 0.00& 0.01& 0.03&$ 28.7$\\
OB180962&8263&{\rm  5.6}&19.37& 0.87& 0.28&$-2.11$& 1.46&$-0.68$&$-2.94$&$-2.62$&$ 1.7$& 0.32&01000&\citet{ob180567}\\
KB182071& &0.9& & & & $-3.04$& 0.00& 0.07& 0.02& 0.01&$ 26.4$\\
OB180977&8277&{\rm 11.5}&18.55& 0.66& 0.24&$-0.49$& 1.31&$-0.83$&$-2.72$&$-4.38$&$ 0.5$& 0.36&11000&\citet{kb190253}\\
KB180728& &2.4& & & & $-2.42$& 0.02& 0.03& 0.09& 0.04&$ 28.6$\\
OB181011&8285&{\rm  2.8}&17.33& 0.77& 0.27&$+1.04$& 1.09&$-1.28$&$-1.92$&$-2.01$&$ 0.1$& 0.03&01010&\citet{ob181011}\\
KB182122& &0.2& & & & $-2.04$& 0.01& 0.01& 0.06& 0.01&$ 29.2$\\
OB181011&8285&{\rm  2.8}&17.33& 0.77& 0.27&$+1.04$& 1.09&$-1.28$&$-1.92$&$-1.82$&$ 0.1$& 0.03&01010&\citet{ob181011}\\
KB182122& &0.2& & & & $-2.04$& 0.01& 0.01& 0.06& 0.02&$ 29.2$\\
OB181185&8311&{\rm  4.8}&18.12& 0.68& 0.28&$+2.47$& 1.20&$-2.16$&$-2.46$&$-4.16$&$-0.8$& 0.09&11001&\citet{ob181185}\\
KB181024& &0.4& & & & $-2.00$& 0.00& 0.00& 0.01& 0.01&$ 27.7$\\
OB181269&8344&{\rm  8.3}&17.44& 0.64& 0.26&$+2.61$& 1.85&$-0.84$&$-3.23$&$-3.24$&$-1.7$& 1.98&11000&\citet{ob181269}\\
KB182418& &0.6& & & & $-1.82$& 0.01& 0.08& 0.01& 0.02&$ 18.3$\\
OB181428&8340&{\rm  5.6}&15.76& 0.87& 0.44&$+1.99$& 1.38&$-0.15$&$-2.14$&$-2.77$&$-1.7$& 0.16&01000&\citet{ob181428}\\
KB180423& &0.4& & & & $+2.11$& 0.03& 0.00& 0.01& 0.01&$ 18.5$\\
OB181647&8374&{\rm  0.6}&19.44& 0.85& 0.17&$-1.35$& 1.72&$-0.96$&$-2.29$&$-2.00$&$-4.3$& 0.03&11000&\citet{2018prime}\\
KB182060& &0.1& & & & $-3.37$& 0.02& 0.03& 0.09& 0.03&$  4.7$\\
KB180029&8295&{\rm  3.3}&18.45& 0.78& 0.48&$-0.09$& 2.24&$-1.57$&$-3.35$&$-4.74$&$-1.5$& 7.50&11001&\citet{kb180029}\\
        & &0.5& & & & $+1.95$& 0.04& 0.05& 0.07& 0.05&$ 29.0$\\
KB180087&8282&{\bf  7.0}&13.96& 1.37& 0.43&$-0.03$& 0.66&$-0.28$&$-0.96$&$-2.68$&$-1.0$& 0.01&11200&\citet{2018subpr}\\
        & &0.0& & & & $+2.14$& 0.01& 0.03& 0.00& 0.09&$ 29.2$\\
KB180247&8308&{\rm  8.8}&18.01& 0.52& 0.43&$+0.66$& 1.03&$-1.19$&$-2.60$&$-2.15$&$-1.7$& 0.07&01000&\citet{2018subpr}\\
        & &1.2& & & & $+2.33$& 0.02& 0.02& 0.05& 0.03&$ 27.2$\\
KB180748&8372&{\rm  9.2}&17.27& 0.79& 0.39&$-0.81$& 0.64&$-1.47$&$-1.96$&$-2.69$&$-3.8$& 0.00&11300&\citet{kb180748}\\
        & &0.8& & & & $-1.97$& 0.01& 0.03& 0.04& 0.03&$  5.0$\\
KB181025&8275&{\bf  4.3}&19.33& 1.11& 0.25&$+2.46$& 0.98&$-2.15$&$-2.26$&$-4.08$&$ 0.6$& 0.03&13500&\citet{kb181025}\\
        & &0.0& & & & $-2.08$& 0.01& 0.01& 0.00& 0.14&$ 28.2$\\
KB181292&8407&{\rm 10.8}&13.99& 1.81& 0.74&$-5.23$& 1.76&$-0.54$&$-2.13$&$-2.45$&$-3.9$& 1.42&11000&\citet{kb181292}\\
        & &1.9& & & & $-0.28$& 0.03& 0.01& 0.06& 0.03&$-14.8$\\
KB181743&8249&{\rm  2.3}&19.60& 0.70& 0.12&$+5.69$& 1.45&$-0.60$&$-2.68$&$-2.92$&$ 1.7$& 0.10&13830&\citet{kb181743}\\
        & &0.5& & & & $-4.79$& 0.02& 0.89& 0.08& 0.08&$ 20.9$\\
KB181990&8230&{\rm  7.1}&16.79& 0.60& 0.35&$+6.77$& 1.66&$-1.37$&$-2.86$&$-2.45$&$-0.7$& 0.61&01400&\citet{kb181990}\\
        & &0.9& & & & $+1.91$& 0.02& 0.02& 0.05& 0.04&$ 15.8$\\
OB190299&8560&{\bf  1.7}&19.45& 0.89& 0.25&$+4.70$& 1.47&$-1.25$&$-2.46$&$-2.00$&$ 0.1$& 0.00&11000&\citet{ob190299}\\
KB192735& &0.0& & & & $+2.57$& 0.02& 0.02& 0.00& 0.03&$ -0.6$\\
OB190362&8564&{\rm  6.8}&16.99& 0.72& 0.43&$+2.11$& 1.35&$-1.01$&$-2.46$&$-2.13$&$ 0.7$& 0.03&01000&\citet{ob190362}\\
KB190075& &1.3& & & & $+4.41$& 0.01& 0.02& 0.01& 0.09&$  2.8$\\
OB190468&8494&{\rm  4.4}&19.43& 0.83& 0.36&$+3.83$& 1.88&$-1.92$&$-3.28$&$-2.45$&$ 0.1$& 1.05&01010&\citet{ob190468}\\
KB192696& &0.6& & & & $+2.34$& 0.02& 0.02& 0.03& 0.02&$ 16.0$\\
OB190468&8494&{\rm  4.4}&19.43& 0.83& 0.36&$+3.83$& 1.88&$-1.92$&$-3.28$&$-1.97$&$ 0.1$& 1.05&01010&\citet{ob190468}\\
KB192696& &0.6& & & & $+2.34$& 0.02& 0.02& 0.03& 0.02&$ 16.0$\\
OB190960&8686&{\rm 11.3}&18.85& 0.92& 0.11&$+6.10$& 1.79&$-2.21$&$-3.48$&$-4.83$&$ 0.2$& 2.95&11201&\citet{ob190960}\\
KB191591& &0.8& & & & $-4.30$& 0.01& 0.01& 0.02& 0.04&$ 26.5$\\
OB191053&8691&{\rm  3.9}&17.61& 0.70& 0.22&$+3.06$& 1.54&$-0.46$&$-2.68$&$-4.91$&$-1.1$& 0.30&01000&\citet{ob191053}\\
KB191504& &0.4& & & & $-2.05$& 0.01& 0.01& 0.05& 0.05&$ 24.6$\\
KB190371&8592&{\rm  7.7}&17.62& 0.64& 0.30&$-1.37$& 0.81&$-0.85$&$-2.19$&$-1.10$&$ 3.5$& 0.01&15400&\citet{kb190371}\\
        & &0.7& & & & $+2.81$& 0.01& 0.01& 0.01& 0.01&$ 14.4$\\
KB190842&8626&{\rm  8.0}&20.03& 1.01& 0.21&$+0.11$& 1.64&$-2.18$&$-3.37$&$-4.39$&$ 1.3$& 1.02&11000&\citet{kb190842}\\
        & &1.8& & & & $-2.02$& 0.01& 0.01& 0.10& 0.13&$ 25.9$\\
KB191042&8637&{\rm 10.8}&19.99& 1.03& 0.20&$+3.02$& 1.05&$-1.11$&$-2.89$&$-3.20$&$ 0.8$& 0.10&01000&\citet{af2}\\
        & &2.1& & & & $-2.46$& 0.01& 0.02& 0.07& 0.04&$ 27.7$\\
KB191715&8697&{\rm  7.6}&19.62& 0.79& 0.15&$+1.90$& 1.64&$-1.25$&$-3.36$&$-2.40$&$-1.7$& 0.86&11040&\citet{kb191715}\\
        & &0.6& & & & $-2.91$& 0.00& 0.01& 0.01& 0.02&$ 22.9$\\
KB191953&8702&{\rm  5.7}&18.39& 0.63& 0.29&$+1.85$& 1.21&$-2.63$&$-2.62$&$-2.71$&$-1.9$& 0.08&02400&\citet{kb191953}\\
        & &0.5& & & & $-1.67$& 0.01& 0.05& 0.01& 0.21&$ 20.8$\\
KB210119&9306&{\bf  2.1}&18.92& 0.77& 0.05&$+2.57$& 1.73&$-1.10$&$-2.74$&$-2.03$&$ 3.4$& 0.08&11000&\citet{kb210119}\\
        & &0.0& & & & $-6.16$& 0.01& 0.02& 0.00& 0.03&$  3.6$\\
KB210171&9326&{\rm  7.6}&17.32& 0.87& 0.21&$+0.27$& 1.62&$-2.25$&$-2.83$&$-4.32$&$ 2.9$& 0.54&11000&\citet{kb210171}\\
        & &1.1& & & & $-2.71$& 0.00& 0.01& 0.05& 0.08&$ 15.5$\\
KB210192&9316&{\rm  4.9}&17.68& 0.58& 0.29&$-0.16$& 1.49&$-2.00$&$-2.70$&$-3.43$&$ 3.0$& 0.14&01000&\citet{kb210119}\\
        & &0.6& & & & $-1.82$& 0.01& 0.04& 0.04& 0.04&$ 11.2$\\
KB210240&9313&{\rm  3.1}&17.44& 0.68& 0.49&$-0.39$& 1.63&$-2.52$&$-2.56$&$-3.19$&$ 3.1$& 0.16&13010&\citet{kb210240}\\
        & &0.4& & & & $-1.43$& 0.02& 0.03& 0.05& 0.07&$ 10.2$\\
KB210240&9313&{\rm  3.1}&17.44& 0.68& 0.49&$-0.39$& 1.63&$-2.52$&$-2.56$&$-2.74$&$ 3.1$& 0.16&13010&\citet{kb210240}\\
        & &0.4& & & & $-1.43$& 0.02& 0.03& 0.05& 0.12&$ 10.2$\\
KB210712&9350&{\rm  2.2}&20.40& 0.69& 0.17&$-0.66$& 2.00&$-0.84$&$-3.37$&$-3.25$&$ 2.2$& 1.38&01000&\citet{kb210712}\\
        & &0.3& & & & $-3.29$& 0.01& 0.01& 0.06& 0.08&$ 24.1$\\
KB210748&9345&{\rm  2.9}&19.19& 0.51& 0.22&$+5.44$& 1.60&$-0.39$&$-2.82$&$-2.92$&$ 1.2$& 0.26&11000&\citet{kb211391}\\
        & &1.1& & & & $-3.03$& 0.06& 0.04& 0.15& 0.16&$ 21.4$\\
KB210909&9354&{\rm  8.2}&17.26& 0.81& 0.56&$+0.75$& 1.21&$-1.22$&$-2.49$&$-2.50$&$ 0.2$& 0.15&01000&\citet{kb210712}\\
        & &1.1& & & & $+1.27$& 0.02& 0.03& 0.04& 0.06&$ 26.0$\\
KB210912&9417&{\rm  3.8}&17.33& 1.19& 0.24&$-1.14$& 1.83&$-0.21$&$-2.57$&$-4.98$&$-2.2$& 1.16&13800&\citet{kb210912}\\
        & &1.0& & & & $-4.09$& 0.02& 0.02& 0.10& 0.19&$ 25.6$\\
KB211077&9378&{\rm  1.8}&18.56& 0.82& 0.26&$-4.15$& 1.40&$-1.97$&$-2.26$&$-2.81$&$-0.2$& 0.09&01010&\citet{kb211077}\\
        & &0.1& & & & $-2.61$& 0.01& 0.02& 0.03& 0.03&$ 29.2$\\
KB211077&9378&{\rm  1.8}&18.56& 0.82& 0.26&$-4.15$& 1.40&$-1.97$&$-2.26$&$-2.76$&$-0.2$& 0.09&01010&\citet{kb211077}\\
        & &0.1& & & & $-2.61$& 0.01& 0.02& 0.03& 0.04&$ 29.2$\\
KB211253&9374&{\rm 15.2}&19.10& 0.80& 0.64&$+0.26$& 0.98&$-2.25$&$-2.89$&$-2.31$&$ 0.1$& 0.10&15000&\citet{kb211391}\\
        & &2.4& & & & $-1.08$& 0.04& 0.04& 0.05& 0.05&$ 28.8$\\
KB211303&9385&{\rm  6.1}&19.81& 0.85& 0.14&$-0.03$& 1.40&$-1.66$&$-3.02$&$-3.19$&$ 0.6$& 0.29&01000&\citet{4subjovi}\\
        & &0.7& & & & $-3.03$& 0.01& 0.01& 0.04& 0.03&$ 29.2$\\
KB211391&9385&{\rm  5.3}&20.96& 0.88& 0.15&$+2.66$& 1.50&$-1.92$&$-3.27$&$-4.44$&$ 0.2$& 0.40&11300&\citet{kb211391}\\
        & &0.6& & & & $-2.80$& 0.02& 0.04& 0.03& 0.03&$ 29.3$\\
KB211554&9395&{\rm  7.3}&18.65& 0.87& 0.42&$-1.89$& 0.71&$-1.29$&$-2.15$&$-2.83$&$-1.1$& 0.01&11000&\citet{4subjovi}\\
        & &1.5& & & & $-2.54$& 0.06& 0.09& 0.08& 0.16&$ 28.9$\\
KB211689&9409&{\rm  6.1}&19.77& 1.18& 0.21&$+0.37$& 1.35&$-2.22$&$-2.84$&$-3.68$&$-1.4$& 0.22&15000&\citet{kb210171}\\
        & &0.8& & & & $-2.99$& 0.02& 0.02& 0.02& 0.08&$ 27.2$\\
KB211898&9422&{\rm  5.1}&17.59& 0.75& 0.51&$+1.00$& 1.35&$-1.80$&$-2.51$&$-2.84$&$-1.9$& 0.15&11030&\citet{kb211898}\\
        & &0.5& & & & $+1.35$& 0.04& 0.05& 0.05& 0.09&$ 23.3$\\
KB212294&9453&{\rm  8.7}&19.42& 0.88& 0.17&$+1.91$& 0.85&$-2.22$&$-2.52$&$-3.25$&$-2.4$& 0.02&15000&\citet{kb210119}\\
        & &1.0& & & & $-2.60$& 0.02& 0.07& 0.14& 0.03&$ 13.1$\\
\hline                                                                
\enddata                                                              
\tablecomments{$M_{\rm cr}$ is in units of $0.075\,M_\odot$.          
Boldface $\mu_\rel$ indicates a lower limit, in which case            
$M_{\rm cr}$ is also a lower limit and $\rho$ is an upper limit.}     
\label{tab:tab1}                                                      
\end{deluxetable}

\begin{deluxetable}{lrrrrrrrrrrl}                                     
\tablecolumns{12} \rotate \tablewidth{0pc}                            
\tablecaption{\textsc {Events without $\mu_\rel$ Measurements}}       
\tablehead{\colhead{Name} &                                           
\colhead{$t_0$} &                                                     
\colhead{$I_0$} &                                                     
\colhead{(V-I)$_0$} &                                                 
\colhead{$A_K$} &                                                     
\colhead{$l/b$} &                                                     
\colhead{$\log t_\e$} &                                               
\colhead{$\log u_0$} &                                                
\colhead{$\log q$} &                                                  
\colhead{$v_{\oplus,\perp}$} &                                        
\colhead{codes}  &                                                    
\colhead{Reference} \cr                                               
}                                                                     
\startdata                                                            
OB160263&7470&15.82& 0.99& 0.00&$-0.95$& 1.21&$-0.24$&$-1.51$&$ 4.4$&15000&\citet{ob160263}\\
KB161515& & & & $-4.06$& 0.03& 0.02& 0.01&$  0.9$\\
KB161836&7488&20.24& 0.82& 0.24&$-0.12$& 1.73&$-1.25$&$-2.35$&$ 3.1$&01000&\citet{kb161836}\\
        & & & & $-1.95$& 0.02& 0.03& 0.08&$ 10.3$\\
KB162397&7550&19.96& 0.82& 0.20&$+4.81$& 1.77&$-1.39$&$-2.40$&$-0.8$&01000&\citet{kb162397}\\
        & & & & $+3.14$& 0.06& 0.06& 0.10&$ 29.1$\\
KB170428&7944&19.40& 1.06& 0.15&$+2.59$& 1.65&$-0.69$&$-4.30$&$-0.5$&01000&\citet{kb171194}\\
        & & & & $-3.55$& 0.01& 0.02& 0.07&$ 28.1$\\
OB170482&7874&18.02& 0.37& 0.28&$-0.20$& 1.60&$-1.29$&$-3.87$&$ 2.7$&01000&\citet{ob170482}\\
KB170084& & & & $-2.80$& 0.02& 0.01& 0.06&$ 19.1$\\
KB171146&7925&19.01& 0.77& 0.18&$-2.44$& 1.41&$-0.54$&$-2.70$&$-0.2$&11000&\citet{kb171038}\\
        & & & & $-4.14$& 0.02& 0.06& 0.08&$ 29.3$\\
OB180516&8228&17.58& 0.55& 0.26&$-0.57$& 1.40&$-0.98$&$-3.89$&$ 3.5$&01000&\citet{kb190253}\\
KB180808& & & & $-3.59$& 0.01& 0.01& 0.05&$ 14.1$\\
OB181119&8316&18.29& 0.70& 0.15&$-1.38$& 1.59&$-0.36$&$-2.74$&$-1.8$&01000&\citet{2018subpr}\\
KB181870& & & & $-4.43$& 0.04& 0.05& 0.11&$ 26.8$\\
OB181126&8298&19.48& 0.79& 0.26&$-1.53$& 1.72&$-2.08$&$-4.13$&$-0.9$&04300&\citet{2018prime}\\
KB182064& & & & $-2.88$& 0.03& 0.03& 0.28&$ 29.0$\\
OB181212&8394&17.63& 0.73& 0.13&$+2.72$& 1.71&$-1.89$&$-2.91$&$-2.5$&11000&\citet{2018prime}\\
KB182299& & & & $-3.17$& 0.00& 0.00& 0.02&$ -4.6$\\
OB181367&8358&18.03& 0.69& 0.14&$+1.29$& 1.36&$-1.59$&$-2.48$&$-2.7$&01000&\citet{2018prime}\\
KB180914& & & & $-2.64$& 0.00& 0.01& 0.02&$ 12.4$\\
KB180030&8272&13.93& 1.14& 0.46&$-0.11$& 1.45&$-0.05$&$-2.56$&$-0.5$&01000&\citet{2018subpr}\\
        & & & & $+1.88$& 0.00& 0.00& 0.05&$ 28.5$\\
KB181976&8183&16.99& 0.87& 0.19&$-5.80$& 1.62&$-0.84$&$-2.50$&$ 6.4$&01000&\citet{kb181976}\\
        & & & & $-3.48$& 0.01& 0.01& 0.13&$ -6.6$\\
KB181996&8348&15.95& 0.99& 0.53&$+6.30$& 1.67&$-1.74$&$-2.82$&$ 0.1$&01000&\citet{kb181976}\\
        & & & & $+1.38$& 0.01& 0.02& 0.11&$ 16.2$\\
KB182004&8239&17.98& 0.69& 0.23&$-0.30$& 1.50&$-0.63$&$-3.43$&$ 2.5$&01000&\citet{2018prime}\\
        & & & & $-2.23$& 0.01& 0.02& 0.11&$ 19.3$\\
KB182602&8270&15.88& 1.06& 0.32&$+6.43$& 1.99&$-0.29$&$-2.78$&$-0.7$&11000&\citet{2018subpr}\\
        & & & & $+2.83$& 0.06& 0.08& 0.07&$ 28.9$\\
KB182718&8355&21.10& 1.37& 0.21&$-0.28$& 2.21&$-1.23$&$-1.71$&$-3.3$&12000&\citet{2018prime}\\
        & & & & $-2.02$& 0.08& 0.08& 0.07&$ 13.0$\\
OB191492&8763&20.08& 0.70& 0.17&$+1.91$& 1.72&$-1.33$&$-3.74$&$-3.5$&01000&\citet{kb190253}\\
KB193004& & & & $-2.63$& 0.03& 0.04& 0.12&$ -7.8$\\
KB190253&8591&17.29& 0.67& 0.37&$+0.13$& 1.78&$-1.27$&$-4.39$&$ 2.6$&11000&\citet{kb190253}\\
        & & & & $-1.43$& 0.07& 0.03& 0.08&$ 13.8$\\
KB190414&8611&18.28& 0.70& 0.41&$+7.18$& 1.85&$-2.36$&$-2.23$&$-0.7$&14700&\citet{kb190414}\\
        & & & & $+1.71$& 0.05& 0.05& 0.26&$ 21.6$\\
KB190953&8638&16.68& 0.55& 0.27&$+1.54$& 1.31&$-0.83$&$-4.38$&$ 0.7$&01000&\citet{kb190253}\\
        & & & & $-2.08$& 0.02& 0.03& 0.04&$ 28.0$\\
KB191367&8668&20.71& 1.06& 0.11&$+1.93$& 1.59&$-1.08$&$-4.30$&$ 0.0$&01000&\citet{kb171194}\\
        & & & & $-4.99$& 0.04& 0.05& 0.12&$ 28.9$\\
KB191552&8715&18.30& 0.63& 0.28&$+2.61$& 2.04&$-0.67$&$-2.33$&$-1.8$&01000&\citet{af2}\\
        & & & & $-1.72$& 0.04& 0.04& 0.05&$ 16.0$\\
KB191806&8716&19.95& 0.76& 0.20&$+1.41$& 2.13&$-1.59$&$-4.71$&$-2.5$&01000&\citet{kb171194}\\
        & & & & $-3.35$& 0.03& 0.03& 0.12&$ 16.1$\\
KB192974&8753&18.00& 0.74& 0.50&$+0.33$& 1.44&$-0.78$&$-3.20$&$-3.1$&01000&\citet{af2}\\
        & & & & $-1.33$& 0.04& 0.04& 0.13&$ -3.1$\\
KB210320&9316&18.84& 0.70& 0.31&$+2.10$& 1.11&$-2.23$&$-3.54$&$ 3.4$&01000&\citet{4subjovi}\\
        & & & & $-4.30$& 0.01& 0.02& 0.07&$ 10.6$\\
KB211105&9376&18.61& 0.51& 0.39&$+2.61$& 1.54&$-0.96$&$-2.70$&$-0.7$&01000&\citet{kb210712}\\
        & & & & $+2.30$& 0.02& 0.03& 0.04&$ 29.1$\\
KB211372&9389&18.39& 0.75& 0.51&$-0.21$& 1.85&$-1.12$&$-3.36$&$-1.4$&01000&\citet{kb211391}\\
        & & & & $+1.85$& 0.05& 0.06& 0.08&$ 29.1$\\
KB212478&9482&18.97& 0.76& 0.24&$+1.16$& 1.53&$-1.00$&$-2.22$&$-2.9$&11000&\citet{kb210712}\\
        & & & & $-2.27$& 0.01& 0.01& 0.03&$ -1.7$\\
\hline                                                                
\enddata                                                              
\label{tab:tab2}                                                      
\end{deluxetable}

\begin{deluxetable}{llllll}                                           
\tablecolumns{6} 
\tablewidth{0pc}                                                      
\tablecaption{\textsc {Meaning and Distributions of 5 Codes}}         
\tablehead{\colhead{Entry} &                                          
\colhead{Notes?} &                                                    
\colhead{Stat.\ Sample?} &                                            
\colhead{$\log q$ Degen.} &                                           
\colhead{Multiplicity} &                                              
\colhead{Spitzer?} \cr                                                
}                                                                     
\startdata                                                            
\hline                                                                
 0&No$\,( 51)$ & &$<0.1\,( 92)$ &2L1S$\,(103)$ &No$\,(103)$ \\
 1&Yes$\,( 64)$ &Likely$\,( 92)$ & &3L1S (2 plan.)$\,(  8)$ &Yes$\,( 12)$ \\
 2& &Maybe$\,(  4)$ &0.1--0.15$\,(  6)$ &3L1S (bin+plan.)$\,(  1)$ & \\
 3& &Needs AO$\,(  8)$ &0.15--0.2$\,(  6)$ &2L2S$\,(  2)$ & \\
 4& &Unlikely$\,(  5)$ &0.2--0.25$\,(  4)$ &3L2S (bin+plan.)$\,(  1)$ & \\
 5& &No$\,(  6)$ &0.25--0.3$\,(  1)$ & & \\
 6& & &0.3--0.35$\,(  0)$ & & \\
 7& & &0.35--0.4$\,(  1)$ & & \\
 8& & &$>0.4\,(  5)$ & & \\
\enddata                                                              
\label{tab:tab3}                                                      
\end{deluxetable}

\begin{deluxetable}{lrrl}                                   
\tablecolumns{4} \tablewidth{0pc}                            
\tablecaption{\textsc {$\Delta\theta$ (2030) and $\Delta K$ for MOA (2007-12) Events}} 
\tablehead{\colhead{Name} &                                           
\colhead{$\Delta\theta$} &                                                     
\colhead{$\Delta K$} &                                                
\colhead{Reference} \cr                                               
}                                                                     
\startdata    
MB10477 & 209&$ 0.0$& \cite{mb10477}\\
OB08379 & 163&$-0.4$& \cite{mb08379}\\ 
OB07368 &  78&$ 1.9$& \cite{ob07368}\\
MB11262 & 215&$ 4.8$& \cite{mb11262}\\ 
OB120563&  58&$-0.9$& \cite{ob120563}\\ 
MB12505 &  77&$ 3.7$& \cite{mb12505}\\  
MB09387 &  78&$ 3.8$& \cite{mb09387}\\
OB08355 &  67&$ 3.2$& \cite{ob08355}\\ 
MB11291 &  39&$ 3.3$& \cite{mb11291}\\ 
\hline                                                                
\enddata                                                              
\label{tab:moa}                                                      
\end{deluxetable}  

\end{document}